\documentclass[aps,pra,reprint,twocolumn,superscriptaddress,showpacs,showkeys,superscriptaddress,longbibliography]{revtex4-1}
\usepackage{amsmath,amssymb,amstext}
\usepackage[usenames,dvipsnames]{color}
\usepackage{graphicx}
\usepackage{bm,bbold,braket}
\usepackage{natbib}
\usepackage{txfonts, comment,stmaryrd}
\usepackage{dcolumn}
\usepackage{color}
\usepackage{xcolor}
\colorlet{rn}{red}
\colorlet{an}{blue}
\usepackage[english]{babel}
\usepackage{lipsum}
\newcommand{\dketstate}[1]{|\mathcal{D}_{#1}\rangle}
\newcommand{\dbrastate}[1]{\langle\mathcal{D}_{#1}|}
\newcommand{\state}[3]{\ket{{#1}_1{#2}_e{#3}_{\mathrm{ph}}}}
\newcommand{\statebra}[3]{\bra{{#1}_1{#2}_e{#3}_{\mathrm{ph}}}}

\usepackage[colorlinks,bookmarks=false,citecolor=blue,linkcolor=red,urlcolor=blue]{hyperref}
\begin{document}

\title{Cavity polariton blockade for non-local entangling gates with trapped atoms}

%\title{Cavity Polariton Blockade for W-state Preparation, $CZ$, $C_2Z$ Gate via a Driven Cavity}

\author{Vineesha Srivastava}
\affiliation{University of Strasbourg and CNRS, CESQ and ISIS (UMR 7006), aQCess, 67000 Strasbourg, France}

\author{Sven Jandura}
\affiliation{University of Strasbourg and CNRS, CESQ and ISIS (UMR 7006), aQCess, 67000 Strasbourg, France}

\author{Gavin K. Brennen}
\affiliation{Center for Engineered Quantum Systems, School of Mathematical and Physical Sciences, Macquarie University, 2109 NSW, Australia}

\author{Guido Pupillo}
\affiliation{University of Strasbourg and CNRS, CESQ and ISIS (UMR 7006), aQCess, 67000 Strasbourg, France}

% \author{...}
% \affiliation{...}

\date{\today}

\begin{abstract}
We propose a scheme for realizing multi-qubit entangled W-state and non-local $CZ$ and $C_2Z$ gates via a cavity polariton blockade mechanism  with a system of atomic qubits coupled to a common cavity mode. The polariton blockade is achieved by tuning the system, an $N-$qubit register, such that no two atoms are simultaneously excited to the qubit excited state, and there is an effective coupling only between the ground state and a singly-excited W state of the qubit register. The control step requires only an external drive of the cavity mode and a global qubit pulse and no individual qubit addressing. We analytically obtain the state preparation error for an $N-$qubit W state which scales as $\sqrt{(1-1/N)}/\sqrt{C}$ where $C$ is the single particle cooperativity. We additionally show the application of the polariton blockade mechanism in realizing a non-local $CZ$ and $C_2Z$ gate by using a different set of computational qubit states, and characterize the gate errors which scale as $\sim 1/\sqrt{C}$. 
\end{abstract}



\maketitle
%\tableofcontents

\section{\label{sec:level1}Introduction}

In quantum computing, perfecting single qubit and two-qubit gates have driven significant progress in the NISQ era~\cite{preskillQuantumComputingNISQ2018,bhartiNoisyIntermediatescaleQuantum2022,kimEvidenceUtilityQuantum2023}, particularly in neutral atom systems~\cite{grahamMultiqubitEntanglementAlgorithms2022,omranGenerationManipulationSchrodinger2019,songGenerationMulticomponentAtomic2019}. Scaling qubit architectures while exploiting  native multi-qubit interactions beyond two qubits could offer new pathways for efficient quantum operations. For example, all-to-all connectivity in qubit architectures drastically reduces circuit depth of quantum circuits~\cite{holmesImpactQubitConnectivity2020} and could offer substantially lower overhead in quantum error correction algorithms to enable fault tolerance~\cite{bravyiHighthresholdLowoverheadFaulttolerant2024,chandraNonlocalResourcesError2024,gottesmanFaultTolerantQuantumComputation2014,LowoverheadFaulttolerantQuantum,pecorariHighrateQuantumLDPC2025,breuckmannQuantumLowDensityParityCheck2021}. In parallel, deterministic generation of multi-qubit entangled states transitioned from  fundamental tests in quantum mechanics, to applications in entanglement-enhanced quantum sensing~\cite{degenQuantumSensing2017, pezze2018quantum, srivastavaEntanglementenhancedQuantumSensing2024}, quantum algorithms for large-scale quantum computing~\cite{grahamMultiqubitEntanglementAlgorithms2022} and distributed quantum computing~\cite{ zhongDeterministicMultiqubitEntanglement2021}. In this regard, cavity QED systems provide a promising route to scalability by offering non-local connectivity between distant qubits, enabling multi-qubit interactions which are otherwise challenging in conventional architectures. 

%Non-local cavity-mediated quantum gates which have been previously proposed or realized with neutral atom spin qubits~\cite{pellizzari1995decoherence, yi2003conditional, borregaard2015heralded}, rely on local addressing of individual qubits, and sometimes involve atomic systems with complicated level structure. . In previous proposals of entanglement generation with cavity QED systems, several schemes are limited to the case of two-atom entanglement~\cite{zheng2000efficient, kastoryano2011dissipative}, and several theoretical proposals of realizing multi-particle entangled states~\cite{duan2003efficient, xiao2007generation} fail to provide a quantitative description of the effect of loss mechanisms- spontaneous emission and cavity photon loss.

In this spirit of exploiting non-local interactions offered by cavity QED setups, several protocols for entanglement generation and quantum gates have been proposed or realized with neutral atoms or ions by mediating interactions between qubits via a quantized bosonic mode, using motional modes of trapped ions~\cite{ciracQuantumComputationsCold1995,garcia-ripollSpeedOptimizedTwoQubit2003,molmerMultiparticleEntanglementHot1999,sackettExperimentalEntanglementFour2000} or optical cavity modes for neutral atom spin qubits~\cite{pellizzariDecoherenceContinuousObservation1995, beigeQuantumComputingUsing2000,borregaardHeraldedQuantumGates2015,lewalleMultiQubitQuantumGate2023,rametteAnyToAnyConnectedCavityMediated2022,sorensenMeasurementInducedEntanglement2003,zhengEfficientSchemeTwoAtom2000,zhengUnconventionalGeometricQuantum2004}. However, for neutral atom spin qubits, only a few of these proposals can be extended to multi-qubit operations~\cite{borregaardHeraldedQuantumGates2015, lewalleMultiQubitQuantumGate2023, duanEfficientEngineeringMultiatom2003}. To exploit both \textit{non-locality} and \textit{multi-qubit} interactions, we recently proposed two practical and deterministic protocols for realizing non-local multi-qubit quantum gate operations: a geometric phase gate and an adiabatic phase gate~\cite{janduraNonlocalMultiqubitQuantum2024}. Both protocols utilize interactions mediated by a common cavity mode by solely driving the cavity mode, requiring only a single control pulse on the cavity mode. These gates operate in distinct parameter regimes: the geometric phase gate works in the regime of a strong cavity drive and a strong detuning of the cavity mediated transition, while the adiabatic phase gate works in the regime of a weak cavity drive. Also recently, we have demonstrated the utility of the geometric phase gate, combined with optimal control techniques, for the preparation of useful probe states for quantum sensing. These states achieve a significant entanglement-enhanced advantage in quantum sensing beyond the standard quantum limit, even in the presence of noise~\cite{srivastavaEntanglementenhancedQuantumSensing2024}. 

In the pioneering work~\cite{barontiniDeterministicGenerationMultiparticle2015}, the authors introduced a deterministic protocol to generate multi-qubit entangled states by employing Quantum Zeno Dynamics (QZD)~\cite{beigeQuantumComputingUsing2000, pupilloScalableQuantumComputation2004}. This is based on nondestructive measurement~\cite{volzMeasurementInternalState2011} in a cavity QED setup with a single-mode cavity that couples to $N$ atomic qubits: Let each atomic qubit be comprised of computational states $|0\rangle$ and $|1\rangle$; All the atoms are initialized in the qubit excited state $|1\rangle$ and acted upon by a resonant pulse on the qubit transition resulting in a coherent evolution of the system which is combined with a simultaneous and continuous measurement performed by probing the cavity resonantly. The measurement is such that the cavity probe is resonant with the cavity mode when all the atoms are in the qubit state $|0\rangle$, and the measurement effectively probes the ground state of the qubit register $|\mathcal{D}_0\rangle = |0\rangle^{\otimes N}$. The measurement back-action on the state $|\mathcal{D}_0\rangle$ prevents it from being populated, due to Quantum Zeno Dynamics. Instead, states very similar to the so-called W state are prepared, with the latter denoted as $|\mathcal{D}_1\rangle = (1/\sqrt{N})(|10\dots 0\rangle + |010 \dots 0\rangle + \dots + |00\dots 1\rangle)$ in the following. These states are robust to particle loss and can be used as a resource for some tasks like distributed sensing~\cite{gottesmanLongerBaselineTelescopesUsing2012}. While the results of Ref.~\cite{barontiniDeterministicGenerationMultiparticle2015} constitute a significant breakthrough in the experimental manipulation of many-particle quantum states, %the practical use of W states in, e.g., quantum information remains limited. It 
it is an interesting open question whether the QZD scheme can be generalized to new protocols ensuring a high-fidelity of preparation of the desired multi-particle state. %as well as of other many-particle entangled states relevant for different applications in quantum technologies, including for quantum computing. 
In addition, it would be highly interesting both theoretically and experimentally whether QZD-like protocols could be devised that allow for performing deterministic multi-qubit quantum operations -- including full quantum gates -- of use for quantum computing and sensing.
Very recent breakthrough experiments with cold neutral atoms trapped in optical tweezers in fiber based Fabry-Perot optical cavity have demonstrated that realizing quantum gates and operations with high-fidelity is becoming possible thanks to the possibility to trap multiple atomic qubits inside an optical cavity in a regime of strong light-matter coupling~\cite{grinkemeyerErrorDetectedQuantumOperations2024}. These works open the way to non-local entanglement generation using the delocalized photon field and are possible key  components of future architectures for distributed quantum computing and sensing~\cite{mainDistributedQuantumComputing2025, rohde2025quantuminternettechnicalversion, Zhang_2021}. %[https://arxiv.org/abs/2501.12107, https://doi.org/10.1038/s41586-024-08404-x, Distributed Quantum Sensing
%Zheshen Zhang, Quntao Zhuang OR ANOTHER REVIEW]. 
%Although the protocol generates more multi-qubit entangled states by following slightly different but close trajectories, the prepared states are not very well characterized. 

% 
In this work, we propose a new protocol that generalizes the idea of QZD-based state preparation to account for the formation of mixed light-matter polariton states for qubits coupled to the cavity mode, and use it to demonstrate theoretically a viable pathway to generating multi-qubit entangled W states as well as two-qubit controlled-Z (CZ) and three-qubit C$_2$Z  quantum gates. Due to coupling to the delocalized cavity mode, the latter gates can be non-local. %Inspired by recent experimental results in trapping multiple atoms inside an optical cavity in the strong coupling regime~\cite{grinkemeyerErrorDetectedQuantumOperations2024}, we present the protocol with the setup of neutral atom spin qubits coupled to a common cavity mode, with an external cavity drive and a global qubit drive. 
Our protocol relies on a new cavity polariton blockade mechanism for multiqubit entanglement generation, which impedes the formation of polariton modes with more than one excitation, due to strong measurement induced excitation blockade. The latter is a combination of Quantum Zeno Dynamics and energy detuning of a selectively probed cavity polariton state from the coherent global qubit drive. Interestingly, the protocol only requires global drives of the cavity for generating a multi-particle entangled state and of the cavity and of a single global laser on the qubit transition to drive the two-qubit and three-qubit quantum gates. %We present a detailed analysis of the state preparation protocol for a multi-qubit entangled W state as well as for protocol to implement non-local CZ and C$_2$Z gates. %Further in contrast to the work in Ref.~\cite{barontiniDeterministicGenerationMultiparticle2015}, 
We present a full quantum-mechanical treatment of the system dynamics and derive analytical expressions for the W state preparation error, as well as the $CZ$ and $C_2Z $ gate errors. These errors are evaluated while taking fully into account the relevant physical losses in experiments, arising from a finite cavity resonance linewidth $\kappa$ and atomic linewidth $\gamma$, using optimal values for drive strength ratios and detuning parameters. We assume coupling of atoms with the cavity mode with coupling strength $g$ in the strong coupling regime, such that the single-particle cooperativity $C = g^2/(\kappa\gamma) \gg 1$. To our knowledge, this is the first time that a full analytical treatment is carried out in the presence of losses for deterministic quantum gates -- with the exception of the protocols in Ref.~\cite{janduraNonlocalMultiqubitQuantum2024}. This work opens the way to the realization of multi-particle non-local entangled states generation and quantum gates based on a new polariton blockade mechanism. While we provide precise predictions for experiments with neutral atoms trapped in cavities, the present work can be relevant to other physical platforms, such as exciton polaritons in the solid state~\cite{Boulier2020, blochStronglyCorrelatedElectron2022, bellessaMaterialsExcitonsPolaritons2024}, depending on achievable light-matter couplings, intrinsic non-linearities and polariton lifetimes in those systems.  

The paper is organized as follows: In Sec.~\ref{sec::Model}, we introduce the system Hamiltonian for N atomic spin qubits coupled to a common cavity mode controlled with two laser drives -- one acting on the cavity mode and the other acting globally on the qubits. In Sec.~\ref{sec::eff-H} we first derive an effective Hamiltonian in the regime where the cavity drive and the losses (cavity decay and spontaneous emission) are treated perturbatively with respect to the cavity-qubit coupling. We then establish the \textit{cavity polariton blockade}, where a cavity polariton state -- an eigenstate of the atoms-cavity coupling Hamiltonian -- becomes resonant to the cavity drive at a specific detuning. The cavity drive then induces dressing of the states it couples to, and can no longer be treated perturbatively in the corresponding subspace formed by the dressed states (blockaded subspace). This creates an energy barrier (blockade-like) or energy leakage (QZD like) in the system depending on whether the strength of detuning of the global drive exceeds the loss rates. By treating the coupling from the global qubit drive between the blockaded subspace with other cavity polariton states perturbatively, one can suppress population in the blockaded subspace. Here, we choose the blockaded subspace such that simultaneous excitation of two qubits to the $|1\rangle$ state is suppressed. The resulting effective Hamiltonian, after tracing out the cavity mode in vacuum and in the absence of losses, is analogous to a driven two-level system with the logical states $\dketstate{0}$ (all qubits in $|0\rangle$) and the W state $\dketstate{1}$ (equal superposition of all states with one qubit in $|1\rangle$ state and rest in the $|0\rangle$ state), which we term as the effective blockade Hamiltonian.
In Sec.~\ref{sec::W_state}, we describe the W state preparation in the presence of losses and present an analytical expression for state preparation infidelity. We obtain the W state-preparation error scaling as $\sqrt{1-1/N}/\sqrt{C}$, hence saturating with the total number of atoms $N$. In Sec.~\ref{sec::C_gates}, we adapt the cavity polariton approach for the implementation of time-optimal $CZ$ and $C_2Z$ gates. By introducing a new computational state $|1'\rangle$, we use $|1\rangle$ state as an auxiliary state to realize a blockade-like interaction, enabling a CZ or C$_2$Z gate with the computational states $|0\rangle$ and $|1'\rangle$. We also semi-analytically present the $CZ$ and $C_2Z$ gate errors, which scale as $1/\sqrt{C}$. This scaling of errors with $C$ for our protocols is consistent with the expected error scaling for deterministic protocols~\cite{sorensenMeasurementInducedEntanglement2003}.
\section{Model}\label{sec::Model}
\begin{figure*}[t]
    \centering
    \includegraphics[width=\textwidth]{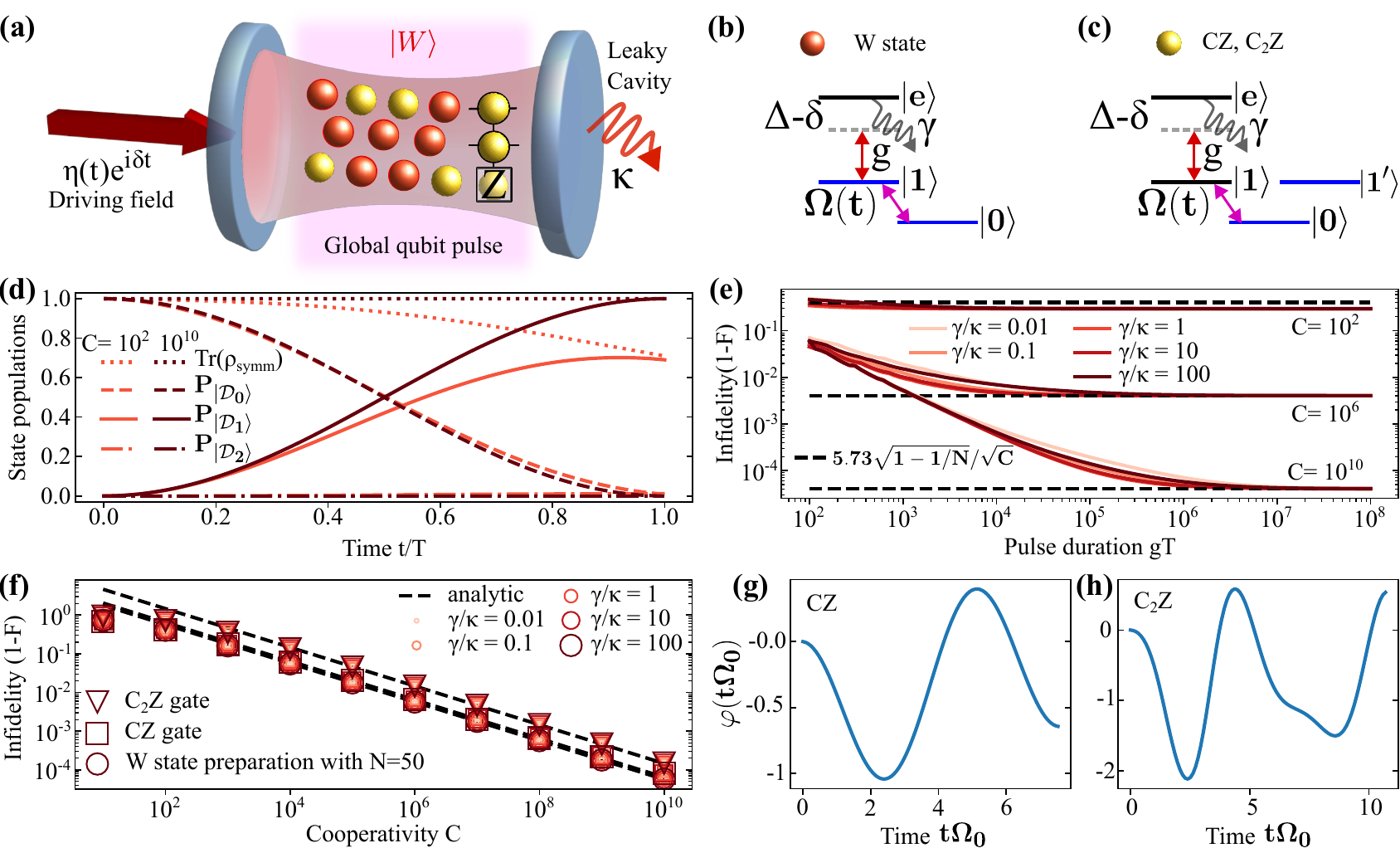}
    \caption{(a) Schematic of atoms trapped inside a cavity and coupled to a common cavity mode, which is externally driven by a classical field $\eta(t)$. An additional global pulse addresses all the qubits. A multi-qubit entangled W state can be prepared with arbitrarily selected atoms atoms(in red) modeled as three-level systems (as shown in (b)) and a $C_2Z$ gate can be implemented with atoms (in yellow) modeled as four-level systems(as shown in (c)). (b,c) Level schematic for atoms implementing W state preparation and $CZ$ or $ C_2Z$ gate. The $|1\rangle \leftrightarrow |e\rangle$ coupling is mediated by the cavity with coupling strength $g$. An additional (global) laser drive couples the states $|0\rangle$ and $|1\rangle$ with Rabi frequency $\Omega(t)$. The computational qubit states are highlighted in blue. (d) State population dynamics obtained numerically by simulating the dynamics under the full Hamiltonian in Eq.~\eqref{eq::full_H}, plotted for states $\dketstate{0}, \dketstate{1}$, and $\dketstate{2}$ denoted by $P_{\dketstate{0}}, P_{\dketstate{1}}$ and $P_{\dketstate{2}}$ respectively for a system with $N=2$. The populations $P_{\dketstate{0}}$ (dashed lines), $P_{\dketstate{1}}$ (solid lines), and $P_{\dketstate{2}}$ (dash-dot lines) at each time add up to the trace of the reduced atomic density matrix (dotted lines) $\mathrm{Tr}(\rho_{\mathrm{symm}}) \leq 1$  where $\rho_{\mathrm{symm}}$ corresponds to the  subspace spanned by states $\{\dketstate{n} \forall n =0,1,\dots N\}$. (e) Infidelity ($1-F$) as a function of the total pulse duration $gT$ for W state preparation with $N=2$ for $C= 10^2, 10^6, 10^{10}$ and $\gamma/\kappa= 0.01, 0.1, 1, 10, 100$. The infidelity converges to the analytical estimate(dashed lines) $5.73\sqrt{1-1/N}/\sqrt{C}$ (See text Sec.~\ref{sec:fid_calculation}) obtained in the limit $T \rightarrow \infty$. (f) Infidelity ($1-F$) as a function of single particle cooperativity for W state preparation with $N=50$, $CZ$ gate and $C_2Z$ gate. The dashed lines represent the analytically calculated errors, and numerical points obtained by simulating the dynamics with the full Hamiltonian (Eq.~\eqref{eq::full_H}) are plotted for $\gamma/\kappa = 0.01, 0.1, 1, 10, 100$ for a fixed pulse duration of $gT= 10^8$. (g,h) Time optimal pulses for implementing $CZ$ gate and $C_2Z$ gate from~\cite{janduraTimeOptimalTwoThreeQubit2022, everedHighfidelityParallelEntangling2023}}
    \label{fig::summary}
\end{figure*}
In this section we describe our system model and Hamiltonian. We first describe the various system components and parameters, and write the Hamiltonian of the system in the laboratory frame. Next we apply a rotating wave approximation and further split the Hamiltonian intro three components set by different energy scales which becomes relevant for the derivation of the effective blockade Hamiltonian using perturbation theory in Sec.~\ref{sec::eff-H}. In Sec.~\ref{subsec::cav_polaritons}, we introduce the cavity polariton states in terms of a convenient symmetric basis for the system Hamiltonian. 

We consider a system of $N$ atoms coupled to an optical cavity which supports a single mode with frequency $\omega_c$ as shown in Fig.~\ref{fig::summary}(a). Each atom is modeled as a three-level system [Fig.~\ref{fig::summary}(b)] with two computational qubit states $|0\rangle$ and $|1\rangle$, and an excited state $|e\rangle$ with finite lifetime $1/\gamma$. We define the energies of the states $|0\rangle$, $|1\rangle$ and $|e\rangle$ as $\omega_0$, $\omega_1$ and $\omega_e$ respectively ($\hbar= 1$). The cavity mode creation and annihilation operators are $\hat a^{\dagger}$ and $\hat a$ respectively, and the cavity excitation has a finite lifetime $1/\kappa$. The atomic levels $|1\rangle$ and $|e\rangle$ are coupled via the cavity mode with coupling strength $g$. An external cavity probe- a drive laser with frequency $\omega_L$ drives the cavity mode with amplitude $\eta(t)$. In addition, there is a free-space coupling between the states $|0\rangle$ and $|1\rangle$ with Rabi-frequency $\Omega (t)$, which can be realized by a global laser pulse on the qubits with frequency $\omega_{\mathrm{gl}}$, given by $\Omega(t)\cos(\omega_{\mathrm{gl}}t)|1\rangle \langle 0| + \mathrm{h.c}$. We define $\hat n_{s}= \sum_{j=1}^{N} |s_j\rangle \langle s_j|$ which denotes the number operator for atoms in state $|s\rangle$ for $s\in \{0,1,e\}$ and $j$ denotes the atom-index. 

The full Hamiltonian 
$\hat H_{\mathrm{full}}$ reads
\begin{align}
\begin{split}
    \hat H_{\mathrm{full}} &= \omega_0 \hat n_0 + \omega_1 \hat n_1 + (\omega_e - i\gamma/2)\hat n_e + (\omega_c - i\kappa/2)\hat a^{\dagger}\hat a\\ &+ g \sum_{j=1}^{N} \left(|e_j\rangle \langle 1_j|  + |1_j\rangle \langle e_j|  \right)(\hat a^{\dagger}+ \hat a)\\
    &+ 2|\eta(t)|\sin(\omega_L t + \arg[\eta(t)](\hat a^{\dagger} + \hat a)\\
    &+ \sum_{j=1}^{N} \left(\Omega(t)\cos(\omega_{\mathrm{gl}}t)|1_j\rangle \langle 0_j| + \mathrm{h.c.}\right)
    \end{split}
\end{align}

We define the detuning between the frequency of the cavity drive laser and of the $|1\rangle \leftrightarrow |e\rangle$ transition as $\Delta = (\omega_e- \omega_1)- \omega_L$, the detuning between the laser and the cavity mode frequency as $\delta = \omega_c- \omega_L$, and the detuning of the $|0\rangle \leftrightarrow |1\rangle$ qubit transition and the global laser as $\delta_{gl}= (\omega_1- \omega_0)- \omega_{gl}$. 

We proceed by defining the following unitary operator 
\begin{equation}\label{rotating_unitary}
    \hat U(t)= \exp{\left[i(\omega_L(\hat a^{\dagger} \hat a + \hat n_e)+  \omega_1 (\hat n_1+ \hat n_e)+ \omega_0 \hat n_0)t\right]}.
\end{equation}
In the rotating frame given by $\hat U(t)$, $\hat H_{\mathrm{full}}$ is transformed as $\hat H = \hat U \hat H_{\mathrm{full}} \hat U^{\dagger} + i\frac{d\hat U}{dt} \hat U^{\dagger}$. In the rotating wave-approximation valid for $g, |\eta| \ll \omega_L$, $|\Omega| \ll \omega_{\mathrm{gl}}$, we then obtain the following non-Hermitian Hamiltonian for the system given by, 
\begin{align}
\begin{split}
    \hat H &= \hat H^{(\Delta, \delta, g)} + \hat H^{(\kappa, \gamma, \eta)} + \hat H^{(\Omega)}\\    
    \hat H^{(\Delta, \delta, g)} &= \delta \hat a^{\dagger}\hat a + \Delta \hat n_e + g (\hat S^{-} \hat a^{\dagger} + \hat S^{+} \hat a)\\
    \hat H^{(\kappa,\gamma,\eta)} &= -\frac{i}{2}\kappa \hat a^{\dagger}\hat a - \frac{i}{2}\gamma \hat n_e + i\eta(t) \left(\hat a^{\dagger}- \hat a\right)\\
    \hat H^{(\Omega)} &=  \sum_{j=1}^{N} \left(\frac{\Omega(t)}{2}|1_j\rangle \langle 0_j| + \frac{\Omega^*(t)}{2}|0_j\rangle \langle 1_j|\right),
    \end{split}
    \label{eq::full_H}
\end{align}
where $\hat S^{-}= \sum_{j=1}^{N}|1_j\rangle \langle e_j|$ and $\hat S^{+} = \sum_{j=1}^{N}|e_j\rangle \langle 1_j|$ are collective operators. The Hamiltonian $\hat H$ in Eq.~\eqref{eq::full_H} consists of three components that represent distinct physical processes. The first component, $\hat H^{(\Delta, \delta, g)}$, includes the Tavis-Cummings interaction Hamiltonian, which describes the coupling of atoms to the shared cavity mode~\cite{tavisExactSolutionNMoleculeRadiationField1968}. The second component $\hat H^{(\kappa, \gamma, \eta)}$, describes the cavity drive and the loss mechanisms, with non-Hermitian contributions from cavity decay (rate $\kappa$) and spontaneous emission from the excited state $\ket{e}$ (rate $\gamma$). The third component, $\hat H^{(\Omega)}$, represents the free-space laser coupling (transversal drive) between the qubit states $|0\rangle$ and $|1\rangle$, driven by a time-dependent Rabi frequency $\Omega(t)$. In Eq.~\eqref{eq::full_H} we define $\Omega(t)= \Omega_0\exp[{i(\varphi(t\Omega_0)+ \delta_{gl}t)}]$, where $\Omega_0$ is the Rabi-frequency of the laser coupling the qubit transition and $\varphi(t\Omega_0)$ is a phase depending on the dimensionless time $t\Omega_0$.
Both photon loss from the cavity and the decay of population from the state $|e\rangle$ are treated as non-Hermitian terms in Eq.~\eqref{eq::full_H}, corresponding to $(-i\kappa/2) \hat a^{\dagger}\hat a$ and $(-i\gamma/2) \hat n_e$, respectively. For the cavity decay, this corresponds to a conditional evolution under the condition that no photon is lost. For the latter term, this implies the assumption that all population decays outside the computational basis states. We come back to this point in Sec.~\ref{sec::W_state}.

\subsection{Cavity polariton states}\label{subsec::cav_polaritons}

Cavity polariton states refer to the eigenstates of the atoms-cavity coupling Hamiltonian $\hat H^{(\Delta, \delta, g)}$, which are the hybrid atom-photon states. 

To diagonalise $\hat H^{(\Delta, \delta, g)}$, we identify the operators  $\hat n = \sum_{j=1}^{N} (|1_j\rangle\langle 1_j| + |e_j\rangle\langle e_j|)$ and $\hat k = \sum_{j=1}^{N}|e_j\rangle\langle e_j|+ \hat a^{\dagger}\hat a$ such that $\left[\hat H^{(\Delta, \delta, g)}, \hat n\right]= \left[\hat H^{(\Delta, \delta, g)}, \hat k\right]=0$. This suggests that $\hat H^{(\Delta, \delta, g)}$ is block-diagonal in eigenstates of $\hat n$ and $\hat k$. Also, with $\left[\hat n, \hat k\right] = 0$, both $\hat n$ and $\hat k$ can be diagonalised simultaneously. We define a convenient symmetric basis defined by the states $\state{a}{b}{m}$ with $0 \leq a+b \leq N$, $m_{ph}=0, 1, \dots \infty$. The state $\state{a}{b}{m}$ corresponds to a symmetric superposition of all states with $a$ atoms in state $|1\rangle$, $b$ atoms in state $|e\rangle$ and $m$ photons in the cavity. Thus, $\hat n \state{a}{b}{m}= (a+b) \state{a}{b}{m}$, $\hat k \state{a}{b}{m}= (b+m)\state{a}{b}{m}$  with $a
+b= n = 0, 1, \dots, N$, and $b+m = k = 0, 1, \dots \infty$. %The  Hamiltonian $\hat H$ can be expanded in this basis as shown in Appendix~\ref{app::H_in_symm_basis}.

Here and in the remainder of the section, we restrict our analysis to the subspace of $\hat H^{(\Delta, \delta, g)}$ spanned by the basis states $\left\{|\psi\rangle : \hat n |\psi\rangle = n|\psi\rangle; \hat k |\psi\rangle = k |\psi\rangle \right\}$ with $n=0,1,2$ and $k=0,1$, which suffices for the discussion of the intended blockade mechanism. 
In the $k=0$ subspace of $\hat H^{(\Delta, \delta, g)}$, we have the eigenstates $\state{n}{0}{0}\equiv |\mathcal{D}_n\rangle \otimes |0\rangle_{\mathrm{cav}}$ with zero energy. Here $\dketstate{n}$ refers to the qubit state which is a symmetric superposition of computational states with $n$ qubits in $|1\rangle$ and the rest in $|0\rangle$.

\begin{figure*}
    \centering
    \includegraphics[width=0.9\textwidth]{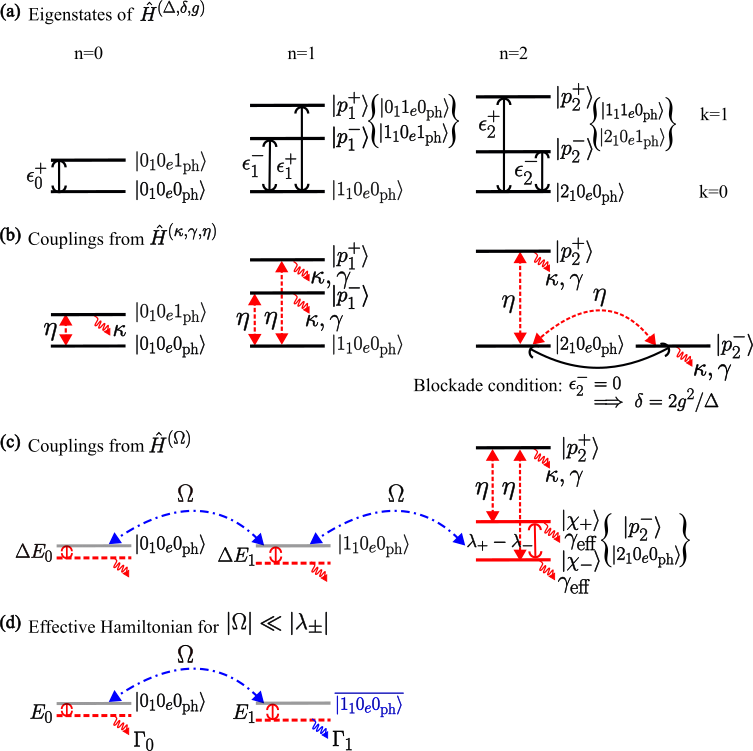}
    \caption{Level schematic overview of the blockade mechanism. (a) Eigenstates and eigenenergies of $\hat H^{(\Delta, \delta, g)}$ truncated to the subspace spanned by states in $n= 0,1,2$, $k=0,1$ (See text Sec.~\ref{subsec::cav_polaritons}). (b) Couplings from $\hat H^{(\kappa, \gamma, \eta)}$ corresponding to the cavity drive with strength $\eta$ are denoted by red arrows. The blockade condition is achieved by setting $\epsilon_2^-=0$, which makes the cavity drive resonant to the $\state{2}{0}{0} \leftrightarrow |p_2^-\rangle$ transition. (c) In the $n=0$ and $n=1$ subspaces, weak $\eta$ coupling shifts the respective states $\state{0}{0}{0}$ and $\state{1}{0}{0}$ in energy (red dashed lines) which also acquire linewidths to the first order in $\kappa, \gamma$. In the $n=2$ subspace, the states $\state{2}{0}{0}$ and $|p_2^-\rangle$ are dressed by the $\eta$ interaction into new states $|\chi_{\pm}\rangle$ (red solid lines) with eigenvalues $\lambda_{\pm}$ (See text Sec.~\ref{subsec::shifts_H_eta}). The couplings from $\hat H^{(\Omega)}$ are shown by blue dash-dot arrows. (d) The effective Hamiltonian restricted to the states $\state{0}{0}{0}$ and $\overline{\state{1}{0}{0}}$ (dressed state due to coupling to $n=2$ subspace via $\hat H^{(\Omega)}$) is obtained in the limit $|\Omega|\ll |\lambda_{\pm}|$(See text Sec.~\ref{subsec::H_Omega}).}
    \label{fig::level_schematic}
\end{figure*}

The $k=1$ subspace of $\hat H^{(\Delta, \delta, g)}$ is written as 
\begin{equation}\label{eq:H_Delta_delta_g_n_k_1}
    \hat H^{(\Delta, \delta, g)}_{n, k=1}=
    \begin{bmatrix}
    \delta & \sqrt{n}g\\
    \sqrt{n}g & \Delta
    \end{bmatrix}
\end{equation}
in the basis spanned by $\{\state{n}{0}{1}, \state{n-1}{1}{0} \}$. As a reminder, $\state{n}{0}{1}$ corresponds to the state with equal superposition of all basis states with $n$ atoms in state $|1\rangle$, no atoms in state $|e\rangle$, and one cavity photon. The state $\state{n-1}{1}{0}$ refers to the equal superposition of all states with $n-1$ atoms in state $|1\rangle$, one atom in state $|e\rangle$, and cavity in vacuum state. The eigenstates of $\hat H^{(\Delta, \delta, g)}_{n,k=1}$ are then the polariton states $|p_n^{\pm}\rangle$ (superposition of states $\state{n}{0}{1}$ and $\state{n-1}{1}{0}$) with eigenenergies $\epsilon_n^{\pm}$, given by

\begin{eqnarray}
    |p_n^+\rangle &=& \cos(\theta/2) \state{n}{0}{1} + \sin(\theta/2) \state{n-1}{1}{0},\label{eq::p_plus_state}\\
    |p_n^-\rangle &=& -\sin(\theta/2) \state{n}{0}{1} + \cos(\theta/2) \state{n-1}{1}{0} ]\label{eq::p_minus_state},
\end{eqnarray}
where $\cos(\theta) = (\delta- \Delta)/(\sqrt{(\delta - \Delta)^2 + 4ng^2})$.
The eigenenergies are given by

\begin{equation}\label{energy_polaritons}
    \epsilon_n^{\pm} = \frac{1}{2}(\delta + \Delta) \pm \frac{1}{2}\sqrt{(\delta - \Delta)^2 + 4ng^2}.
\end{equation}
We hence obtain 
\begin{equation}
    \hat H^{(\Delta, \delta, g)}_{n, k=1}= \epsilon_n^{+} |p_n^+\rangle \langle p_n^+| + \epsilon_n^{-}|p_n^{-}\rangle \langle p_n^{-}|.
\end{equation}
 Note that for $n=0$, we have only the $|p_0^{+}\rangle = \state{n}{0}{1}$ state with $\epsilon_0^{+}= \delta$.
Figure~\ref{fig::level_schematic}(a) visualizes the energy spectrum of $\hat H^{(\Delta, \delta, g)}$ with eigenstates $\state{n}{0}{0}$ in the $k=0$ subspace and states $|p_n^{\pm}\rangle$ in the $k=1$ subspace.

In the following Sec.~\ref{sec::eff-H} we derive an effective blockade Hamiltonian, with Eq.~\eqref{eq::full_H} as the starting point and by assuming the three components of the Hamiltonian $\hat H$ in Eq.~\eqref{eq::full_H} as being associated with different timescales in the system. We start in the diagonalized basis of $\hat H^{(\Delta, \delta, g)}$ formed by cavity polariton states introduced in Sec.~\ref{subsec::cav_polaritons}, then add $\hat H^{(\kappa, \gamma, \eta)}$ and $\hat H^{(\Omega)}$ as perturbative couplings by assuming a hierarchy of timescales: $\mathbb{T}(\hat H^{(\Delta, \delta, g)}) \ll \mathbb{T}(\hat H^{(\kappa, \gamma, \eta)}) \ll \mathbb{T}(\hat H^{(\Omega)})$.
  
\section{Effective Blockade Hamiltonian}\label{sec::eff-H}
  In this section, we detail the cavity polariton blockade mechanism which prevents two qubits to be simultaneously excited to the $|1\rangle$ state. We demonstrate this blockade mechanism by first deriving the blockade condition which sets the cavity probe resonant with the $N$-atom-cavity system when exactly two atoms are in the state $|1\rangle$. Second, we describe the dynamics under the blockade mechanism by deriving an effective non-Hermitian Hamiltonian $\hat H_{\mathrm{eff}}$ restricted to the subspace with states $|\mathcal{D}_0\rangle$, initial state with all qubits in $|0\rangle$ and $\overline{|\mathcal{D}_1\rangle}$ (see Eq.~\eqref{H_eff_emperical} below), which is a state close to the W state $\dketstate{1}$ -- resulting from the blockade condition -- from the total Hamiltonian $\hat H =  \hat H^{(\Delta, \delta, g)} + \hat H^{(\kappa, \gamma, \eta)} + \hat H^{(\Omega)}$ of Sec.~\ref{sec::Model}. 
  
  We work in the regime $\Delta, \delta, g \gg \eta, \kappa, \gamma$ and $\Omega_0 \ll \eta$ to derive the blockade condition and the effective Hamiltonian. In this regime, the cavity mode, driven with strength $\eta$, is excited much more slowly than the atom-cavity coupling dynamics, with timescales comparable to the losses characterized by $\kappa$ and $\gamma$. Meanwhile, the qubit transition occurs even more slowly than the dynamics governed by the cavity drive. This separation of timescales allows us to treat $\hat{H}^{(\kappa, \gamma, \eta)}$ as a perturbation to $\hat{H}^{(\Delta, \delta, g)}$, with $\hat{H}^{(\Omega)}$ serving as an additional perturbation to the effective system. 
  
  The derivation has the following three steps: (i) We establish the blockade condition which results in the cavity polariton state with two qubits in $|1\rangle$ to resonantly interact with the cavity drive. More precisely, this condition leads to the transition between two cavity polariton states in the $n=2$ subspace resonant with the cavity drive. (ii) Next, we add the Hamiltonian term with the cavity drive $\hat H^{(\kappa,\gamma,\eta)}$. This has two effects - a) In the $n=2$ subspace, owing to the resonance condition set by the blockade condition, the cavity drive further dresses the resonant cavity polariton states into new dressed states. And (b) in other $n$ subspaces ($n \neq 2$), this coupling can be treated perturbatively because of the limit $\Delta, \delta, g \gg \eta, \kappa, \gamma$, which results in effective energy shifts on the cavity polariton states. We calculate the dressed state energies for the former states, and calculate the energy shifts to the latter polariton states. (iii) Finally, we add the coupling term $\hat H^{(\Omega)}$ and obtain the effective Hamiltonian. Steps (i), (ii) and (iii) are detailed in Secs.~\ref{subsec::cav_polariton_blockade_condition},~\ref{subsec::shifts_H_eta} and ~\ref{subsec::H_Omega} below, respectively. 

The resulting effective Hamiltonian $\hat H_{\mathrm{eff}}$ defined on the qubit subspace has the form given by
\begin{align}\label{H_eff_emperical}
\begin{split}
    \hat H_{\mathrm{eff}} &= \left(E_0 - i\frac{\Gamma_0}{2}\right)\dketstate{0} \dbrastate{0} + \left(E_1 - i\frac{\Gamma_1}{2}\right)\overline{\dketstate{1}}\, \overline{\dbrastate{1}} \\ &+ \frac{\sqrt{N}\Omega(t)}{2}\overline{\dketstate{1}} \dbrastate{0} + \frac{\sqrt{N}\Omega^*(t)}{2}\dketstate{0} \overline{\dketstate{1}},
    \end{split}
\end{align}
where $E_{0}, E_{1}$ and $\Gamma_0, \Gamma_{1}$ are the effective energies and the linewidths corresponding to the states $\dketstate{0}$ and $\overline{\dketstate{1}}$, respectively.
In Eq.~\eqref{H_eff_emperical}, $\dketstate{0} = |00\dots 0\rangle$ corresponds to all qubits in the $|0\rangle$ state. The state  
\begin{equation}\label{eq::Dicke12}
    \overline{\dketstate{1}} = \dketstate{1} + \mathcal{O}(\kappa, \gamma)\dketstate{2}
\end{equation}
is our target state -- a many-particle (and possibly non-local) W state in the presence of atom and photon losses. For $\kappa, \gamma = 0$, it corresponds to the W state, $|W\rangle = \dketstate{1}= \left(\frac{1}{\sqrt{N}}\sum_{j=1}^{N} |100\dots\rangle + |010\dots\rangle + \dots |00\dots 1\rangle\right)$, which is a symmetric Dicke state with one atom in state $|1\rangle$. Similarly, the state $\dketstate{2}$ corresponds to a symmetric Dicke state with two atoms in state $|1\rangle$, and thus the second term in the r.h.s. of Eq. ~\eqref{eq::Dicke12} represents the first order corrections to $\dketstate{1}$ in the presence of finite losses with rates $\kappa, \gamma \neq 0$. 

An explicit form for the state $\overline{\dketstate{1}}$ is obtained by evolving the state $\dketstate{0}$ with Hamiltonian $\hat H_{\mathrm{eff}}$, with $\Omega(t)$ chosen such that the $\dketstate{0}\leftrightarrow \dketstate{1}$ transition is driven resonantly for a time $T = \pi/(\sqrt{N}\Omega_0)$. We find that the final state $|\psi(T)\rangle$ after the time evolution $T$ is given by
\begin{align}
\begin{split}
    |\psi (T)\rangle &= -i\sin\left(\frac{\pi \Omega'}{2\Omega_0}\right)\frac{\Omega_0}{\Omega'}\overline{\dketstate{1}} \\
    &+ \left(\cos\left(\frac{\pi\Omega'}{2\Omega_0}\right)- \sin\left(\frac{\pi\Omega'}{2\Omega}\right)\frac{(\Gamma_0 - \Gamma_1)}{2\sqrt{N}\Omega'}\right)\dketstate{0}
\end{split}
\end{align}
where $\Omega'= \Omega_0\sqrt{1 - (\Gamma_0- \Gamma_1)^2 /(4 N\Omega_0^2})$.
The final state $|\psi(T)\rangle$ obtained above has a non-vanishing component along $\dketstate{0}$ because $\Gamma_0 \neq \Gamma_1 \neq 0$. However as $\kappa, \gamma \rightarrow 0$, $|\psi(T)\rangle \rightarrow \dketstate{1} $. We take $T \propto \Omega_0^{-1}$ and as we will see in the following, the blockade regime is set in the limit $\Omega_0 \rightarrow 0$ and hence the effective Hamiltonian is derived in the limit $T \rightarrow \infty$.\\

In the following Secs.~\ref{subsec::cav_polariton_blockade_condition}-\ref{subsec::H_Omega}, we discuss in detail the steps (i)-(iii) above leading to the effective blockade dynamics. We then use the effective dynamics to illustrate the W state preparation and the realization of the CZ and C$_2$Z gates in Secs.~\ref{sec::W_state} and~\ref{sec::C_gates}, respectively.

\subsection{Cavity polariton blockade condition}\label{subsec::cav_polariton_blockade_condition}

In this section we start with the diagonalised Hamiltonian $\hat H^{(\Delta, \delta, g)}$ discussed in Sec.~\ref{subsec::cav_polaritons} and establish the cavity polariton blockade condition. 

Consider first the energy spectrum of the Hamiltonian $\hat H^{(\Delta, \delta, g)}$ in the $n=0,1,2$ and $k=0,1$ subspace as shown in Fig.~\ref{fig::level_schematic}(a). Note that $\hat H^{(\kappa, \gamma, \eta)}$ couples the states in $k$ with states in $k+1$ within the same $n$ subspace. The state $\state{n}{0}{0}$ is hence coupled to the states $|p_n^{\pm}\rangle$ via $\hat H^{(\kappa, \gamma, \eta)}$ (See Fig.~\ref{fig::level_schematic}(b)).

The cavity polariton blockade condition makes the $\eta$ coupling mediated by cavity drive term $i\eta(t)(\hat a^{\dagger}- \hat a)$ from $\hat H^{(\kappa, \gamma, \eta)}$ resonant with the atom-cavity system with two qubits in $|1\rangle$ state. That is, the transition between the cavity polariton states $\state{2}{0}{0} \equiv \dketstate{2}\otimes |0\rangle_{\mathrm{cav}}$ and $|p_2^-\rangle$ in the $n=2$ subspace is made resonant with the cavity drive. This is achieved by tuning the cavity drive detuning $\delta$ such that 
\begin{equation}\label{eq::blockade_condition}
    \delta= 2g^2/\Delta.
\end{equation}
This is similar to setting $\epsilon_2^{-} =0$ in Eq.~\eqref{energy_polaritons}. As a result, light enters the cavity and is transmitted when $\delta$ is chosen according to Eq.~\eqref{eq::blockade_condition}. 

In the following Sec.~\ref{subsec::shifts_H_eta}, we discuss the implications of this condition when the Hamiltonian $\hat H^{(\kappa,\gamma, \eta)}$ is introduced. 

\subsection{Dressed states and energy shifts due to perturbative couplings from $\hat H^{(\kappa,\gamma, \eta)} $}\label{subsec::shifts_H_eta}

In this section, we add couplings from the Hamiltonian $\hat H^{(\kappa,\gamma, \eta)}$ consisting of the cavity drive term and the loss rates. This coupling is treated perturbatively in the $n \neq 2$ subspaces and non-perturbatively in the $n=2$ subspace because of a resonant coupling introduced by the cavity polariton blockade condition set by Eq.~\eqref{eq::blockade_condition}. We start first by calculating the energies and linewidths of the dressed polaritons states in the $n=2$ subspace. We then calculate the energy shifts on the polariton states in the $n=0$ and $n=1$ subspaces. 

Let the dressed polariton states formed by the $\eta$ coupling between $\state{2}{0}{0}$ and $|p_2^-\rangle$ in the $n=2$ subspace be denoted by $|\chi_{\pm}\rangle$ (See Fig.~\ref{fig::level_schematic}(c)). These states correspond to the eigenstates of the Hamiltonian $\hat H^{(\Delta,\delta,g)} + \hat H^{(\kappa, \gamma, \eta)}$ in the $n=2$ subspace. The corresponding eigenvalues $\lambda_{\pm}$ are obtained as the eigenvalues of the matrix 
\begin{align}\label{eq:lambda_matrix}
\begin{split}
&
\begin{bmatrix}
    \statebra{2}{0}{0} \hat H^{(\kappa, \gamma, \eta)} \state{2}{0}{0} & \statebra{2}{0}{0} \hat H^{(\kappa,\gamma, \eta)}| p_2^{-}\rangle \\
    \langle p_2^{-}| \hat H^{(\kappa, \gamma, \eta)} \state{2}{0}{0} & \langle p_2^-| \hat H^{(\kappa, \gamma, \eta)}|p_2^{-}\rangle
\end{bmatrix}\\
&= \begin{bmatrix}
        0 & i\eta\Delta/\sqrt{\Delta^2 + 2g^2}\\
        -i\eta\Delta/\sqrt{\Delta^2 + 2g^2}& -i(\kappa\Delta^2 + 2\gamma g^2)/(2(\Delta^2 + 2g^2))
    \end{bmatrix}.
\end{split}
\end{align}

In Eq.~\eqref{eq:lambda_matrix}, we have used $|p_2^-\rangle$ obtained by setting $n=2$ and $\delta = 2 g^2/\Delta$ in Eq.~\eqref{eq::p_minus_state}. It is obtained as 
\begin{equation}\label{eq:p2-}
    |p_2^-\rangle = \frac{-\Delta \state{2}{0}{1} + \sqrt{2}g \state{1}{1}{0}}{\sqrt{\Delta^2 + 2g^2}}.
\end{equation}

Defining $\eta_{\mathrm{eff}} = \eta/(\sqrt{\Delta + 2g^2/\Delta})$ and $\gamma_{\mathrm{eff}}= (\kappa\Delta + 2\gamma g^2)/(\Delta + 2g^2/\Delta)$, we obtain,

\begin{equation}\label{eq:lambda_vals}
    \lambda_{\pm}= \pm \sqrt{\eta_{\mathrm{eff}}^2 -\frac{\gamma_{\mathrm{eff}}^2}{16}} -i\frac{\gamma_{\mathrm{eff}}}{4}.
\end{equation}
We will analyse these eigenvalues in the next section.

In the $n=0$ and $n=1$ subspaces, $|\eta| \ll |\epsilon_{n=0,1}^{\pm}|$ results in weak couplings mediated by $\hat H^{(\kappa, \gamma, \eta)}$ between the states $\state{0}{0}{0}$ and $\state{1}{0}{0}$ in $k=0$ to the corresponding polariton states $|p_{0}^+\rangle$ and $|p_1^{\pm}\rangle$ in $k=1$ respectively. These perturbative couplings shift the states $\state{0}{0}{0}$ and $\state{1}{0}{0}$ downward in energy, which up to third order in $\hat H^{(\kappa, \gamma, \eta)}$ are calculated as $\Delta E_{0}$ and $\Delta E_{1}$ respectively. They are given by (see Appendix~\ref{app::shifts_n0_n1})

\begin{align}
    \Delta E_{0}& = -\frac{\eta^2}{\delta} -i \frac{\kappa \eta^2 }{2\delta^2} \label{eq:shift_E0}\\
    \Delta E_{1}& = -\frac{\eta^2}{\delta - g^2/\Delta} -\frac{i}{2}\left(\frac{\eta^2(\kappa\Delta^2 + \gamma g^2)}{\Delta^2(\delta - g^2/\Delta)^2}\right).\label{eq:shift_E1}
\end{align}
The energy corrections are obtained up to the first order in $\kappa, \gamma$. Note from Eqs.~\eqref{eq:shift_E0} and ~\eqref{eq:shift_E1} that  the states $\state{0}{0}{0}$ and $\state{1}{0}{0}$ also acquire a linewidth owing to the weak coupling to decaying states $|p_0^+\rangle$ and $|p_1^{\pm}\rangle$ respectively. The corrections to the states can be neglected as the energy contributions of the residual states are in second order of $\kappa, \gamma$. We will consider these energy shifts in the next section to write down the effective Hamiltonian.    

\subsection{Coupling from $\hat H^{(\Omega)}$ and effective Hamiltonian} \label{subsec::H_Omega}

In this section, we add the qubit coupling term $\hat H^{(\Omega)}$ which couples the states among different $n$ subspaces. As shown in Fig.~\ref{fig::level_schematic}(c), $\hat H^{(\Omega)}$ couples the state $\state{0}{0}{0}$ with state $\state{1}{0}{0}$, and the state $\state{1}{0}{0}$ is further coupled to the states $|\chi_{\pm}\rangle$ in $n=2$ subspace (recall from Eq.~\eqref{eq:lambda_vals} the corresponding eigenvalues $\lambda_{\pm}$).  

The goal of this section is to obtain the effective Hamiltonian as in Eq.~\eqref{H_eff_emperical}, which is done in the following steps. (i) First we establish the limit  $\Omega_0 \ll |\lambda_{\pm}|$ such that the coupling between $\state{1}{0}{0}$ and $|\chi_{\pm}\rangle$ is either strongly detuned or the states $|\chi_{\pm}\rangle$ are strongly decaying, preventing excitation of the system from $\state{1}{0}{0}$ to $|\chi_{\pm}\rangle$. We have from Eq.~\eqref{eq:lambda_vals} the following two cases.
\begin{equation}\label{eq::cases}
\eta_{\mathrm{eff}}
\begin{cases}
    (\text{case 1})\; \ge \gamma_{\mathrm{eff}}/4 \\
    \,\implies |\lambda_{-}|=|\lambda_+| 
 = \eta_{\mathrm{eff}} \propto \sqrt{\Omega_0}\\
  (\text{case 2})\; < \gamma_{\mathrm{eff}}/4\\
  \implies |\lambda_-| > |\lambda_+| = \gamma_{\mathrm{eff}}/4 - \sqrt{\gamma_{\mathrm{eff}}^2/16 - \eta_{\mathrm{eff}}^2} \\
 \implies |\lambda_{+}|\ge  \frac{\gamma_{\mathrm{eff}}}{4}\left(1- \left(1-  \frac{8\eta^2_{\mathrm{eff}}}{\gamma^2_{\mathrm{eff}}}\right)\right)= \frac{2\eta^2_{\mathrm{eff}}}{\gamma_{\mathrm{eff}}} \propto \frac{\Omega_0}{\gamma_{\mathrm{eff}}}.
\end{cases}
\end{equation}
In writing the proportionality in the two cases above, we assumed $\eta \propto \sqrt{\Omega_0}$. Later in Sec.~\ref{sec::W_state} and Sec.~\ref{sec::C_gates}, we show that an optimal choice of $\eta^2/\Omega_0$ results in a minimum operational infidelity of W state preparation and $CZ$, $C_2Z$ gates, respectively. 
 Hence from Eq.~\eqref{eq:lambda_vals} and Eq.~\eqref{eq::cases} $\Omega_0/|\lambda_{\pm}| \rightarrow 0 $ as $\Omega_0, \kappa, \gamma \rightarrow 0$. Note here that in the case of $\eta_{\rm eff} \leq \gamma_{\rm eff}/4$, $\lambda_{\pm}$ are purely imaginary and hence correspond to only the broadening of the states $|\chi_{\pm}\rangle$. In this limit, the blockade effect can be seen to be arising from a decay induced QZD-like effect instead of that arising from far-detuned transitions.

(ii) Secondly, in this limit, that is when $\Omega_0\ll |\lambda_{\pm}|$, we have a weak coupling between $\state{1}{0}{0}$ and $|\chi_{\pm}\rangle$ in $n=2$ subspace mediated by $\hat H^{(\Omega)}$, and as a result $\state{1}{0}{0}$ experiences an additional energy correction $\Delta E'_1$, and is weakly dressed giving state corrections which are first order in $\kappa, \gamma$. Let the dressed state be denoted by $\overline{\state{1}{0}{0}}$ (See Figs.~\ref{fig::level_schematic}(c)-(d)).

To obtain the energy and state corrections, we define the Hamiltonian in $n=2$ subspace as $\hat H_{n=2}$ given by
\begin{equation}\label{eq:H_n_2}
    \hat H_{n=2}= \hat H^{(\Delta, \delta, g)}_{n=2} + \hat H^{(\kappa, \gamma, \eta)}_{n=2}= 
    \begin{bmatrix}
        0& -i\eta&0\\
        i\eta& \delta- i\kappa/2& \sqrt{2}g\\
        0&\sqrt{2}g& \Delta- i\gamma/2
    \end{bmatrix}.
\end{equation}
The Hamiltonian matrix in Eq.~\eqref{eq:H_n_2} is written in the basis $\{\state{2}{0}{0}, \state{2}{0}{1}, \state{1}{1}{0}\}$. 
The energy corrections to state $\state{1}{0}{0}$ and the dressed state $\overline{\state{1}{0}{0}}$ are then obtained as 

\begin{align}\label{eq:shift_E1_prime}
\begin{split}
    \Delta E'_{1}&= \statebra{1}{0}{0} \hat H^{(\Omega)} \left(\hat H_{n=2}\right)^{-1}\hat H^{(\Omega)} \state{1}{0}{0}\\
    &= -\frac{\Omega_0^2}{2}(N-1)\statebra{2}{0}{0} \left(\hat H_{n=2}\right)^{-1}\state{2}{0}{0}\\
    &= \frac{-i\Omega_0^2(N-1)}{2\eta^2}\left(\frac{\kappa}{2}+ \frac{\gamma g^2}{\Delta^2}\right),
    \end{split}
\end{align}

\begin{align}\label{eq:correction_state}
\begin{split}
    \overline{\state{1}{0}{0}} &= \state{1}{0}{0}- \left(\hat H_{n=2}\right)^{-1} \hat H^{(\Omega)}\state{1}{0}{0} \\
    &= \state{1}{0}{0}- \frac{i \sqrt{2(N-1)}\Omega^*}{4\eta^2}\left(\kappa + \frac{2g^2\gamma}{\Delta^2}\right)\state{2}{0}{0}.
    \end{split}
\end{align}

In evaluating Eqs.~\eqref{eq:shift_E1_prime} and~\eqref{eq:correction_state}, we have used $\delta = 2g^2/\Delta$ from Eq.~\eqref{eq::blockade_condition} and $\hat H^{(\Omega)}\state{1}{0}{0}= (\Omega^*\sqrt{N-1}/\sqrt{2}) \state{2}{0}{0}$. Note that $\hat H^{(\Omega)}\state{1}{0}{0}$ also has a component along $\state{0}{0}{0}$, which is not relevant for the calculation of $\Delta E'_1$.

Finally by combining the energy shifts $\Delta E_1$ from Eq.~\eqref{eq:shift_E1} obtained from perturbative couplings mediated by $\hat H^{(\kappa, \gamma, \eta)}$, and $\Delta E'_1$ from Eq.\eqref{eq:shift_E1_prime} obtained from perturbative couplings mediated by $\hat H^{(\Omega)}$, we can write the energy of state $\overline{\state{1}{0}{0}}$ as
\begin{align}\label{E1_val}
\begin{split}
    &E_{1} - \frac{i}{2}\Gamma_1 = \Delta E_{1}+ \Delta E'_{1}\\
    &= -\frac{\eta^2 \Delta}{g^2} -\frac{i}{2}\left(\frac{\eta^2\Delta^2\kappa}{g^4} + \frac{\eta^2\gamma}{g^2} + \frac{(N-1)\Omega_0^2}{\eta^2}\left(\frac{\kappa}{2} + \frac{\gamma g^2 }{\Delta^2}\right)\right)
\end{split}
\end{align}

%Since the energy and state corrections to state $\state{0}{0}{0}$  due to perturbative coupling to states in $n=2$ subspace via $\hat H^{(\Omega)}$ are of higher orders in $\kappa, \gamma$, they can be neglected. 
We obtain the energy of $\state{0}{0}{0}$ state as obtained in Eq.~\eqref{eq:shift_E0} as 
\begin{equation}\label{E0_val}
    E_0  -\frac{i}{2}\Gamma_0= \Delta E_0=  -\frac{\eta^2\Delta}{2g^2} -\frac{i}{2}\left(\frac{\eta^2\Delta^2\kappa}{4g^4}\right)
\end{equation}

Equation~\eqref{E1_val} and Eq.~\eqref{E0_val} summarize the main results of this section, providing the energies $E_0$ and $E_1$ of the effective two level system with states $\dketstate{0}$ and $\overline{\dketstate{1}}$, respectively, which scale as $\frac{\eta^2\Delta}{g^2}$. These equations also describe the corresponding effective linewidths $\Gamma_0$ and $\Gamma_1$, which depend on the loss rates $\kappa$ and $\gamma$. Using the obtained values, we can now write the effective Hamiltonian restricted to the states $\state{0}{0}{0}$ and $\overline{\state{1}{0}{0}}$ as

\begin{align}\label{eq:H_eff_emperical_symm_basis}
\begin{split}
    \hat H'_{\mathrm{eff}} &= \left(E_0 - i\frac{\Gamma_0}{2}\right)\state{0}{0}{0} \statebra{0}{0}{0} \\ &+ \left(E_1 - i\frac{\Gamma_1}{2}\right)\overline{\state{1}{0}{0}}\, \overline{\state{1}{0}{0}} \\ &+ \frac{\sqrt{N}\Omega(t)}{2}\overline{\state{1}{0}{0}} \statebra{0}{0}{0} + \mathrm{h.c}.
    \end{split}
\end{align}

By tracing out the cavity field, which remains in the vacuum state $|0\rangle_{\mathrm{cav}}$ throughout the effective dynamics, and using the definitions $\state{0}{0}{0}= \dketstate{0} \otimes|0\rangle_{\mathrm{cav}} $ and $\overline{\state{1}{0}{0}}= \overline{\dketstate{1}} \otimes|0\rangle_{\mathrm{cav}}$, we obtain the effective Hamiltonian $\hat H_{\mathrm{eff}}$ in Eq.~\eqref{H_eff_emperical}. This Hamiltonian describes a driven two-level system with states $\state{0}{0}{0}$ and $\overline{\state{1}{0}{0}}$.

\section{Non-local W state preparation}\label{sec::W_state}  
In this section, we exploit the effective blockade dynamics to deterministically prepare the state $\overline{\dketstate{1}}$, which approaches the W state $\dketstate{1}$ in the limit $\kappa/g,\gamma/g \rightarrow 0$ (see Eq.~\eqref{eq::Dicke12}). Additionally, we derive an analytical expression for the state-preparation infidelity when $\kappa, \gamma \neq 0$. 
We recall the effective Hamiltonian $\hat H_{\mathrm{eff}}$ from Eq.~\eqref{H_eff_emperical} and Eqs.~\eqref{E0_val}, ~\eqref{E1_val} as 
\begin{align}
\begin{split}
    \hat H_{\mathrm{eff}} &= \left(-\frac{\eta^2\Delta}{2g^2} - \frac{i}{2}\Gamma_0\right)\dketstate{0} \dbrastate{0} + \left(-\frac{\eta^2\Delta}{g^2} -\frac{i}{2}\Gamma_1\right)\overline{\dketstate{1}}\, \overline{\dbrastate{1}}\\
    &+ \frac{\sqrt{N}\Omega(t)}{2}\overline{\dketstate{1}} \dbrastate{0} + \frac{\sqrt{N}\Omega^*(t)}{2}\dketstate{0} \overline{\dbrastate{1}},
\end{split}
\end{align}

where 
\begin{align}
    \Gamma_0 &= \frac{\kappa\eta^2\Delta^2}{4g^4}\label{eq::Gamma_0}\\
    \Gamma_1 &= \eta^2\left(\frac{\kappa\Delta^2}{g^4} + \frac{\gamma}{g^2}\right) + \frac{(N-1)\Omega_0^2}{\eta^2}\left(\frac{\kappa}{2} + \frac{\gamma g^2}{\Delta^2}\right)\label{eq::Gamma_1}.
\end{align}
By going into a rotating frame given by $\hat U= \exp\left[i\left(-\frac{\eta^2\Delta}{2g^2}(\dketstate{0}\dbrastate{0} + \overline{\dketstate{1}}\, \overline{\dketstate{1}}) - \delta_{gl}\overline{\dketstate{1}}\,\overline{\dbrastate{1}}\right)t\right]$ and by choosing $\eta^2\Delta/(2g^2)= \delta_{gl}$ for resonant transfer, with $\Omega(t)= \Omega_0e^{i\delta_{\mathrm{gl}}t}$ (with $\varphi(\Omega_0t)=0$), we obtain

\begin{align}
\begin{split}
    \hat H_{\mathrm{eff}} &=  - \frac{i}{2}\Gamma_0 \dketstate{0} \dbrastate{0} +  -\frac{i}{2}\Gamma_1 \overline{\dketstate{1}}\, \overline{\dbrastate{1}}\\
    &+ \frac{\sqrt{N}\Omega_0}{2}\overline{\dketstate{1}} \dbrastate{0} + \frac{\sqrt{N}\Omega_0}{2}\dketstate{0} \overline{\dbrastate{1}}.
    \end{split}
    \label{eq::H_eff}
\end{align}
 
In the absence of loss ($\Gamma_0= \Gamma_1= 0$), starting with initial state $\dketstate{0}$ and by choosing a pulse of duration $T = \pi/ (\sqrt{N}\Omega_0)$, the state $\overline{\dketstate{1}}= \dketstate{1}$ is  prepared with unit fidelity. In the following, we obtain an analytical expression for the state preparation error in the presence of loss ($\Gamma_0, \Gamma_1 \neq 0$).

\subsection{W state preparation fidelity calculation}\label{sec:fid_calculation}
In this section, we obtain the state-preparation error $1-F$ of state $\overline{\dketstate{1}}$ as
\begin{align}
    1- F &= \frac{\pi}{2\sqrt{N}\Omega_0} \left(\Gamma_0+ \Gamma_1\right) \label{infid_W_Gamma}\\
    &= \frac{\pi}{2\sqrt{N}} \left[ \frac{\eta^2}{\Omega_0} \left(\frac{5\kappa\Delta^2}{4g^4} + \frac{\gamma}{g^2}\right) + \frac{(N-1)\Omega_0}{\eta^2}\left(\frac{\kappa}{2}+ \frac{\gamma g^2}{\Delta^2}\right)\right].\label{eq::infid_W_analytical}
\end{align}

In writing Eq.~\eqref{eq::infid_W_analytical}, the values of $\Gamma_0$ and $\Gamma_1$ are substituted from Eq.~\eqref{eq::Gamma_0} and Eq.~\eqref{eq::Gamma_1} respectively. Note that since we consider the population decay from the $|e\rangle$ state decaying outside of the computational subspace in our model, the fidelity estimate that we obtain here corresponds to a lower bound on the actual fidelity.

To derive Eq.~\eqref{infid_W_Gamma}, we rewrite $\hat H_{\mathrm{eff}}= \hat H_0 + \hat H_{\mathrm{nh}}$ with $\hat H_0 = (\sqrt{N}\Omega_0/2)\overline{\dketstate{1}} \dbrastate{0} + \mathrm{h.c}.$ and $\hat H_{\mathrm{nh}}= -i(\Gamma_0/2)\dketstate{0}\dbrastate{0} -i(\Gamma_1/2)\overline{\dketstate{1}}\,\overline{\dbrastate{1}}$. Let $\hat U_0(t)$ and $\hat U(t)$ be the time-evolution operators for time $t$ under $\hat H_0$ and $\hat H_{\mathrm{eff}}$ respectively. For a duration $T$, up to first order in $\Gamma_0, \Gamma_1$, we have
\begin{equation}\label{eq:UT}
    \hat U (T) = \hat U_0(T) - i \int_{0}^{T} \hat U_0 (T)\hat U^{\dagger}_0(t) \hat H_{\mathrm{nh}} \hat U_0(t) dt.
\end{equation}

We define the fidelity $F$ of the $\overline{\dketstate{1}}$ state preparation as the squared overlap between the final state $|\psi(T)\rangle = \hat U(T)\dketstate{0}$ and the target state $\overline{\dketstate{1}}$. The infidelity $1- F$ is given by
\begin{equation}\label{eq::infid_overlap}
    1- F = 1- \left|\overline{\dbrastate{1}}\hat U(T) \dketstate{0}\right|^2.
\end{equation}
Using $\hat U(T)$ from Eq.~\eqref{eq:UT} and defining $|\psi_0(t)\rangle = \hat U_0(t)\dketstate{0} $, $1-F$ in Eq.~\eqref{eq::infid_overlap} is obtained as
\begin{equation}\label{eq:infid_integrals}
    1-F= \Gamma_0 \int_0^{T} dt \left|\dbrastate{0}\psi_0(t)\rangle \right|^2 + \Gamma_1 \int_0^{T} dt \left|\overline{\dbrastate{1}}\psi_0(t)\rangle \right|^2.
\end{equation}

The integrals $\int_{0}^{T} dt \left|\langle s| \psi_0(t) \right|^2$ in Eq.~\eqref{eq:infid_integrals} denote the time spent in state $|s\rangle \in \{\dketstate{0}, \overline{\dketstate{1}}\}$ during the unitary evolution $\hat U_0$ under $\hat H_0$ of initial state $\dketstate{0}$ for time $T$. In this case, both states $\dketstate{0}$ and $\overline{\dketstate{1}}$ are occupied for equal times $T/2$. 
Hence, with $T= \pi/(\sqrt{N}\Omega_0)$, we obtain $1-F$ as in Eq.~\eqref{infid_W_Gamma}.

The optimal values of $\eta^2/\Omega_0$ and $\Delta$ that minimize $1-F$ in Eq.~\eqref{eq::infid_W_analytical} are derived analytically as follows. First, $1-F$ is expressed as a function of the variable $x = \Delta^2$, such that $1-F = f(x) = ax + x/b + c$. By setting its derivative $\dot{f} = 0$, the minimum value of $f(x)$ is found to be $(1-F)_{\rm min} = 2\sqrt{ab}+ c$, occurring at $x = \sqrt{b/a} = (\Delta^2)_{\text{opt}}$. This result allows $(1-F)_{\rm min}$ to be further expressed in terms of $\eta^2/\Omega_0$ as $(1-F)_{\rm min} = a' (\eta^2/\Omega_0) + b'/(\eta^2/\Omega_0) + c'$. Following a similar procedure, minimizing $(1-F)_{\rm min}$ with respect to $\eta^2/\Omega_0$ yields the optimal value $1-F_{\rm opt.}= 2\sqrt{a'b'} + c'$ at $\left(\eta^2/\Omega_0 \right)_{\rm opt.} = \sqrt{b'/a'}$. These steps provide the optimal parameters

\begin{align}
    \Delta_{\mathrm{opt.}}^{W} &= \left(\frac{8}{5}\right)^{1/4} \sqrt{\frac{\gamma}{\kappa}}g \label{opt_Delta}, \\
  \left(\frac{\eta^2}{\Omega_0}\right)_{\mathrm{opt.}}^{W} &= \sqrt{\frac{N-1}{2}\frac{\kappa}{\gamma}}g.  \label{opt_eta}
\end{align}
% Note that we set $\delta_{\mathrm{gl}} \propto \eta^2 \propto \Omega_0$ which does not break the validity of the blockade regime $\Delta, \delta, g \gg \eta, \kappa, \gamma; \, \Omega_0 \ll \eta$ we work in.
With the optimal values from Eq.~\eqref{opt_Delta} and Eq.~\eqref{opt_eta}, and by defining cooperativity $C= g^2/(\kappa\gamma)$, we finally obtain the state preparation error from Eq.~\eqref{eq::infid_W_analytical} as
\begin{equation}\label{eq::exact_opt_err_W}
    1-F^{W}_{\mathrm{opt.}} = \pi\sqrt{(1-1/N)(\sqrt{5/8}+7/8)}/\sqrt{C} \approx \frac{5.73\sqrt{1-1/N}}{\sqrt{C}}.
\end{equation}

In order to verify our effective model (Eq.~\eqref{eq::H_eff}) and the analytic expression for the state preparation error (Eq.~\eqref{eq::exact_opt_err_W}), in Fig.~\ref{fig::summary}(d), we numerically simulate the Schr\"{o}dinger evolution under the full Hamiltonian (Eq.~\eqref{eq::full_H}), starting from the initial state $\state{0}{0}{0} = \dketstate{0}\otimes |0\rangle_{\rm cav}$. We plot the state populations dynamics of the states $\dketstate{0}$, $\dketstate{1}$ and $\dketstate{2}$ along with the trace of density operator in the symmetric subspace (after tracing out the cavity mode) given by $\mathrm{Tr}(\rho_{\mathrm{symm.}}) = \sum_{n=0}^{N} |\dbrastate{n}\mathcal{D}_n\rangle|^2$. These results are computed for $N=2$ with single particle cooperativities $C= 10^2, 10^{10}$, keeping $\gamma/\kappa= 1$. Our results show that the state preparation infidelity - quantified by the final state population in $\dketstate{1}$ (Eq.~\eqref{eq::infid_overlap}) is primarily due to $\mathrm{Tr}\left(\rho_{\mathrm{symm}}\right) < 1$. This leakage out of the symmetric subspace, arising from non-zero decay rates $\Gamma_0$ and $\Gamma_1$ (due to non-zero $\kappa$, $\gamma$), is as predicted by our effective model in Eq.~\eqref{eq::H_eff}. Furthermore, in Fig.~\ref{fig::summary}(e), we plot the state preparation infidelity $1-F$ as a function of the total pulse duration $T$ for $N=2$ for different values of cooperativities $C$ and different $\gamma/\kappa$ ratios. These numerical results are compared against the analytical infidelity obtained in Eq.~\eqref{eq::exact_opt_err_W}. In the limit $T\rightarrow \infty$, the numerical infidelity converges to the analytical estimate, scaling as $\propto 1/\sqrt{C}$ and independent of $\gamma/\kappa$ for large $C$. Finally, in Fig.~\ref{fig::summary}(f) we extend this analysis to a larger system with $N=50$. Here we compare the analytic infidelity (Eq.~\eqref{eq::exact_opt_err_W}) in the $T \rightarrow \infty$ limit and compare it against the CZ ($N=2$) and C$_2$Z ($N=3$) gate errors which are discussed in Sec.~\ref{sec::C_gates}. We find an excellent agreement between the numerical results from with the full Hamiltonian dynamics (Eq.~\eqref{eq::full_H}) and the analytic errors derived for large single-particle cooperativities $C$, which further validates our effective blockade dynamics. 

\begin{figure*}[ht]
    \centering
    \includegraphics[width=\textwidth]{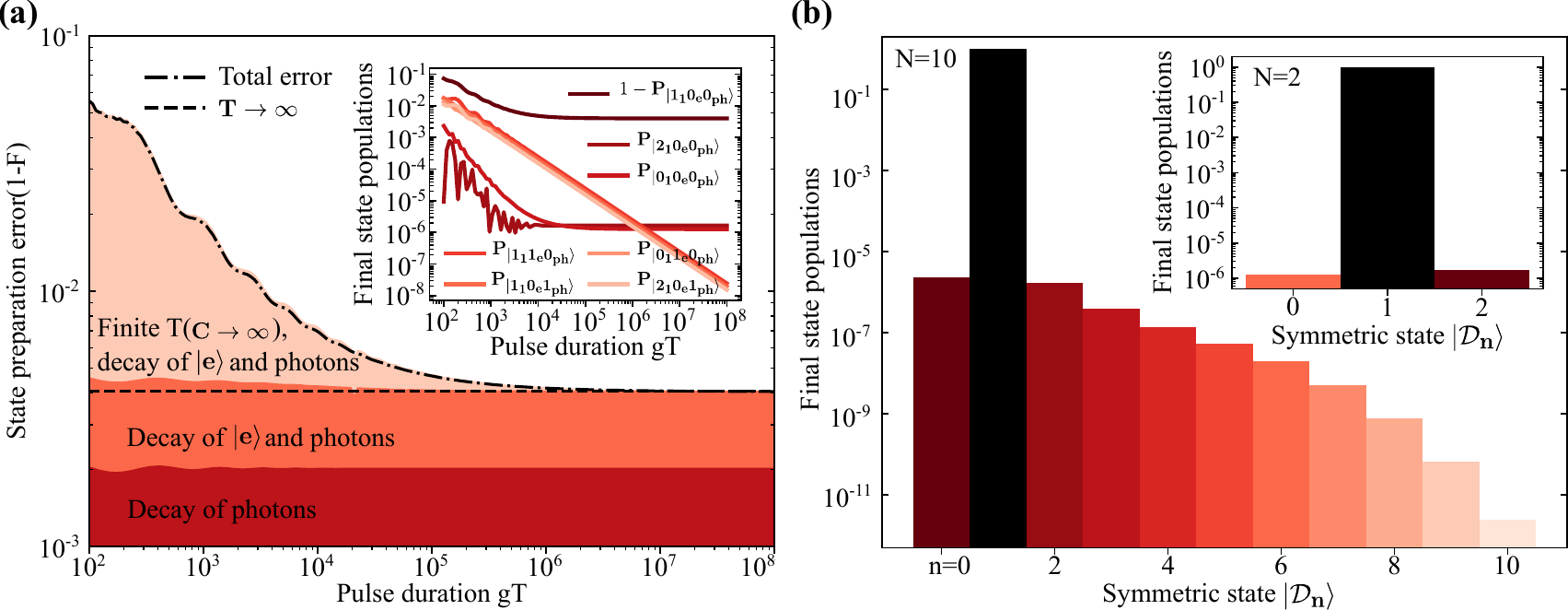}
    \caption{(a) W state preparation error for $N=2$ as a function of the total operation time for $\kappa/g = 10^{-3}$, $\gamma/g = 10^{-3}$. The error due to the decay from $|e\rangle$ state, the error due to loss of photons and the error due to finite time (calculated in the limit $C \rightarrow \infty$) adds up to give the total error (dash-dot line). The dashed line is the analytical error given by $4.05/\sqrt{C}$ calculated in the limit $T \rightarrow \infty$. (a, inset) Final state population (in log-scale) in relevant states $\state{a}{b}{m}$ as a function of the pulse duration $gT$ for the same parameters as in (a). The final state as $T \rightarrow \infty$ has non-vanishing components along the state $\state{0}{0}{0}$ and $\state{2}{0}{0}$ apart from the near-unity population in the target $\state{1}{0}{0}$ state. (b) Final state populations (in log-scale) in the atomic symmetric Dicke states $\dketstate{n}$ for $N=10$ and $N=2$ (inset) for $\kappa/g = 10^{-3}$, $\gamma/g = 10^{-3}$.}
    \label{fig::W_summary}
\end{figure*}

Next, in order to show that our model accurately captures all error sources and correctly predicts the final state populations, we compute separately the contributions from decay of the $|e\rangle$ state (non-zero spontaneous emission rate $\gamma$), cavity decay rate (non-zero $\kappa$) and non-adiabatic drive effects, and then verify that their sum agrees with the total infidelity computed numerically. In Fig.~\ref{fig::W_summary}(a), we show the error distribution as a function of $T$ for $N=2$, where the error due to decay of $|e\rangle$ state is computed as $\gamma \int_{0}^{T} \langle \psi(t) | \hat n_e |\psi(t)\rangle dt $ and the cavity decay error as $\kappa \int_{0}^{T} \langle \psi(t) | \hat a^{\dagger}\hat a |\psi(t)\rangle dt $; an additional error, due to non-adiabatic effects at short times, is obtained by setting $C \rightarrow \infty$. The sum of these three errors is in excellent agreement with the independently computed infidelity (indicated by the dashed-dotted line), confirming that our model accounts accurately for all the error sources for over finite $T$ and as $T\rightarrow \infty$. The inset of Fig.~\ref{fig::W_summary}(a) shows the final population $|\langle \psi_s| \psi(T) \rangle|^2$ in different states $|\psi_s\rangle$ for a range of total pulse duration $gT$ for $N=2$. The final population in all states $\rightarrow 0$ as $T\rightarrow \infty$ except for the target state $\state{1}{0}{0}\equiv \dketstate{1}\otimes |0\rangle_{\mathrm{cav.}}$ and states $\state{0}{0}{0}, \state{2}{0}{0}$ which have order one correction of amplitude in $\kappa, \gamma$ in the final state $|\psi(T)\rangle$.  Fig.~\ref{fig::W_summary}(b) shows the final state populations in all the atomic-symmetric Dicke states $\dketstate{n} \forall n= 0, \dots N$ completing the atomic symmetric subspace with density operator $\rho_{\mathrm{symm.}} = \mathrm{Tr}_{\mathrm{cav.}}\left(|\psi(T)\rangle \langle \psi(T)|\right)$, for $N=2$ and $N=10$. The final population in $\dketstate{n}$ then corresponds to $\dbrastate{n}\rho_{\mathrm{symm.}} \dketstate{n}$. The population leak into the states $\dketstate{0}$ and $\dketstate{2}$ is consistent with our analysis, being proportional to terms that are second order in $\kappa$ and $\gamma$.

\section{Non-local $CZ$ and $C_2Z$ gate implementation}\label{sec::C_gates}
In this section, we exploit the effective Hamiltonian derived in Eq.~\eqref{H_eff_emperical} to implement a $CZ$ and a $C_2Z$ gate with $N=2$ and $N=3$ distant atoms respectively. For this, each atom is modeled as a four-level system with states $\{|0\rangle, |1'\rangle, |1\rangle, |e\rangle\}$. We introduce an additional state $|1'\rangle$ with energy $\omega'_1$ such that the computational subspace is now spanned by the states $\{|0\rangle, |1'\rangle\}$ (See Fig.~\ref{fig::summary}(c)). All other energies and couplings remain the same as described in Sec.~\ref{sec::Model}.

We obtain the full Hamiltonian with the $|1'\rangle$ state as 
\begin{align}
    \begin{split}
    \hat H_{\mathrm{full}} &= \omega'_1 \hat n_{1'}+ \omega_0 \hat n_0 + \omega_1 \hat n_1 + (\omega_e - i\gamma/2)\hat n_e \\ &+ g \sum_{j=1}^{N} \left(|e_j\rangle \langle 1_j|  + |1_j\rangle \langle e_j|  \right)(\hat a^{\dagger}+ \hat a) + \hat H_{\mathrm{drive}}\\
    &+ \sum_{j=1}^{N} \left(\Omega(t)\cos(\omega_{\mathrm{gl}}t)|1_j\rangle \langle 0_j| + \mathrm{h.c.}\right),
    \end{split}
\end{align}
with $\hat n_{1'}= \sum_{j=1}^{N} |1'_j\rangle \langle 1'_j|$. Transforming $\hat H_{\mathrm{full}}$ under the rotating frame given by 
\begin{equation}
    \hat U(t)= \exp{\left[i(\omega_L(\hat a^{\dagger} \hat a + \hat n_e)+  \omega_1 (\hat n_1+ \hat n_e)+ \omega_0 \hat n_0 + \omega'_{1}\hat n_{1'})t\right]}
\end{equation}
and in the rotating-wave approximation, we obtain the same starting Hamiltonian as in Eq.~\eqref{eq::full_H}, and hence the same approach can be followed to arrive at an effective Blockade Hamiltonian as derived in Sec.~\ref{sec::eff-H}. All additional eigenstates of $\hat H^{(\Delta, \delta, g)}$ with atoms in states $|0\rangle, |1'\rangle$- with no atoms in $|1\rangle$- acquire an energy shift similar to that for $\state{0}{0}{0}$. 

In Sections~\ref{subsec::CZ} and~\ref{subsec::C2Z} below, we write down the effective Hamiltonians for implementing a $CZ$ and $C_2Z$ gate respectively. We simulate the gates using the time-optimal pulses found in Refs.~\cite{janduraTimeOptimalTwoThreeQubit2022, everedHighfidelityParallelEntangling2023}, and obtain optimal gate parameters for minimizing infidelity for both $CZ$ and $C_2Z$ gates. We also show that gate errors scale as $1/\sqrt{C}$ for both gates. We would like to note here that other optimal solutions for this gate protocol can be obtained more formally, for example by directly finding optimal pulses with the full Hamiltonian, which could in principle perform better for this system than the  pulses from Refs.~\cite{janduraTimeOptimalTwoThreeQubit2022, everedHighfidelityParallelEntangling2023}.

\subsection{$CZ$ gate}\label{subsec::CZ}
In this section, we exploit the effective blockade hamiltonian obtained in Section~\ref{sec::eff-H} to implement a controlled-Z (CZ) gate between two distant atoms.
For CZ gate, we consider the two- atom computational basis states $\{|1'1'\rangle, |1'0\rangle, |00\rangle\}$. For initial atom states $\{|1'1'\rangle , |1'0\rangle, |00\rangle\}$, the effective Hamiltonian acts in the subspace spanned by $\{|1'1'\rangle\}$, $\{|1'0\rangle, |1'1\rangle\}$ and $\{|00\rangle, |W\rangle\}$, respectively where $|W\rangle = (|01\rangle + |10\rangle)/\sqrt{2}$. In each of the three decoupled subspaces, we can write the effective Hamiltonian in the same way as in Eq.~\eqref{H_eff_emperical}, but with $N$ replaced by $N_0$, the number of atoms initialized in state $|0\rangle$. We denote the effective Hamiltonians in each of these subspaces as $\hat H_{1'1'}$, $\hat H_{1' 0}$, and $\hat H_{00}$.
Let the effective decay of the state with $N_0$ atoms initialized in state $|0\rangle$ be $\Gamma_1^{(N_0)}$ with 
\begin{equation}
    \Gamma_1^{(N_0)}= \frac{\eta^2\Delta^2\kappa}{g^4} + \frac{\eta^2\gamma}{g^2} + (N_0-1)\Omega_0^2\left(\frac{\kappa}{2\eta^2} + \frac{g^2 \gamma}{\eta^2\Delta^2}\right).
    \label{eq::Gamma_1_N0}
\end{equation}
We obtain up to single-qubit operations,
\begin{align}
\begin{split}
    \hat H_{1'1'} &= -\frac{i\Gamma_0}{2} |1'1'\rangle\langle 1'1'|\\
    \hat H_{1'0} &= \left(\frac{\Omega(t)}{2}|1'1\rangle \langle 1'0| + \mathrm{h.c}\right) + \frac{-i\Gamma_0}{2}|1'0\rangle \langle 1'0|\\ &+ - \frac{i\Gamma^{(1)}_1}{2}|1'1\rangle \langle 1'1|\\
    \hat H_{00} &= \left(\frac{\sqrt{2}\Omega(t)}{2}|W\rangle \langle 00| + \mathrm{h.c}\right)+-\frac{i\Gamma_0}{2} |00\rangle \langle 00| \\ &+ - \frac{i\Gamma^{(2)}_1}{2} |W\rangle \langle W|
    \end{split}
\end{align}
Let $e^{i\xi_{N_0}}$ be the phases acquired from the evolution of each of the computational basis states $\{|1'1'\rangle, |1'0\rangle, |00\rangle\}$ with $N_0=0, 1, 2$ respectively. The evolution implements a $CZ$ gate when $\xi_{0}= 0$, $\xi_{1}= \theta$ and $\xi_{2}= 2\theta + \pi$ for a single-qubit phase $\theta$. Hence up to single qubit phase gates acting on $|0\rangle$ state, a $CZ$ operation can be realized exactly for $\Gamma_0, \Gamma_1= 0$ by evolving the qubits under the effective Hamiltonian. We note that approach is similar to that used in~\cite{janduraTimeOptimalTwoThreeQubit2022} to implement a time-optimal $CZ$ gate. In the presence of losses, by following a similar treatment as in Sec.~\ref{sec:fid_calculation}, we can write the error of the $CZ$ gate operation as 
\begin{equation}\label{cz_infid}
    1- F= \frac{1}{4\Omega} (\Gamma_0(\tau_{1'1'} + 2\tau_{1'0} + \tau_{00}) + 2\Gamma^{(1)}_{1}\tau_{1'1} + \Gamma^{(2)}_{1}\tau_{W}),
\end{equation}
where $\tau_{q}$ is the dimensionless time spent in the state $|q\rangle$. The prefactor $2$ with $\tau_{1'0}$ and $\tau_{1'1}$ is to take into account all states $|1'0\rangle, |01'\rangle$ and $|1'1\rangle, |11'\rangle$, respectively. 
By inserting the $\Gamma$ values from Eq.~\eqref{eq::Gamma_0} and Eq.~\eqref{eq::Gamma_1_N0}, we find optimal values of $\Delta$ and $\Omega_0/\eta^2$ which minimize the gate error for fixed $g$, $\gamma$, $\kappa$. The optimal values are given by
\begin{widetext}
\begin{align}\label{opt_CZ_param}
    (\Delta)^{CZ}_{\mathrm{opt.}} &= \left(\frac{8(\tau_{1'1}+ \tau_{1'0})}{\tau_{1'1'} + 2\tau_{1'0}+ \tau_{00}+ 4\tau_{1'1}+ 4\tau_{W}}\right)^{1/4} \sqrt{\frac{\gamma}{\kappa}}g\\
    \left(\frac{\Omega_0}{\eta^2}\right)^{CZ}_{\mathrm{opt.}} &= \sqrt{\frac{\frac{1}{16}(\tau_{1'1'} + 2\tau_{1'0}+ \tau_{00}+ 4\tau_{1'1}+ 4\tau_{W})\frac{\Delta^2 \kappa}{g^4} + \frac{1}{4}(\tau_{1'1}+\tau_{1'0})\frac{\gamma}{g^2}}{\tau_{W}  (\frac{\kappa}{2} + \frac{\gamma g^2}{\Delta^2})}}
\end{align}
\end{widetext}
The optimal values $(\Delta)^{CZ}_{\mathrm{opt.}}$ and $\left(\frac{\Omega_0}{\eta^2}\right)^{CZ}_{\mathrm{opt.}}$ can be computed by numerically obtaining the values of $\tau_q$. For this, we solve the Schrodinger dynamics for the effective Hamiltonians $\hat H_{1'1'}, \hat H_{1',0}, \hat H_{00}$ with $\Gamma_0= \Gamma^{(N_0)}_1=0$ for total pulse duration $T$. We use $\Omega(t)= \Omega_0 \exp[i(\varphi(t\Omega_0)]$ with $\Omega_0 T = 7.612$, corresponding to the time-optimal solution for the blockade CZ gate for Rydberg qubits~\cite{janduraTimeOptimalTwoThreeQubit2022}. The phase $\varphi(t\Omega_0)$ is taken from the time-optimal pulse plotted in Fig.~\ref{fig::summary}(g).  We obtain $\tau_{q}= \Omega_0\int_{0}^{T} |\langle q|\psi_{s}(t)\rangle|^2 dt$ where $|\psi_{s}(t)\rangle$ is the state at time $t$ for initial state $|s\rangle$ in a given subspace. That is, $|s\rangle$ corresponds to the states $|1'1'\rangle, |1'0\rangle, |00\rangle$ for $|q\rangle$ associated with $\hat H_{1'1'}, \hat H_{1'0}, \hat H_{00}$ respectively. 

By substituting the obtained optimal parameters, we get the gate error for $\Gamma_0 \neq \Gamma_1\neq 0$ from Eq.~\eqref{cz_infid} as 
\begin{equation}\label{boundcz}
    1- F^{CZ}_{\mathrm{opt.}} = 6.45 \frac{1}{\sqrt{C}}
\end{equation}

\begin{figure*}[ht]
    \centering
    \includegraphics[width=\textwidth]{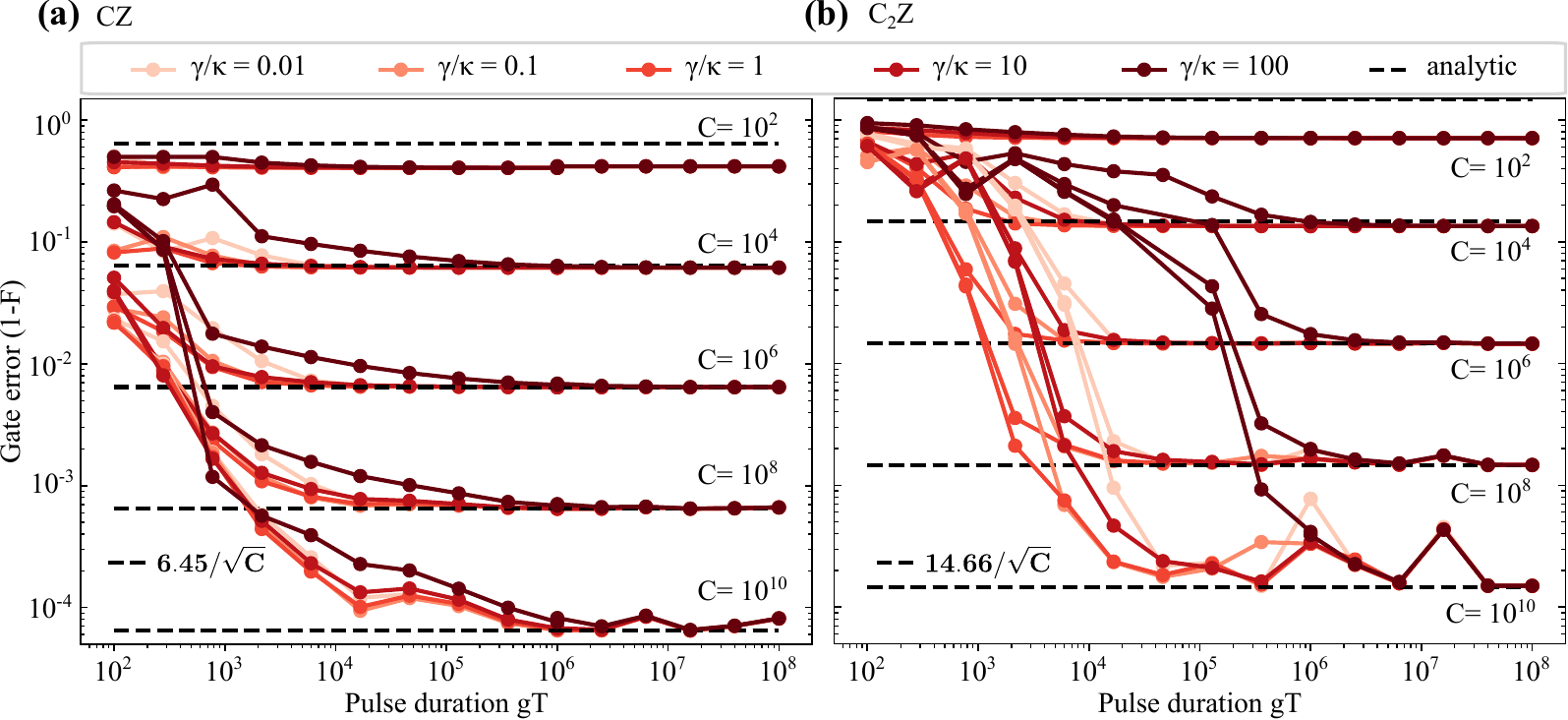}
    \caption{Gate error for as a function of the total operation time for (a) $CZ$ gate and (b) $C_2Z$ gate for $C= 10^2, 10^4, 10^6, 10^8, 10^{10}$ and $\gamma/\kappa= 0.01, 0.1, 1, 10, 100$. The infidelity converges to the analytical estimate (dashed lines) $1- F \propto 1/\sqrt{C}$ (See text Sec.~\ref{subsec::CZ},~\ref{subsec::C2Z}) obtained in the limit $T \rightarrow \infty$.}
    \label{fig::CZ_C2Z}
\end{figure*}

In Fig.~\ref{fig::summary}(f) and Fig.~\ref{fig::CZ_C2Z}(a), we numerically obtain the gate error by simulating the dynamics of the state $|\psi_{in}\rangle= (|1'1'\rangle + |1'0\rangle + |01'\rangle + |00\rangle)/2$ under the full Hamiltonian (Eq.~\eqref{eq::full_H}) for time $T$ using the time-optimal pulse to obtain the final state $|\psi(T)\rangle$. We use $\Omega(t)= \Omega_0 \exp[{i(\varphi(t\Omega_0)+ \delta_{gl}t)}]$ with $\delta_{gl}= \eta^2\Delta/(2g^2)$ (as discussed in Sec.~\ref{sec::W_state}), $\Omega_0 T= 7.612$ and the laser phase $\varphi(\Omega_0 t)$ corresponding to the time-optimal pulse as shown in Fig.~\ref{fig::summary}(g). The expected final state under a $CZ$ gate is $|\psi_f\rangle = (|1'1'\rangle + e^{i\theta} |1'0\rangle + e^{i\theta} |01'\rangle + e^{i(2\theta+ \pi)}|00\rangle)/2$ where $\theta$ is  the single qubit phase. The single-qubit phase $\theta$ is optimized to minimize the gate error computed as $1-F= |\langle \psi_f| \psi(T) \rangle|^2 $. In Fig.~\ref{fig::summary}(d), the obtained gate error is plotted as a function of cooperativity $C$ for $gT= 10^8$. The numerical results (square markers) are independent of $\gamma/\kappa$ ratio and for large $C$ match excellently with the analytical estimate (dashed line) as obtained in Eq.\eqref{boundcz}. In Fig.~\ref{fig::CZ_C2Z}(a), the gate error is plotted as a function of the total pulse duration $gT$ for different values of cooperativities and linewidth ratios $\gamma/\kappa$. The dashed lines corresponds to the analytical error obtained in Eq.~\eqref{boundcz}. 
We see that for large cooperativities, in the limit $T \rightarrow \infty$, the gate error converges to our analytical estimate which depends only on $C= g^2/(\kappa\gamma)$ and is independent of the ratio $\gamma/\kappa$.

In the next section, we naturally extend the application of the effective blockade Hamiltonian to implement a non-local $C_2Z$ gate which acts on arbitrarily initialized $N=3$ atoms.
\subsection{$C_2Z$ gate}\label{subsec::C2Z}
In this section, similar to the $CZ$ gate, we show the implementation of a $C_2Z$ gate in the blockade regime with the time-optimal pulse shown in Fig.~\ref{fig::summary}(h). Here, we consider three-atom computational states $\{|1'1'1'\rangle, |1'1'0\rangle, |1'00\rangle, |000\rangle\}$. The effective blockade Hamiltonians (up to single qubit gates) are obtained in each of the decoupled subspaces corresponding to these states respectively as 
\begin{align}
\begin{split}
    \hat H_{1'1'1'} &=  - \frac{i\Gamma_0}{2}|1'1'1'\rangle \langle 1'1'1'|\\
    \hat H_{1'1'0} &= \frac{\Omega(t)}{2}\left(|1'1'0\rangle \langle 1'1'1|+ \mathrm{h.c}\right) - \frac{i\Gamma_0}{2}|1'1'0\rangle \langle 1'1'0|\\&- \frac{i\Gamma^{(1)}_1}{2}|1'1'1\rangle \langle 1'1'1|\\
    \hat H_{1'00} &= \frac{\sqrt{2}\Omega(t)}{2}\left(|1'00\rangle \langle 1'W|+ \mathrm{h.c}\right) - \frac{i\Gamma_0}{2}|1'00\rangle \langle 1'00|\\&- \frac{i\Gamma^{(2)}_1}{2}|1'W\rangle \langle 1'W|\\
    \hat H_{000} &= \frac{\sqrt{3}\Omega(t)}{2}\left(|000\rangle \langle W1|+ \mathrm{h.c}\right) - \frac{i\Gamma_0}{2}|000\rangle \langle 000|\\& - \frac{i\Gamma^{(3)}_1}{2}|W1\rangle \langle W1|
    \end{split}
\end{align}
where $|1'W\rangle= |1'\rangle \otimes |W\rangle $ and 
$|W1\rangle = (|0 0 1\rangle + |0 1 0\rangle + |1 0 0\rangle)/\sqrt{3} $.

Following a similar treatment as in Sec.~\ref{sec:fid_calculation}, we can write the $C_2Z$ gate error and the optimal parameters as
\begin{widetext}
\begin{align}
    1 - F^{C_2Z}_{\mathrm{opt.}} &= \frac{1}{8\Omega_0} \left((\tau_{1'1'1'}+ 3\tau_{1'1'0} + 3\tau_{1'00}+ \tau_{1'1'1'})\Gamma_{0} + 3\tau_{1'1'1}\Gamma^{(1)}_1 + 3\tau_{1'W}\Gamma^{(2)}_1 + \tau_{W1}\Gamma^{(3)}_{1} \right)\\
    (\Delta)^{C_2Z}_{\mathrm{opt.}} &= \left(\frac{8(3\tau_{1'1'1} + 3\tau_{1'W} + \tau_{W1})}{(\tau_{1'1'1'}+ 3\tau_{1'1'0}+ 3\tau_{1'00}+ \tau_{000}+ 12\tau_{1'1'1}+ 12\tau_{1'W} + 4\tau_{W1})}\right)^{1/4} \sqrt{\frac{\gamma}{\kappa}} g\\
    \left(\frac{\eta^2}{\Omega_0}\right)^{C_2Z}_{\mathrm{opt.}}&= \sqrt{\frac{\frac{1}{8}(3\tau_{1'W}+ 2\tau_{W1}) (\frac{\kappa}{2} + \frac{\gamma g^2}{\Delta^2})}{\frac{1}{32}\frac{\Delta^2\kappa}{g^4}(\tau_{1'1'1'}+ 3\tau_{1'1'0}+ 3\tau_{1'00}+ \tau_{000} + 12\tau_{1'1'1}+ 12\tau_{1'W}+ 4\tau_{W1}) + \frac{1}{8}\frac{\gamma}{g^2}(3\tau_{1'1'1}+ 3\tau_{1'W}+ \tau_{W1})}}
\end{align}
\end{widetext}
We obtain the numerically calculated values $\tau_q$ as described in Sec.~\ref{subsec::CZ}, here by using $\Omega(t)= \Omega_0 \exp[{i(\varphi(t\Omega_0)+ \delta_{gl}t)}]$ with $\delta_{gl}=0$, $\Omega_0 T = 10.809$ and $\varphi(t\Omega_0)$ corresponding to the time-optimal pulse for $C_2Z$ gate as shown in Fig.~\ref{fig::summary}(h).
On substituting the $\tau_q$ values, we obtain
\begin{equation}\label{boundc2z}
    1- F^{C_2Z}_{\mathrm{opt.}}= 14.66 \frac{1}{\sqrt{C}}.
\end{equation}
In Figs~\ref{fig::summary}(d) and~\ref{fig::CZ_C2Z}(b), we numerically obtain the $C_2Z$ gate error by simulating the dynamics of the state $|\psi_{in}\rangle= (|1'1'1'\rangle + |1'1'0\rangle + |01'1'\rangle + |1'01'\rangle+ |1'00\rangle+ |01'0\rangle+ |001'\rangle+ |000\rangle)/\sqrt{8}$ under the full Hamiltonian (Eq.~\eqref{eq::full_H}) for time $T$ using the time-optimal pulse for $C_2Z$ gate. That is, we use  $\Omega(t)= \Omega_0 \exp[{i(\varphi(t\Omega_0)+ \delta_{gl}t)}]$ with $\delta_{gl}= \eta^2\Delta/(2g^2)$ (as discussed in Sec.~\ref{sec::W_state}), $\Omega_0 T= 10.809$, and $\varphi(t\Omega_0)$ from the time-optimal pulse plotted in Fig.~\ref{fig::summary}(h). Note that here the time-optimal pulse is designed such that the conditional phase is acquired by the state $|1'1'1'\rangle$, i.e. the expected final state under the $C_2Z$ gate operation is $|\psi_f\rangle =( e^{i(3\theta+ \pi)}|1'1'1'\rangle + e^{i2\theta}\left(|1'1'0\rangle + |01'1'\rangle + |1'01'\rangle\right) + e^{i\theta}\left(|1'00\rangle + |01'0\rangle + |001'\rangle \right) + |000\rangle )/\sqrt{8}$~\cite{everedHighfidelityParallelEntangling2023}, where $\theta$ is a single-qubit phase. For the final state $|\psi(T)\rangle$, the gate error or infidelity is calculated as $1-F= |\langle \psi_f| \psi(T) \rangle|^2 $. 

Figure.~\ref{fig::W_summary}(f) verifies the gate error scaling as a function of cooperativity $C$. The numerical points (triangles) obtained for different values of $\gamma/\kappa$ have a good match with the analytical estimate (dashed line) obtained in Eq.~\eqref{boundc2z} for large cooperativities which is independent of $\gamma/\kappa$. In Fig.~\ref{fig::CZ_C2Z}(b), the gate error is plotted as a function of different total pulse duration $gT$ for different ratios $\gamma/\kappa$ and different cooperativities $C$. We see a general trend of the error decreasing with increasing $T$ for all $\gamma/\kappa$ ratios, and converging close to the analytical estimate (dashed line) depending only on $C$ and independent of $\gamma/\kappa$.

In this section, we have seen the application of the blockade mechanism in implementing a non-local $CZ$ and $C_2Z$ gate between distant physical qubits. The effective blockade dynamics result in a differential evolution of the states with different number of atoms initialized in the $|0\rangle$ state corresponding to different initial computational qubit states with qubit subspace spanned by $\{|0\rangle, |1'\rangle\}$. In the next section, we give fidelity estimates for some realistic cavity QED parameters for neutral atom and molecular qubits. 

\section{Realistic fidelities for experiments with neutral atoms and molecules}
In this section, we present some examples of experimental quantum computing platforms where the non-local excitation blockade can be implemented. 

As a first example, we consider neutral atom systems of $^{87}Rb$ atoms coupled to a fiber Fabry-Perot optical cavity~\cite{hungerFiberFabryPerot2010, uphoffFrequencySplittingPolarization2015, grinkemeyerErrorDetectedQuantumOperations2024} similar to the cavity in Ref.~\cite{barontiniDeterministicGenerationMultiparticle2015}. We consider the states $|0\rangle = |5 S_{1/2} \, F=1\, m_F =0\rangle$, $|1\rangle = 5 ^2S_{1/2} \, F=2 \, m_F= 0\rangle$, $|1'\rangle = |5 ^2S_{1/2}\, F= 2\, m_F= 1 \rangle$ and $|e\rangle = |5 ^2P_{3/2}\, F= 3\, m_F= 0\rangle$ such that the $D_2$ transition line in $^{87}Rb$ with wavelength $\lambda = 780\,$nm and $\gamma = 2\pi \times 6 \,$MHz corresponds to the cavity-coupled $|1\rangle \leftrightarrow |e\rangle$ transition. We consider a Fabry-Perot fiber cavity with finesse $\mathcal{F} \approx 2 \times 10^5 $, waist radius $\omega_r \approx 2 \, \mu$m, and $L= 40 \, \mu$m which gives $C = 3\lambda^2\mathcal{F} /(2\pi^3\omega_r^2) \approx 1500$, $g = \sqrt{3\lambda^2c\gamma/(2\pi^2w_r^2L)} \approx 2\pi \times 400$ MHz and $\kappa= \pi C/L \mathcal{F} \approx 2\pi \times 20 \,$MHz. With this system, the W state preparation with $N=10$ atoms is achieved with fidelity $F= 86\%$ in time $T \approx 10^4/g = 4\,\mu$s. A $CZ, C_2Z$ gate is realized in with fidelities $\approx 80 \%, 69 \% $ respectively in time $T = 10^4/g \approx 4 \mu$s.

Secondly, we consider the case of achieving strong coupling by coupling Rydberg-Rydberg transitions with large electric dipole moments to a microwave on-chip resonator~\cite{kaiserCavitydrivenRabiOscillations2022, pritchardHybridAtomphotonQuantum2014}. Here we assume both $|1\rangle$ and $|e\rangle$ states as Rydberg states $|90 ^2P_{3/2}\rangle$ and $|90 ^2 S_{1/2}\rangle$ in Cs respectively. We have $1/\gamma= 820 \,\mu$s and also a non-negligible decay rate $\gamma_1$ of the $|1\rangle$ state given by $1/\gamma_1 = 2 \,$ms. We have the transition frequency $\omega_e- \omega_1 = \omega_{1e}= 2\pi \times 5.03\,$ GHz and coupling strength $g \approx 2\pi \times 4\,$MHz\cite{pritchardHybridAtomphotonQuantum2014}. Assuming an achievable quality factor of microwave resonator of $Q \approx 3 \times 10^8$~\cite{leiHighCoherenceSuperconducting2020}, we have $\kappa= \omega_{1e}/Q = 2\pi \times 17\,$Hz corresponding to cavity photon lifetime $1/\kappa  \approx 9.3\,$ms. With this system, we can hence achieve a cooperativity of $C \approx 5 \times 10^9$. Incorporating the additional  decay $\gamma_1$ of state $|1\rangle$, the effective decay of the target W state is modified as $\Gamma_1'= \Gamma_1 + \left(1+ \frac{\eta^2\Delta^2}{g^4} + \frac{(N-1)\Omega_0^2(1+ g^2/\Delta^2)}{\eta^2}\right)\gamma_1$. The W-state preparation infidelity hence has an extra contribution proportional to $\gamma_1 T/2$, and diverges in the limit $T \rightarrow \infty$ after an initial decrease for finite $T$. With this system, a maximum  fidelity of $F= 98.3\%$ is obtained for a pulse duration of $T \approx 930/g \approx 37\,\mu$s for $N=10$ atoms. A $CZ$ gate  with infidelity $1-F= 4.5 \times 10^{-3}$, is realized in time $T= 280/g \approx 11\,\mu$s, and a $C_2Z$ gate with infidelity $1-F= 7 \times 10^{-3}$ is realized in time $T = 530/g \approx 21\,\mu$s.

We note that the use of Rydberg states can be further leveraged by exciting them to maximal angular momentum states known as circular Rydberg states which have inherently long lifetimes of several seconds. It is possible to similarly couple transitions within circular Rydberg states (with principal quantum numbers of the order $50$) to a high-quality Fabry-Perot microwave resonator with superconducting mirrors~\cite{bruneObservingProgressiveDecoherence1996, signolesCoherentTransferLowAngularMomentum2017}. The Rydberg atoms can be further trapped inside a micro structure such that the spontaneous emission from circular states is inhibited~\cite{mozleyTrappingCoherentManipulation2005, cohenQuantumComputingCircular2021} giving an increased lifetime of $\approx 100 s$.

Another platform of interest is a system of cold polar molecules coupled to a superconducting high-Q stripline cavity. Here, we show the example of CaF molecules. We choose the states in the basis $|N,S,J,I,F,m_F\rangle$ as $|0\rangle=|0,1/2,1/2,1/2,0,0\rangle$, $|1\rangle=|0,1/2,1/2,1/2,1,0\rangle$ and $|e\rangle=|1,1/2,1/2,1/2,1,0\rangle$. Here $\omega_e - \omega_1= \omega_{1e} \approx 2\pi \times 21$ GHz and $\gamma < 10^{-2}\,$Hz~\cite{buhmannSurfaceinducedHeatingCold2008} is negligible. A coupling strength of $g\approx 2\pi \times 10$ kHz for the $|1\rangle \leftrightarrow |e\rangle$ is achievable~\cite{rablHybridQuantumProcessors2006}. With a quality factor $Q \approx 3 \times 10^8$, $\kappa= \omega_{1e}/Q= 2\pi \times 70 Hz$ corresponding to cavity-photon lifetime $1/\kappa \approx 2.3\,$ms. With this system, choosing $\Delta= 2\pi \times 50\,$kHz, a W state with $N=10$ atoms can be prepared with $91\%$ fidelity in time $T \approx 1.9\,$ms. 

\section{Conclusion and Outlook}

We have presented a cavity polariton blockade mechanism in a cavity QED setup which is exploited for generation of a non-local multi-atom W-state and non-local $CZ$ and $C_2Z$ gates. The latter are obtained just by driving the cavity externally with a probe laser along with an additional global pulse acting on the atoms.  
A complete quantum mechanical treatment of the system, including the effects of spontaneous emission and cavity decay, allows to characterize the W-state preparation fidelity and the $CZ, C_2Z$ gate errors as a function of the single particle cooperativity C. The errors are found to scale as $\mathcal{O}(C^{-1/2})$, moreover the error of N-atom W-state preparation saturates with $N$. We present the protocol results with example setups of neutral atoms coupled to a common optical cavity mode, and Rydberg atoms and cold polar molecules coupled to a common microwave mode. The former achieve the W state preparation for moderately sized systems of $N=10$ in fast operation times of a few microseconds; while the latter achieve high state preparation fidelities. 

%For the W-state preparation, each atomic system is modelled as a three-level system with computational states $|0\rangle$ and $|1\rangle$, and for $CZ$ and $C_2Z$ gates, an additional $|1'\rangle$ state is used such that the computational subspace is spanned by the states $|0\rangle$ and $|1'\rangle$. The cavity mode is assumed throughout to couple the states $|1\rangle$ and an excited state $|e\rangle$. 

Moreover, a cavity-QED setup with minimal control knobs- the cavity probe and global qubit pulse- supported by the current experimental progress with neutral atoms in optical cavities~\cite{grinkemeyerErrorDetectedQuantumOperations2024}, can be used as a toolbox to prepare arbitrary many-qubit entangled states by employing optimal control techniques. These techniques can be tailored to prepare optimal states for quantum sensing~\cite{srivastavaEntanglementenhancedQuantumSensing2024}, or increasingly complex entangled states optimized for quantum Fisher information, which is a subject of future work. 

\section{Acknowledgments}
This research has received funding from the European Union’s Horizon 2020 research and innovation programme under the Marie Sk{\l}odowska-Curie project 847471 (QUSTEC) and project 955479 (MOQS), the Horizon Europe programme HORIZON-CL4-2021-DIGITAL-EMERGING-01-30 via the project 101070144 (EuRyQa) and from the French National Research Agency under the Investments of the Future Program projects ANR-21-ESRE-0032 (aQCess) and ANR-22-CE47-0013-02 (CLIMAQS). GKB acknowledges support from the Australian Research Council Centre of Excellence for Engineered Quantum Systems (Grant No. CE 170100009).

\appendix

%\section{Rewriting Hamiltonian in symmetric basis}\label{app::H_in_symm_basis}

\section{Energy shifts due to perturbative couplings from cavity drive}\label{app::shifts_n0_n1}

In this section, we present the calculation of the perturbative energy shifts on the states $\state{0}{0}{0}$ and $\state{1}{0}{0}$, due to the couplings governed by the non-Hermitian Hamiltonian $\hat H^{(\kappa, \gamma, \eta)}$ in the $k=0$ and $k=1$ subspaces of $\hat H^{(\Delta, \delta, g)}$. 

Using time-independent perturbation theory, we calculate the energy shifts up to third order in $\hat H^{(\kappa, \gamma, \eta)}$, which also correspond to effective linewidths up to first order in $\kappa$ and $\gamma$. The shifts on states $\state{0}{0}{0}$ and $\state{1}{0}{0}$ are denoted as $\Delta E_0$ and $\Delta E_1$ respectively, which are obtained as 

\begin{widetext}
\begin{align}\label{energy_shift_01}
\begin{split}
    \Delta E_{0} &= \statebra{0}{0}{0} \hat H^{(\kappa,\gamma,\eta)} \state{0}{0}{0} \\
    &+ \frac{\statebra{0}{0}{0}\hat H^{(\kappa,\gamma,\eta)}|p_0^{+}\rangle \langle p_0^{+} | \hat H^{(\kappa,\gamma,\eta)}\state{0}{0}{0}}{-\epsilon_0^{+}}
    + \frac{\statebra{0}{0}{0}\hat H^{(\kappa,\gamma,\eta)} \state{0}{0}{0} \langle p_0^{+} | \hat H^{(\kappa,\gamma,\eta)}|p_0^{+}\rangle \langle p_0^{+} | \hat H^{(\kappa,\gamma,\eta)}\state{0}{0}{0}} {(-\epsilon_0^+)^2}\\
    \Delta E_{1} &= \statebra{1}{0}{0} \hat H^{(\kappa,\gamma,\eta)} \state{1}{0}{0}
    + \statebra{1}{0}{0}\hat H^{(\kappa,\gamma,\eta)} \left(\frac{|p_1^{+}\rangle\langle p_1^{+}|}{-\epsilon_1^{+}} + \frac{|p_1^{-}\rangle \langle p_1^{-}|}{-\epsilon_1^{-}}\right)\hat H^{(\kappa,\gamma,\eta)} \state{1}{0}{0}\\
    &+ \statebra{1}{0}{0}\hat H^{(\kappa,\gamma,\eta)} \left(\frac{|p_1^{+}\rangle\langle p_1^{+}|}{-\epsilon_1^{+}} + \frac{|p_1^{-}\rangle \langle p_1^{-}|}{-\epsilon_1^{-}}\right)\hat H^{(\kappa,\gamma,\eta)}\left(\frac{|p_1^{+}\rangle\langle p_1^{+}|}{-\epsilon_1^{+}} + \frac{|p_1^{-}\rangle \langle p_1^{-}|}{-\epsilon_1^{-}}\right)\hat H^{(\kappa,\gamma,\eta)} \state{1}{0}{0}. %\\
    % &= -\frac{i}{2}\kappa -\eta^2 (\statebra{1}{0}{0} \hat a )\left(\hat H_{n=1, k=1} ^{(\Delta, \delta, g)}\right)^{-1} (\hat a^{\dagger} \state{1}{0}{0}) 
    % -\eta^2 (\statebra{1}{0}{0} \hat a )\left(\hat H_{n=1, k=1} ^{(\Delta, \delta, g)}\right)^{-1} \hat H^{(\kappa, \gamma, \delta)}_{n=1, k=1} \left(\hat H_{n=1, k=1} ^{(\Delta, \delta, g)}\right)^{-1} (\hat a^{\dagger} \state{1}{0}{0})
    \end{split}
\end{align}
\end{widetext}

To simplify Eq.~\eqref{energy_shift_01}, we use 
\begin{align}
    &\left(\frac{|p_1^{+}\rangle\langle p_1^{+}|}{\epsilon_1^{+}} + \frac{|p_1^{-}\rangle \langle p_1^{-}|}{\epsilon_1^{-}}\right) = \left(\hat H^{(\Delta, \delta, g)}_{n=1, k=1}\right)^{-1}, \\
    &\text{with   } \hat H^{(\Delta, \delta, g)}_{n=1, k=1} = 
    \begin{bmatrix}
    \delta & g\\
    g & \Delta
    \end{bmatrix},\\
    &\text{and    }
    \hat H^{(\kappa, \gamma, \eta)}_{n=1, k=1} = 
    \begin{bmatrix}
        -i\frac{\kappa}{2}& 0\\
        0 & -i\frac{\gamma}{2}
    \end{bmatrix},
\end{align}
where the matrices are written in the basis $\{|1_1 0_e 1_{ph}\rangle, |0_1 1_e 0_{ph}\rangle\}$. On calculating Eq.~\eqref{energy_shift_01} using the above matrices, one obtains the energy shifts as in Eq.~\eqref{eq:shift_E0} and Eq.~\eqref{eq:shift_E1}.

\bibliography{references}

\begin{thebibliography}{56}%
\makeatletter
\providecommand \@ifxundefined [1]{%
 \@ifx{#1\undefined}
}%
\providecommand \@ifnum [1]{%
 \ifnum #1\expandafter \@firstoftwo
 \else \expandafter \@secondoftwo
 \fi
}%
\providecommand \@ifx [1]{%
 \ifx #1\expandafter \@firstoftwo
 \else \expandafter \@secondoftwo
 \fi
}%
\providecommand \natexlab [1]{#1}%
\providecommand \enquote  [1]{``#1''}%
\providecommand \bibnamefont  [1]{#1}%
\providecommand \bibfnamefont [1]{#1}%
\providecommand \citenamefont [1]{#1}%
\providecommand \href@noop [0]{\@secondoftwo}%
\providecommand \href [0]{\begingroup \@sanitize@url \@href}%
\providecommand \@href[1]{\@@startlink{#1}\@@href}%
\providecommand \@@href[1]{\endgroup#1\@@endlink}%
\providecommand \@sanitize@url [0]{\catcode `\\12\catcode `\$12\catcode
  `\&12\catcode `\#12\catcode `\^12\catcode `\_12\catcode `\%12\relax}%
\providecommand \@@startlink[1]{}%
\providecommand \@@endlink[0]{}%
\providecommand \url  [0]{\begingroup\@sanitize@url \@url }%
\providecommand \@url [1]{\endgroup\@href {#1}{\urlprefix }}%
\providecommand \urlprefix  [0]{URL }%
\providecommand \Eprint [0]{\href }%
\providecommand \doibase [0]{http://dx.doi.org/}%
\providecommand \selectlanguage [0]{\@gobble}%
\providecommand \bibinfo  [0]{\@secondoftwo}%
\providecommand \bibfield  [0]{\@secondoftwo}%
\providecommand \translation [1]{[#1]}%
\providecommand \BibitemOpen [0]{}%
\providecommand \bibitemStop [0]{}%
\providecommand \bibitemNoStop [0]{.\EOS\space}%
\providecommand \EOS [0]{\spacefactor3000\relax}%
\providecommand \BibitemShut  [1]{\csname bibitem#1\endcsname}%
\let\auto@bib@innerbib\@empty
%</preamble>
\bibitem [{\citenamefont {Preskill}(2018)}]{preskillQuantumComputingNISQ2018}%
  \BibitemOpen
  \bibfield  {author} {\bibinfo {author} {\bibfnamefont {John}\ \bibnamefont
  {Preskill}},\ }\bibfield  {title} {\enquote {\bibinfo {title} {Quantum
  {{Computing}} in the {{NISQ}} era and beyond},}\ }\href {\doibase
  10.22331/q-2018-08-06-79} {\bibfield  {journal} {\bibinfo  {journal}
  {Quantum}\ }\textbf {\bibinfo {volume} {2}},\ \bibinfo {pages} {79} (\bibinfo
  {year} {2018})},\ \Eprint {http://arxiv.org/abs/1801.00862} {arXiv:1801.00862
  [quant-ph]} \BibitemShut {NoStop}%
\bibitem [{\citenamefont {Bharti}\ \emph {et~al.}(2022)\citenamefont {Bharti},
  \citenamefont {{Cervera-Lierta}}, \citenamefont {Kyaw}, \citenamefont {Haug},
  \citenamefont {{Alperin-Lea}}, \citenamefont {Anand}, \citenamefont
  {Degroote}, \citenamefont {Heimonen}, \citenamefont {Kottmann}, \citenamefont
  {Menke}, \citenamefont {Mok}, \citenamefont {Sim}, \citenamefont {Kwek},\
  and\ \citenamefont
  {{Aspuru-Guzik}}}]{bhartiNoisyIntermediatescaleQuantum2022}%
  \BibitemOpen
  \bibfield  {author} {\bibinfo {author} {\bibfnamefont {Kishor}\ \bibnamefont
  {Bharti}}, \bibinfo {author} {\bibfnamefont {Alba}\ \bibnamefont
  {{Cervera-Lierta}}}, \bibinfo {author} {\bibfnamefont {Thi~Ha}\ \bibnamefont
  {Kyaw}}, \bibinfo {author} {\bibfnamefont {Tobias}\ \bibnamefont {Haug}},
  \bibinfo {author} {\bibfnamefont {Sumner}\ \bibnamefont {{Alperin-Lea}}},
  \bibinfo {author} {\bibfnamefont {Abhinav}\ \bibnamefont {Anand}}, \bibinfo
  {author} {\bibfnamefont {Matthias}\ \bibnamefont {Degroote}}, \bibinfo
  {author} {\bibfnamefont {Hermanni}\ \bibnamefont {Heimonen}}, \bibinfo
  {author} {\bibfnamefont {Jakob~S.}\ \bibnamefont {Kottmann}}, \bibinfo
  {author} {\bibfnamefont {Tim}\ \bibnamefont {Menke}}, \bibinfo {author}
  {\bibfnamefont {Wai-Keong}\ \bibnamefont {Mok}}, \bibinfo {author}
  {\bibfnamefont {Sukin}\ \bibnamefont {Sim}}, \bibinfo {author} {\bibfnamefont
  {Leong-Chuan}\ \bibnamefont {Kwek}}, \ and\ \bibinfo {author} {\bibfnamefont
  {Al{\'a}n}\ \bibnamefont {{Aspuru-Guzik}}},\ }\bibfield  {title} {\enquote
  {\bibinfo {title} {Noisy intermediate-scale quantum algorithms},}\ }\href
  {\doibase 10.1103/RevModPhys.94.015004} {\bibfield  {journal} {\bibinfo
  {journal} {Reviews of Modern Physics}\ }\textbf {\bibinfo {volume} {94}},\
  \bibinfo {pages} {015004} (\bibinfo {year} {2022})}\BibitemShut {NoStop}%
\bibitem [{\citenamefont {Kim}\ \emph {et~al.}(2023)\citenamefont {Kim},
  \citenamefont {Eddins}, \citenamefont {Anand}, \citenamefont {Wei},
  \citenamefont {{van den Berg}}, \citenamefont {Rosenblatt}, \citenamefont
  {Nayfeh}, \citenamefont {Wu}, \citenamefont {Zaletel}, \citenamefont
  {Temme},\ and\ \citenamefont {Kandala}}]{kimEvidenceUtilityQuantum2023}%
  \BibitemOpen
  \bibfield  {author} {\bibinfo {author} {\bibfnamefont {Youngseok}\
  \bibnamefont {Kim}}, \bibinfo {author} {\bibfnamefont {Andrew}\ \bibnamefont
  {Eddins}}, \bibinfo {author} {\bibfnamefont {Sajant}\ \bibnamefont {Anand}},
  \bibinfo {author} {\bibfnamefont {Ken~Xuan}\ \bibnamefont {Wei}}, \bibinfo
  {author} {\bibfnamefont {Ewout}\ \bibnamefont {{van den Berg}}}, \bibinfo
  {author} {\bibfnamefont {Sami}\ \bibnamefont {Rosenblatt}}, \bibinfo {author}
  {\bibfnamefont {Hasan}\ \bibnamefont {Nayfeh}}, \bibinfo {author}
  {\bibfnamefont {Yantao}\ \bibnamefont {Wu}}, \bibinfo {author} {\bibfnamefont
  {Michael}\ \bibnamefont {Zaletel}}, \bibinfo {author} {\bibfnamefont
  {Kristan}\ \bibnamefont {Temme}}, \ and\ \bibinfo {author} {\bibfnamefont
  {Abhinav}\ \bibnamefont {Kandala}},\ }\bibfield  {title} {\enquote {\bibinfo
  {title} {Evidence for the utility of quantum computing before fault
  tolerance},}\ }\href {\doibase 10.1038/s41586-023-06096-3} {\bibfield
  {journal} {\bibinfo  {journal} {Nature}\ }\textbf {\bibinfo {volume} {618}},\
  \bibinfo {pages} {500--505} (\bibinfo {year} {2023})}\BibitemShut {NoStop}%
\bibitem [{\citenamefont {Graham}\ \emph {et~al.}(2022)\citenamefont {Graham},
  \citenamefont {Song}, \citenamefont {Scott}, \citenamefont {Poole},
  \citenamefont {Phuttitarn}, \citenamefont {Jooya}, \citenamefont {Eichler},
  \citenamefont {Jiang}, \citenamefont {Marra}, \citenamefont {Grinkemeyer},
  \citenamefont {Kwon}, \citenamefont {Ebert}, \citenamefont {Cherek},
  \citenamefont {Lichtman}, \citenamefont {Gillette}, \citenamefont {Gilbert},
  \citenamefont {Bowman}, \citenamefont {Ballance}, \citenamefont {Campbell},
  \citenamefont {Dahl}, \citenamefont {Crawford}, \citenamefont {Blunt},
  \citenamefont {Rogers}, \citenamefont {Noel},\ and\ \citenamefont
  {Saffman}}]{grahamMultiqubitEntanglementAlgorithms2022}%
  \BibitemOpen
  \bibfield  {author} {\bibinfo {author} {\bibfnamefont {T.~M.}\ \bibnamefont
  {Graham}}, \bibinfo {author} {\bibfnamefont {Y.}~\bibnamefont {Song}},
  \bibinfo {author} {\bibfnamefont {J.}~\bibnamefont {Scott}}, \bibinfo
  {author} {\bibfnamefont {C.}~\bibnamefont {Poole}}, \bibinfo {author}
  {\bibfnamefont {L.}~\bibnamefont {Phuttitarn}}, \bibinfo {author}
  {\bibfnamefont {K.}~\bibnamefont {Jooya}}, \bibinfo {author} {\bibfnamefont
  {P.}~\bibnamefont {Eichler}}, \bibinfo {author} {\bibfnamefont
  {X.}~\bibnamefont {Jiang}}, \bibinfo {author} {\bibfnamefont
  {A.}~\bibnamefont {Marra}}, \bibinfo {author} {\bibfnamefont
  {B.}~\bibnamefont {Grinkemeyer}}, \bibinfo {author} {\bibfnamefont
  {M.}~\bibnamefont {Kwon}}, \bibinfo {author} {\bibfnamefont {M.}~\bibnamefont
  {Ebert}}, \bibinfo {author} {\bibfnamefont {J.}~\bibnamefont {Cherek}},
  \bibinfo {author} {\bibfnamefont {M.~T.}\ \bibnamefont {Lichtman}}, \bibinfo
  {author} {\bibfnamefont {M.}~\bibnamefont {Gillette}}, \bibinfo {author}
  {\bibfnamefont {J.}~\bibnamefont {Gilbert}}, \bibinfo {author} {\bibfnamefont
  {D.}~\bibnamefont {Bowman}}, \bibinfo {author} {\bibfnamefont
  {T.}~\bibnamefont {Ballance}}, \bibinfo {author} {\bibfnamefont
  {C.}~\bibnamefont {Campbell}}, \bibinfo {author} {\bibfnamefont {E.~D.}\
  \bibnamefont {Dahl}}, \bibinfo {author} {\bibfnamefont {O.}~\bibnamefont
  {Crawford}}, \bibinfo {author} {\bibfnamefont {N.~S.}\ \bibnamefont {Blunt}},
  \bibinfo {author} {\bibfnamefont {B.}~\bibnamefont {Rogers}}, \bibinfo
  {author} {\bibfnamefont {T.}~\bibnamefont {Noel}}, \ and\ \bibinfo {author}
  {\bibfnamefont {M.}~\bibnamefont {Saffman}},\ }\bibfield  {title} {\enquote
  {\bibinfo {title} {Multi-qubit entanglement and algorithms on a neutral-atom
  quantum computer},}\ }\href {\doibase 10.1038/s41586-022-04603-6} {\bibfield
  {journal} {\bibinfo  {journal} {Nature}\ }\textbf {\bibinfo {volume} {604}},\
  \bibinfo {pages} {457--462} (\bibinfo {year} {2022})}\BibitemShut {NoStop}%
\bibitem [{\citenamefont {Omran}\ \emph {et~al.}(2019)\citenamefont {Omran},
  \citenamefont {Levine}, \citenamefont {Keesling}, \citenamefont {Semeghini},
  \citenamefont {Wang}, \citenamefont {Ebadi}, \citenamefont {Bernien},
  \citenamefont {Zibrov}, \citenamefont {Pichler}, \citenamefont {Choi},
  \citenamefont {Cui}, \citenamefont {Rossignolo}, \citenamefont {Rembold},
  \citenamefont {Montangero}, \citenamefont {Calarco}, \citenamefont {Endres},
  \citenamefont {Greiner}, \citenamefont {Vuleti{\'c}},\ and\ \citenamefont
  {Lukin}}]{omranGenerationManipulationSchrodinger2019}%
  \BibitemOpen
  \bibfield  {author} {\bibinfo {author} {\bibfnamefont {A.}~\bibnamefont
  {Omran}}, \bibinfo {author} {\bibfnamefont {H.}~\bibnamefont {Levine}},
  \bibinfo {author} {\bibfnamefont {A.}~\bibnamefont {Keesling}}, \bibinfo
  {author} {\bibfnamefont {G.}~\bibnamefont {Semeghini}}, \bibinfo {author}
  {\bibfnamefont {T.~T.}\ \bibnamefont {Wang}}, \bibinfo {author}
  {\bibfnamefont {S.}~\bibnamefont {Ebadi}}, \bibinfo {author} {\bibfnamefont
  {H.}~\bibnamefont {Bernien}}, \bibinfo {author} {\bibfnamefont {A.~S.}\
  \bibnamefont {Zibrov}}, \bibinfo {author} {\bibfnamefont {H.}~\bibnamefont
  {Pichler}}, \bibinfo {author} {\bibfnamefont {S.}~\bibnamefont {Choi}},
  \bibinfo {author} {\bibfnamefont {J.}~\bibnamefont {Cui}}, \bibinfo {author}
  {\bibfnamefont {M.}~\bibnamefont {Rossignolo}}, \bibinfo {author}
  {\bibfnamefont {P.}~\bibnamefont {Rembold}}, \bibinfo {author} {\bibfnamefont
  {S.}~\bibnamefont {Montangero}}, \bibinfo {author} {\bibfnamefont
  {T.}~\bibnamefont {Calarco}}, \bibinfo {author} {\bibfnamefont
  {M.}~\bibnamefont {Endres}}, \bibinfo {author} {\bibfnamefont
  {M.}~\bibnamefont {Greiner}}, \bibinfo {author} {\bibfnamefont
  {V.}~\bibnamefont {Vuleti{\'c}}}, \ and\ \bibinfo {author} {\bibfnamefont
  {M.~D.}\ \bibnamefont {Lukin}},\ }\bibfield  {title} {\enquote {\bibinfo
  {title} {Generation and manipulation of {{Schr{\"o}dinger}} cat states in
  {{Rydberg}} atom arrays},}\ }\href {\doibase 10.1126/science.aax9743}
  {\bibfield  {journal} {\bibinfo  {journal} {Science}\ }\textbf {\bibinfo
  {volume} {365}},\ \bibinfo {pages} {570--574} (\bibinfo {year}
  {2019})}\BibitemShut {NoStop}%
\bibitem [{\citenamefont {Song}\ \emph {et~al.}(2019)\citenamefont {Song},
  \citenamefont {Xu}, \citenamefont {Li}, \citenamefont {Zhang}, \citenamefont
  {Zhang}, \citenamefont {Liu}, \citenamefont {Guo}, \citenamefont {Wang},
  \citenamefont {Ren}, \citenamefont {Hao}, \citenamefont {Feng}, \citenamefont
  {Fan}, \citenamefont {Zheng}, \citenamefont {Wang}, \citenamefont {Wang},\
  and\ \citenamefont {Zhu}}]{songGenerationMulticomponentAtomic2019}%
  \BibitemOpen
  \bibfield  {author} {\bibinfo {author} {\bibfnamefont {Chao}\ \bibnamefont
  {Song}}, \bibinfo {author} {\bibfnamefont {Kai}\ \bibnamefont {Xu}}, \bibinfo
  {author} {\bibfnamefont {Hekang}\ \bibnamefont {Li}}, \bibinfo {author}
  {\bibfnamefont {Yu-Ran}\ \bibnamefont {Zhang}}, \bibinfo {author}
  {\bibfnamefont {Xu}~\bibnamefont {Zhang}}, \bibinfo {author} {\bibfnamefont
  {Wuxin}\ \bibnamefont {Liu}}, \bibinfo {author} {\bibfnamefont {Qiujiang}\
  \bibnamefont {Guo}}, \bibinfo {author} {\bibfnamefont {Zhen}\ \bibnamefont
  {Wang}}, \bibinfo {author} {\bibfnamefont {Wenhui}\ \bibnamefont {Ren}},
  \bibinfo {author} {\bibfnamefont {Jie}\ \bibnamefont {Hao}}, \bibinfo
  {author} {\bibfnamefont {Hui}\ \bibnamefont {Feng}}, \bibinfo {author}
  {\bibfnamefont {Heng}\ \bibnamefont {Fan}}, \bibinfo {author} {\bibfnamefont
  {Dongning}\ \bibnamefont {Zheng}}, \bibinfo {author} {\bibfnamefont {Da-Wei}\
  \bibnamefont {Wang}}, \bibinfo {author} {\bibfnamefont {H.}~\bibnamefont
  {Wang}}, \ and\ \bibinfo {author} {\bibfnamefont {Shi-Yao}\ \bibnamefont
  {Zhu}},\ }\bibfield  {title} {\enquote {\bibinfo {title} {Generation of
  multicomponent atomic {{Schr{\"o}dinger}} cat states of up to 20 qubits},}\
  }\href {\doibase 10.1126/science.aay0600} {\bibfield  {journal} {\bibinfo
  {journal} {Science}\ }\textbf {\bibinfo {volume} {365}},\ \bibinfo {pages}
  {574--577} (\bibinfo {year} {2019})}\BibitemShut {NoStop}%
\bibitem [{\citenamefont {Holmes}\ \emph {et~al.}(2020)\citenamefont {Holmes},
  \citenamefont {Johri}, \citenamefont {Guerreschi}, \citenamefont {Clarke},\
  and\ \citenamefont {Matsuura}}]{holmesImpactQubitConnectivity2020}%
  \BibitemOpen
  \bibfield  {author} {\bibinfo {author} {\bibfnamefont {Adam}\ \bibnamefont
  {Holmes}}, \bibinfo {author} {\bibfnamefont {Sonika}\ \bibnamefont {Johri}},
  \bibinfo {author} {\bibfnamefont {Gian~Giacomo}\ \bibnamefont {Guerreschi}},
  \bibinfo {author} {\bibfnamefont {James~S}\ \bibnamefont {Clarke}}, \ and\
  \bibinfo {author} {\bibfnamefont {A~Y}\ \bibnamefont {Matsuura}},\ }\bibfield
   {title} {\enquote {\bibinfo {title} {Impact of qubit connectivity on quantum
  algorithm performance},}\ }\href {\doibase 10.1088/2058-9565/ab73e0}
  {\bibfield  {journal} {\bibinfo  {journal} {Quantum Science and Technology}\
  }\textbf {\bibinfo {volume} {5}},\ \bibinfo {pages} {025009} (\bibinfo {year}
  {2020})}\BibitemShut {NoStop}%
\bibitem [{\citenamefont {Bravyi}\ \emph {et~al.}(2024)\citenamefont {Bravyi},
  \citenamefont {Cross}, \citenamefont {Gambetta}, \citenamefont {Maslov},
  \citenamefont {Rall},\ and\ \citenamefont
  {Yoder}}]{bravyiHighthresholdLowoverheadFaulttolerant2024}%
  \BibitemOpen
  \bibfield  {author} {\bibinfo {author} {\bibfnamefont {Sergey}\ \bibnamefont
  {Bravyi}}, \bibinfo {author} {\bibfnamefont {Andrew~W.}\ \bibnamefont
  {Cross}}, \bibinfo {author} {\bibfnamefont {Jay~M.}\ \bibnamefont
  {Gambetta}}, \bibinfo {author} {\bibfnamefont {Dmitri}\ \bibnamefont
  {Maslov}}, \bibinfo {author} {\bibfnamefont {Patrick}\ \bibnamefont {Rall}},
  \ and\ \bibinfo {author} {\bibfnamefont {Theodore~J.}\ \bibnamefont
  {Yoder}},\ }\bibfield  {title} {\enquote {\bibinfo {title} {High-threshold
  and low-overhead fault-tolerant quantum memory},}\ }\href {\doibase
  10.1038/s41586-024-07107-7} {\bibfield  {journal} {\bibinfo  {journal}
  {Nature}\ }\textbf {\bibinfo {volume} {627}},\ \bibinfo {pages} {778--782}
  (\bibinfo {year} {2024})}\BibitemShut {NoStop}%
\bibitem [{\citenamefont {Chandra}\ \emph {et~al.}(2024)\citenamefont
  {Chandra}, \citenamefont {Muraleedharan},\ and\ \citenamefont
  {Brennen}}]{chandraNonlocalResourcesError2024}%
  \BibitemOpen
  \bibfield  {author} {\bibinfo {author} {\bibfnamefont {Omprakash}\
  \bibnamefont {Chandra}}, \bibinfo {author} {\bibfnamefont {Gopikrishnan}\
  \bibnamefont {Muraleedharan}}, \ and\ \bibinfo {author} {\bibfnamefont
  {Gavin~K.}\ \bibnamefont {Brennen}},\ }\href {\doibase
  10.48550/arXiv.2409.05818} {\enquote {\bibinfo {title} {Non-local resources
  for error correction in quantum {{LDPC}} codes},}\ } (\bibinfo {year}
  {2024}),\ \Eprint {http://arxiv.org/abs/2409.05818} {arXiv:2409.05818
  [quant-ph]} \BibitemShut {NoStop}%
\bibitem [{\citenamefont
  {Gottesman}(2014)}]{gottesmanFaultTolerantQuantumComputation2014}%
  \BibitemOpen
  \bibfield  {author} {\bibinfo {author} {\bibfnamefont {Daniel}\ \bibnamefont
  {Gottesman}},\ }\href {\doibase 10.48550/arXiv.1310.2984} {\enquote {\bibinfo
  {title} {Fault-{{Tolerant Quantum Computation}} with {{Constant
  Overhead}}},}\ } (\bibinfo {year} {2014}),\ \Eprint
  {http://arxiv.org/abs/1310.2984} {arXiv:1310.2984 [quant-ph]} \BibitemShut
  {NoStop}%
\bibitem [{Low(2022)}]{LowoverheadFaulttolerantQuantum}%
  \BibitemOpen
  \href {\doibase 10.1126/sciadv.abn1717} {\enquote {\bibinfo {title}
  {Low-overhead fault-tolerant quantum computing using long-range
  connectivity},}\ }\bibinfo {howpublished}
  {https://www.science.org/doi/10.1126/sciadv.abn1717} (\bibinfo {year}
  {2022})\BibitemShut {NoStop}%
\bibitem [{\citenamefont {Pecorari}\ \emph {et~al.}(2025)\citenamefont
  {Pecorari}, \citenamefont {Jandura}, \citenamefont {Brennen},\ and\
  \citenamefont {Pupillo}}]{pecorariHighrateQuantumLDPC2025}%
  \BibitemOpen
  \bibfield  {author} {\bibinfo {author} {\bibfnamefont {Laura}\ \bibnamefont
  {Pecorari}}, \bibinfo {author} {\bibfnamefont {Sven}\ \bibnamefont
  {Jandura}}, \bibinfo {author} {\bibfnamefont {Gavin~K.}\ \bibnamefont
  {Brennen}}, \ and\ \bibinfo {author} {\bibfnamefont {Guido}\ \bibnamefont
  {Pupillo}},\ }\bibfield  {title} {\enquote {\bibinfo {title} {High-rate
  quantum {{LDPC}} codes for long-range-connected neutral atom registers},}\
  }\href {\doibase 10.1038/s41467-025-56255-5} {\bibfield  {journal} {\bibinfo
  {journal} {Nature Communications}\ }\textbf {\bibinfo {volume} {16}},\
  \bibinfo {pages} {1111} (\bibinfo {year} {2025})}\BibitemShut {NoStop}%
\bibitem [{\citenamefont {Breuckmann}\ and\ \citenamefont
  {Eberhardt}(2021)}]{breuckmannQuantumLowDensityParityCheck2021}%
  \BibitemOpen
  \bibfield  {author} {\bibinfo {author} {\bibfnamefont {Nikolas~P.}\
  \bibnamefont {Breuckmann}}\ and\ \bibinfo {author} {\bibfnamefont
  {Jens~Niklas}\ \bibnamefont {Eberhardt}},\ }\bibfield  {title} {\enquote
  {\bibinfo {title} {Quantum {{Low-Density Parity-Check Codes}}},}\ }\href
  {\doibase 10.1103/PRXQuantum.2.040101} {\bibfield  {journal} {\bibinfo
  {journal} {PRX Quantum}\ }\textbf {\bibinfo {volume} {2}},\ \bibinfo {pages}
  {040101} (\bibinfo {year} {2021})}\BibitemShut {NoStop}%
\bibitem [{\citenamefont {Degen}\ \emph {et~al.}(2017)\citenamefont {Degen},
  \citenamefont {Reinhard},\ and\ \citenamefont
  {Cappellaro}}]{degenQuantumSensing2017}%
  \BibitemOpen
  \bibfield  {author} {\bibinfo {author} {\bibfnamefont {C.~L.}\ \bibnamefont
  {Degen}}, \bibinfo {author} {\bibfnamefont {F.}~\bibnamefont {Reinhard}}, \
  and\ \bibinfo {author} {\bibfnamefont {P.}~\bibnamefont {Cappellaro}},\
  }\bibfield  {title} {\enquote {\bibinfo {title} {Quantum sensing},}\ }\href
  {\doibase 10.1103/RevModPhys.89.035002} {\bibfield  {journal} {\bibinfo
  {journal} {Reviews of Modern Physics}\ }\textbf {\bibinfo {volume} {89}},\
  \bibinfo {pages} {035002} (\bibinfo {year} {2017})}\BibitemShut {NoStop}%
\bibitem [{\citenamefont {Pezz{\`e}}\ \emph {et~al.}(2018)\citenamefont
  {Pezz{\`e}}, \citenamefont {Smerzi}, \citenamefont {Oberthaler},
  \citenamefont {Schmied},\ and\ \citenamefont {Treutlein}}]{pezze2018quantum}%
  \BibitemOpen
  \bibfield  {author} {\bibinfo {author} {\bibfnamefont {Luca}\ \bibnamefont
  {Pezz{\`e}}}, \bibinfo {author} {\bibfnamefont {Augusto}\ \bibnamefont
  {Smerzi}}, \bibinfo {author} {\bibfnamefont {Markus~K}\ \bibnamefont
  {Oberthaler}}, \bibinfo {author} {\bibfnamefont {Roman}\ \bibnamefont
  {Schmied}}, \ and\ \bibinfo {author} {\bibfnamefont {Philipp}\ \bibnamefont
  {Treutlein}},\ }\bibfield  {title} {\enquote {\bibinfo {title} {Quantum
  metrology with nonclassical states of atomic ensembles},}\ }\href@noop {}
  {\bibfield  {journal} {\bibinfo  {journal} {Reviews of Modern Physics}\
  }\textbf {\bibinfo {volume} {90}},\ \bibinfo {pages} {035005} (\bibinfo
  {year} {2018})}\BibitemShut {NoStop}%
\bibitem [{\citenamefont {Srivastava}\ \emph {et~al.}(2024)\citenamefont
  {Srivastava}, \citenamefont {Jandura}, \citenamefont {Brennen},\ and\
  \citenamefont {Pupillo}}]{srivastavaEntanglementenhancedQuantumSensing2024}%
  \BibitemOpen
  \bibfield  {author} {\bibinfo {author} {\bibfnamefont {Vineesha}\
  \bibnamefont {Srivastava}}, \bibinfo {author} {\bibfnamefont {Sven}\
  \bibnamefont {Jandura}}, \bibinfo {author} {\bibfnamefont {Gavin~K.}\
  \bibnamefont {Brennen}}, \ and\ \bibinfo {author} {\bibfnamefont {Guido}\
  \bibnamefont {Pupillo}},\ }\href {\doibase 10.48550/arXiv.2409.12932}
  {\enquote {\bibinfo {title} {Entanglement-enhanced quantum sensing via
  optimal global control},}\ } (\bibinfo {year} {2024}),\ \Eprint
  {http://arxiv.org/abs/2409.12932} {arXiv:2409.12932 [quant-ph]} \BibitemShut
  {NoStop}%
\bibitem [{\citenamefont {Zhong}\ \emph {et~al.}(2021)\citenamefont {Zhong},
  \citenamefont {Chang}, \citenamefont {Bienfait}, \citenamefont {Dumur},
  \citenamefont {Chou}, \citenamefont {Conner}, \citenamefont {Grebel},
  \citenamefont {Povey}, \citenamefont {Yan}, \citenamefont {Schuster},\ and\
  \citenamefont {Cleland}}]{zhongDeterministicMultiqubitEntanglement2021}%
  \BibitemOpen
  \bibfield  {author} {\bibinfo {author} {\bibfnamefont {Youpeng}\ \bibnamefont
  {Zhong}}, \bibinfo {author} {\bibfnamefont {Hung-Shen}\ \bibnamefont
  {Chang}}, \bibinfo {author} {\bibfnamefont {Audrey}\ \bibnamefont
  {Bienfait}}, \bibinfo {author} {\bibfnamefont {{\'E}tienne}\ \bibnamefont
  {Dumur}}, \bibinfo {author} {\bibfnamefont {Ming-Han}\ \bibnamefont {Chou}},
  \bibinfo {author} {\bibfnamefont {Christopher~R.}\ \bibnamefont {Conner}},
  \bibinfo {author} {\bibfnamefont {Joel}\ \bibnamefont {Grebel}}, \bibinfo
  {author} {\bibfnamefont {Rhys~G.}\ \bibnamefont {Povey}}, \bibinfo {author}
  {\bibfnamefont {Haoxiong}\ \bibnamefont {Yan}}, \bibinfo {author}
  {\bibfnamefont {David~I.}\ \bibnamefont {Schuster}}, \ and\ \bibinfo {author}
  {\bibfnamefont {Andrew~N.}\ \bibnamefont {Cleland}},\ }\bibfield  {title}
  {\enquote {\bibinfo {title} {Deterministic multi-qubit entanglement in a
  quantum network},}\ }\href {\doibase 10.1038/s41586-021-03288-7} {\bibfield
  {journal} {\bibinfo  {journal} {Nature}\ }\textbf {\bibinfo {volume} {590}},\
  \bibinfo {pages} {571--575} (\bibinfo {year} {2021})}\BibitemShut {NoStop}%
\bibitem [{\citenamefont {Cirac}\ and\ \citenamefont
  {Zoller}(1995)}]{ciracQuantumComputationsCold1995}%
  \BibitemOpen
  \bibfield  {author} {\bibinfo {author} {\bibfnamefont {J.~I.}\ \bibnamefont
  {Cirac}}\ and\ \bibinfo {author} {\bibfnamefont {P.}~\bibnamefont {Zoller}},\
  }\bibfield  {title} {\enquote {\bibinfo {title} {Quantum {{Computations}}
  with {{Cold Trapped Ions}}},}\ }\href {\doibase 10.1103/PhysRevLett.74.4091}
  {\bibfield  {journal} {\bibinfo  {journal} {Physical Review Letters}\
  }\textbf {\bibinfo {volume} {74}},\ \bibinfo {pages} {4091--4094} (\bibinfo
  {year} {1995})}\BibitemShut {NoStop}%
\bibitem [{\citenamefont {{Garc{\'\i}a-Ripoll}}\ \emph
  {et~al.}(2003)\citenamefont {{Garc{\'\i}a-Ripoll}}, \citenamefont {Zoller},\
  and\ \citenamefont {Cirac}}]{garcia-ripollSpeedOptimizedTwoQubit2003}%
  \BibitemOpen
  \bibfield  {author} {\bibinfo {author} {\bibfnamefont {J.~J.}\ \bibnamefont
  {{Garc{\'\i}a-Ripoll}}}, \bibinfo {author} {\bibfnamefont {P.}~\bibnamefont
  {Zoller}}, \ and\ \bibinfo {author} {\bibfnamefont {J.~I.}\ \bibnamefont
  {Cirac}},\ }\bibfield  {title} {\enquote {\bibinfo {title} {Speed {{Optimized
  Two-Qubit Gates}} with {{Laser Coherent Control Techniques}} for {{Ion Trap
  Quantum Computing}}},}\ }\href {\doibase 10.1103/PhysRevLett.91.157901}
  {\bibfield  {journal} {\bibinfo  {journal} {Physical Review Letters}\
  }\textbf {\bibinfo {volume} {91}},\ \bibinfo {pages} {157901} (\bibinfo
  {year} {2003})}\BibitemShut {NoStop}%
\bibitem [{\citenamefont {M{\o}lmer}\ and\ \citenamefont
  {S{\o}rensen}(1999)}]{molmerMultiparticleEntanglementHot1999}%
  \BibitemOpen
  \bibfield  {author} {\bibinfo {author} {\bibfnamefont {Klaus}\ \bibnamefont
  {M{\o}lmer}}\ and\ \bibinfo {author} {\bibfnamefont {Anders}\ \bibnamefont
  {S{\o}rensen}},\ }\bibfield  {title} {\enquote {\bibinfo {title}
  {Multiparticle {{Entanglement}} of {{Hot Trapped Ions}}},}\ }\href {\doibase
  10.1103/PhysRevLett.82.1835} {\bibfield  {journal} {\bibinfo  {journal}
  {Physical Review Letters}\ }\textbf {\bibinfo {volume} {82}},\ \bibinfo
  {pages} {1835--1838} (\bibinfo {year} {1999})}\BibitemShut {NoStop}%
\bibitem [{\citenamefont {Sackett}\ \emph {et~al.}(2000)\citenamefont
  {Sackett}, \citenamefont {Kielpinski}, \citenamefont {King}, \citenamefont
  {Langer}, \citenamefont {Meyer}, \citenamefont {Myatt}, \citenamefont {Rowe},
  \citenamefont {Turchette}, \citenamefont {Itano}, \citenamefont {Wineland},\
  and\ \citenamefont {Monroe}}]{sackettExperimentalEntanglementFour2000}%
  \BibitemOpen
  \bibfield  {author} {\bibinfo {author} {\bibfnamefont {C.~A.}\ \bibnamefont
  {Sackett}}, \bibinfo {author} {\bibfnamefont {D.}~\bibnamefont {Kielpinski}},
  \bibinfo {author} {\bibfnamefont {B.~E.}\ \bibnamefont {King}}, \bibinfo
  {author} {\bibfnamefont {C.}~\bibnamefont {Langer}}, \bibinfo {author}
  {\bibfnamefont {V.}~\bibnamefont {Meyer}}, \bibinfo {author} {\bibfnamefont
  {C.~J.}\ \bibnamefont {Myatt}}, \bibinfo {author} {\bibfnamefont
  {M.}~\bibnamefont {Rowe}}, \bibinfo {author} {\bibfnamefont {Q.~A.}\
  \bibnamefont {Turchette}}, \bibinfo {author} {\bibfnamefont {W.~M.}\
  \bibnamefont {Itano}}, \bibinfo {author} {\bibfnamefont {D.~J.}\ \bibnamefont
  {Wineland}}, \ and\ \bibinfo {author} {\bibfnamefont {C.}~\bibnamefont
  {Monroe}},\ }\bibfield  {title} {\enquote {\bibinfo {title} {Experimental
  entanglement of four particles},}\ }\href {\doibase 10.1038/35005011}
  {\bibfield  {journal} {\bibinfo  {journal} {Nature}\ }\textbf {\bibinfo
  {volume} {404}},\ \bibinfo {pages} {256--259} (\bibinfo {year}
  {2000})}\BibitemShut {NoStop}%
\bibitem [{\citenamefont {Pellizzari}\ \emph {et~al.}(1995)\citenamefont
  {Pellizzari}, \citenamefont {Gardiner}, \citenamefont {Cirac},\ and\
  \citenamefont {Zoller}}]{pellizzariDecoherenceContinuousObservation1995}%
  \BibitemOpen
  \bibfield  {author} {\bibinfo {author} {\bibfnamefont {T.}~\bibnamefont
  {Pellizzari}}, \bibinfo {author} {\bibfnamefont {S.~A.}\ \bibnamefont
  {Gardiner}}, \bibinfo {author} {\bibfnamefont {J.~I.}\ \bibnamefont {Cirac}},
  \ and\ \bibinfo {author} {\bibfnamefont {P.}~\bibnamefont {Zoller}},\
  }\bibfield  {title} {\enquote {\bibinfo {title} {Decoherence, {{Continuous
  Observation}}, and {{Quantum Computing}}: {{A Cavity QED Model}}},}\ }\href
  {\doibase 10.1103/PhysRevLett.75.3788} {\bibfield  {journal} {\bibinfo
  {journal} {Physical Review Letters}\ }\textbf {\bibinfo {volume} {75}},\
  \bibinfo {pages} {3788--3791} (\bibinfo {year} {1995})}\BibitemShut {NoStop}%
\bibitem [{\citenamefont {Beige}\ \emph {et~al.}(2000)\citenamefont {Beige},
  \citenamefont {Braun}, \citenamefont {Tregenna},\ and\ \citenamefont
  {Knight}}]{beigeQuantumComputingUsing2000}%
  \BibitemOpen
  \bibfield  {author} {\bibinfo {author} {\bibfnamefont {Almut}\ \bibnamefont
  {Beige}}, \bibinfo {author} {\bibfnamefont {Daniel}\ \bibnamefont {Braun}},
  \bibinfo {author} {\bibfnamefont {Ben}\ \bibnamefont {Tregenna}}, \ and\
  \bibinfo {author} {\bibfnamefont {Peter~L.}\ \bibnamefont {Knight}},\
  }\bibfield  {title} {\enquote {\bibinfo {title} {Quantum {{Computing Using
  Dissipation}} to {{Remain}} in a {{Decoherence-Free Subspace}}},}\ }\href
  {\doibase 10.1103/PhysRevLett.85.1762} {\bibfield  {journal} {\bibinfo
  {journal} {Physical Review Letters}\ }\textbf {\bibinfo {volume} {85}},\
  \bibinfo {pages} {1762--1765} (\bibinfo {year} {2000})}\BibitemShut {NoStop}%
\bibitem [{\citenamefont {Borregaard}\ \emph {et~al.}(2015)\citenamefont
  {Borregaard}, \citenamefont {K{\'o}m{\'a}r}, \citenamefont {Kessler},
  \citenamefont {S{\o}rensen},\ and\ \citenamefont
  {Lukin}}]{borregaardHeraldedQuantumGates2015}%
  \BibitemOpen
  \bibfield  {author} {\bibinfo {author} {\bibfnamefont {J.}~\bibnamefont
  {Borregaard}}, \bibinfo {author} {\bibfnamefont {P.}~\bibnamefont
  {K{\'o}m{\'a}r}}, \bibinfo {author} {\bibfnamefont {E.~M.}\ \bibnamefont
  {Kessler}}, \bibinfo {author} {\bibfnamefont {A.~S.}\ \bibnamefont
  {S{\o}rensen}}, \ and\ \bibinfo {author} {\bibfnamefont {M.~D.}\ \bibnamefont
  {Lukin}},\ }\bibfield  {title} {\enquote {\bibinfo {title} {Heralded
  {{Quantum Gates}} with {{Integrated Error Detection}} in {{Optical
  Cavities}}},}\ }\href {\doibase 10.1103/PhysRevLett.114.110502} {\bibfield
  {journal} {\bibinfo  {journal} {Physical Review Letters}\ }\textbf {\bibinfo
  {volume} {114}},\ \bibinfo {pages} {110502} (\bibinfo {year}
  {2015})}\BibitemShut {NoStop}%
\bibitem [{\citenamefont {Lewalle}\ \emph {et~al.}(2023)\citenamefont
  {Lewalle}, \citenamefont {Martin}, \citenamefont {Flurin}, \citenamefont
  {Zhang}, \citenamefont {Blumenthal}, \citenamefont {{Hacohen-Gourgy}},
  \citenamefont {Burgarth},\ and\ \citenamefont
  {Whaley}}]{lewalleMultiQubitQuantumGate2023}%
  \BibitemOpen
  \bibfield  {author} {\bibinfo {author} {\bibfnamefont {Philippe}\
  \bibnamefont {Lewalle}}, \bibinfo {author} {\bibfnamefont {Leigh~S.}\
  \bibnamefont {Martin}}, \bibinfo {author} {\bibfnamefont {Emmanuel}\
  \bibnamefont {Flurin}}, \bibinfo {author} {\bibfnamefont {Song}\ \bibnamefont
  {Zhang}}, \bibinfo {author} {\bibfnamefont {Eliya}\ \bibnamefont
  {Blumenthal}}, \bibinfo {author} {\bibfnamefont {Shay}\ \bibnamefont
  {{Hacohen-Gourgy}}}, \bibinfo {author} {\bibfnamefont {Daniel}\ \bibnamefont
  {Burgarth}}, \ and\ \bibinfo {author} {\bibfnamefont {K.~Birgitta}\
  \bibnamefont {Whaley}},\ }\bibfield  {title} {\enquote {\bibinfo {title} {A
  {{Multi-Qubit Quantum Gate Using}} the {{Zeno Effect}}},}\ }\href {\doibase
  10.22331/q-2023-09-07-1100} {\bibfield  {journal} {\bibinfo  {journal}
  {Quantum}\ }\textbf {\bibinfo {volume} {7}},\ \bibinfo {pages} {1100}
  (\bibinfo {year} {2023})},\ \Eprint {http://arxiv.org/abs/2211.05988}
  {arXiv:2211.05988 [quant-ph]} \BibitemShut {NoStop}%
\bibitem [{\citenamefont {Ramette}\ \emph {et~al.}(2022)\citenamefont
  {Ramette}, \citenamefont {Sinclair}, \citenamefont {Vendeiro}, \citenamefont
  {Rudelis}, \citenamefont {Cetina},\ and\ \citenamefont
  {Vuleti{\'c}}}]{rametteAnyToAnyConnectedCavityMediated2022}%
  \BibitemOpen
  \bibfield  {author} {\bibinfo {author} {\bibfnamefont {Joshua}\ \bibnamefont
  {Ramette}}, \bibinfo {author} {\bibfnamefont {Josiah}\ \bibnamefont
  {Sinclair}}, \bibinfo {author} {\bibfnamefont {Zachary}\ \bibnamefont
  {Vendeiro}}, \bibinfo {author} {\bibfnamefont {Alyssa}\ \bibnamefont
  {Rudelis}}, \bibinfo {author} {\bibfnamefont {Marko}\ \bibnamefont {Cetina}},
  \ and\ \bibinfo {author} {\bibfnamefont {Vladan}\ \bibnamefont
  {Vuleti{\'c}}},\ }\bibfield  {title} {\enquote {\bibinfo {title}
  {Any-{{To-Any Connected Cavity-Mediated Architecture}} for {{Quantum
  Computing}} with {{Trapped Ions}} or {{Rydberg Arrays}}},}\ }\href {\doibase
  10.1103/PRXQuantum.3.010344} {\bibfield  {journal} {\bibinfo  {journal} {PRX
  Quantum}\ }\textbf {\bibinfo {volume} {3}},\ \bibinfo {pages} {010344}
  (\bibinfo {year} {2022})}\BibitemShut {NoStop}%
\bibitem [{\citenamefont {S{\o}rensen}\ and\ \citenamefont
  {M{\o}lmer}(2003)}]{sorensenMeasurementInducedEntanglement2003}%
  \BibitemOpen
  \bibfield  {author} {\bibinfo {author} {\bibfnamefont {Anders~S.}\
  \bibnamefont {S{\o}rensen}}\ and\ \bibinfo {author} {\bibfnamefont {Klaus}\
  \bibnamefont {M{\o}lmer}},\ }\bibfield  {title} {\enquote {\bibinfo {title}
  {Measurement {{Induced Entanglement}} and {{Quantum Computation}} with
  {{Atoms}} in {{Optical Cavities}}},}\ }\href {\doibase
  10.1103/PhysRevLett.91.097905} {\bibfield  {journal} {\bibinfo  {journal}
  {Physical Review Letters}\ }\textbf {\bibinfo {volume} {91}},\ \bibinfo
  {pages} {097905} (\bibinfo {year} {2003})}\BibitemShut {NoStop}%
\bibitem [{\citenamefont {Zheng}\ and\ \citenamefont
  {Guo}(2000)}]{zhengEfficientSchemeTwoAtom2000}%
  \BibitemOpen
  \bibfield  {author} {\bibinfo {author} {\bibfnamefont {Shi-Biao}\
  \bibnamefont {Zheng}}\ and\ \bibinfo {author} {\bibfnamefont {Guang-Can}\
  \bibnamefont {Guo}},\ }\bibfield  {title} {\enquote {\bibinfo {title}
  {Efficient {{Scheme}} for {{Two-Atom Entanglement}} and {{Quantum Information
  Processing}} in {{Cavity QED}}},}\ }\href {\doibase
  10.1103/PhysRevLett.85.2392} {\bibfield  {journal} {\bibinfo  {journal}
  {Physical Review Letters}\ }\textbf {\bibinfo {volume} {85}},\ \bibinfo
  {pages} {2392--2395} (\bibinfo {year} {2000})}\BibitemShut {NoStop}%
\bibitem [{\citenamefont
  {Zheng}(2004)}]{zhengUnconventionalGeometricQuantum2004}%
  \BibitemOpen
  \bibfield  {author} {\bibinfo {author} {\bibfnamefont {Shi-Biao}\
  \bibnamefont {Zheng}},\ }\bibfield  {title} {\enquote {\bibinfo {title}
  {Unconventional geometric quantum phase gates with a cavity {{QED}}
  system},}\ }\href {\doibase 10.1103/PhysRevA.70.052320} {\bibfield  {journal}
  {\bibinfo  {journal} {Physical Review A}\ }\textbf {\bibinfo {volume} {70}},\
  \bibinfo {pages} {052320} (\bibinfo {year} {2004})}\BibitemShut {NoStop}%
\bibitem [{\citenamefont {Duan}\ and\ \citenamefont
  {Kimble}(2003)}]{duanEfficientEngineeringMultiatom2003}%
  \BibitemOpen
  \bibfield  {author} {\bibinfo {author} {\bibfnamefont {L.-M.}\ \bibnamefont
  {Duan}}\ and\ \bibinfo {author} {\bibfnamefont {H.~J.}\ \bibnamefont
  {Kimble}},\ }\bibfield  {title} {\enquote {\bibinfo {title} {Efficient
  {{Engineering}} of {{Multiatom Entanglement}} through {{Single-Photon
  Detections}}},}\ }\href {\doibase 10.1103/PhysRevLett.90.253601} {\bibfield
  {journal} {\bibinfo  {journal} {Physical Review Letters}\ }\textbf {\bibinfo
  {volume} {90}},\ \bibinfo {pages} {253601} (\bibinfo {year}
  {2003})}\BibitemShut {NoStop}%
\bibitem [{\citenamefont {Jandura}\ \emph {et~al.}(2024)\citenamefont
  {Jandura}, \citenamefont {Srivastava}, \citenamefont {Pecorari},
  \citenamefont {Brennen},\ and\ \citenamefont
  {Pupillo}}]{janduraNonlocalMultiqubitQuantum2024}%
  \BibitemOpen
  \bibfield  {author} {\bibinfo {author} {\bibfnamefont {Sven}\ \bibnamefont
  {Jandura}}, \bibinfo {author} {\bibfnamefont {Vineesha}\ \bibnamefont
  {Srivastava}}, \bibinfo {author} {\bibfnamefont {Laura}\ \bibnamefont
  {Pecorari}}, \bibinfo {author} {\bibfnamefont {Gavin~K.}\ \bibnamefont
  {Brennen}}, \ and\ \bibinfo {author} {\bibfnamefont {Guido}\ \bibnamefont
  {Pupillo}},\ }\bibfield  {title} {\enquote {\bibinfo {title} {Nonlocal
  multiqubit quantum gates via a driven cavity},}\ }\href {\doibase
  10.1103/PhysRevA.110.062610} {\bibfield  {journal} {\bibinfo  {journal}
  {Physical Review A}\ }\textbf {\bibinfo {volume} {110}},\ \bibinfo {pages}
  {062610} (\bibinfo {year} {2024})}\BibitemShut {NoStop}%
\bibitem [{\citenamefont {Barontini}\ \emph {et~al.}(2015)\citenamefont
  {Barontini}, \citenamefont {Hohmann}, \citenamefont {Haas}, \citenamefont
  {Est{\`e}ve},\ and\ \citenamefont
  {Reichel}}]{barontiniDeterministicGenerationMultiparticle2015}%
  \BibitemOpen
  \bibfield  {author} {\bibinfo {author} {\bibfnamefont {Giovanni}\
  \bibnamefont {Barontini}}, \bibinfo {author} {\bibfnamefont {Leander}\
  \bibnamefont {Hohmann}}, \bibinfo {author} {\bibfnamefont {Florian}\
  \bibnamefont {Haas}}, \bibinfo {author} {\bibfnamefont {J{\'e}r{\^o}me}\
  \bibnamefont {Est{\`e}ve}}, \ and\ \bibinfo {author} {\bibfnamefont {Jakob}\
  \bibnamefont {Reichel}},\ }\bibfield  {title} {\enquote {\bibinfo {title}
  {Deterministic generation of multiparticle entanglement by quantum {{Zeno}}
  dynamics},}\ }\href {\doibase 10.1126/science.aaa0754} {\bibfield  {journal}
  {\bibinfo  {journal} {Science}\ }\textbf {\bibinfo {volume} {349}},\ \bibinfo
  {pages} {1317--1321} (\bibinfo {year} {2015})}\BibitemShut {NoStop}%
\bibitem [{\citenamefont {Pupillo}\ \emph {et~al.}(2004)\citenamefont
  {Pupillo}, \citenamefont {Rey}, \citenamefont {Brennen}, \citenamefont
  {Williams},\ and\ \citenamefont
  {Clark}}]{pupilloScalableQuantumComputation2004}%
  \BibitemOpen
  \bibfield  {author} {\bibinfo {author} {\bibfnamefont {Guido}\ \bibnamefont
  {Pupillo}}, \bibinfo {author} {\bibfnamefont {Ana~Maria}\ \bibnamefont
  {Rey}}, \bibinfo {author} {\bibfnamefont {Gavin}\ \bibnamefont {Brennen}},
  \bibinfo {author} {\bibfnamefont {Carl~J.}\ \bibnamefont {Williams}}, \ and\
  \bibinfo {author} {\bibfnamefont {Charles~W.}\ \bibnamefont {Clark}},\
  }\bibfield  {title} {\enquote {\bibinfo {title} {Scalable quantum computation
  in systems with {{Bose-Hubbard}} dynamics},}\ }\href {\doibase
  10.1080/09500340408231798} {\bibfield  {journal} {\bibinfo  {journal}
  {Journal of Modern Optics}\ }\textbf {\bibinfo {volume} {51}},\ \bibinfo
  {pages} {2395--2404} (\bibinfo {year} {2004})}\BibitemShut {NoStop}%
\bibitem [{\citenamefont {Volz}\ \emph {et~al.}(2011)\citenamefont {Volz},
  \citenamefont {Gehr}, \citenamefont {Dubois}, \citenamefont {Est{\`e}ve},\
  and\ \citenamefont {Reichel}}]{volzMeasurementInternalState2011}%
  \BibitemOpen
  \bibfield  {author} {\bibinfo {author} {\bibfnamefont {J{\"u}rgen}\
  \bibnamefont {Volz}}, \bibinfo {author} {\bibfnamefont {Roger}\ \bibnamefont
  {Gehr}}, \bibinfo {author} {\bibfnamefont {Guilhem}\ \bibnamefont {Dubois}},
  \bibinfo {author} {\bibfnamefont {J{\'e}r{\^o}me}\ \bibnamefont
  {Est{\`e}ve}}, \ and\ \bibinfo {author} {\bibfnamefont {Jakob}\ \bibnamefont
  {Reichel}},\ }\bibfield  {title} {\enquote {\bibinfo {title} {Measurement of
  the internal state of a single atom without energy exchange},}\ }\href
  {\doibase 10.1038/nature10225} {\bibfield  {journal} {\bibinfo  {journal}
  {Nature}\ }\textbf {\bibinfo {volume} {475}},\ \bibinfo {pages} {210--213}
  (\bibinfo {year} {2011})}\BibitemShut {NoStop}%
\bibitem [{\citenamefont {Gottesman}\ \emph {et~al.}(2012)\citenamefont
  {Gottesman}, \citenamefont {Jennewein},\ and\ \citenamefont
  {Croke}}]{gottesmanLongerBaselineTelescopesUsing2012}%
  \BibitemOpen
  \bibfield  {author} {\bibinfo {author} {\bibfnamefont {Daniel}\ \bibnamefont
  {Gottesman}}, \bibinfo {author} {\bibfnamefont {Thomas}\ \bibnamefont
  {Jennewein}}, \ and\ \bibinfo {author} {\bibfnamefont {Sarah}\ \bibnamefont
  {Croke}},\ }\bibfield  {title} {\enquote {\bibinfo {title} {Longer-{{Baseline
  Telescopes Using Quantum Repeaters}}},}\ }\href {\doibase
  10.1103/PhysRevLett.109.070503} {\bibfield  {journal} {\bibinfo  {journal}
  {Physical Review Letters}\ }\textbf {\bibinfo {volume} {109}},\ \bibinfo
  {pages} {070503} (\bibinfo {year} {2012})}\BibitemShut {NoStop}%
\bibitem [{\citenamefont {Grinkemeyer}\ \emph {et~al.}(2024)\citenamefont
  {Grinkemeyer}, \citenamefont {{Guardado-Sanchez}}, \citenamefont {Dimitrova},
  \citenamefont {Shchepanovich}, \citenamefont {Mandopoulou}, \citenamefont
  {Borregaard}, \citenamefont {Vuleti{\'c}},\ and\ \citenamefont
  {Lukin}}]{grinkemeyerErrorDetectedQuantumOperations2024}%
  \BibitemOpen
  \bibfield  {author} {\bibinfo {author} {\bibfnamefont {Brandon}\ \bibnamefont
  {Grinkemeyer}}, \bibinfo {author} {\bibfnamefont {Elmer}\ \bibnamefont
  {{Guardado-Sanchez}}}, \bibinfo {author} {\bibfnamefont {Ivana}\ \bibnamefont
  {Dimitrova}}, \bibinfo {author} {\bibfnamefont {Danilo}\ \bibnamefont
  {Shchepanovich}}, \bibinfo {author} {\bibfnamefont {G.~Eirini}\ \bibnamefont
  {Mandopoulou}}, \bibinfo {author} {\bibfnamefont {Johannes}\ \bibnamefont
  {Borregaard}}, \bibinfo {author} {\bibfnamefont {Vladan}\ \bibnamefont
  {Vuleti{\'c}}}, \ and\ \bibinfo {author} {\bibfnamefont {Mikhail~D.}\
  \bibnamefont {Lukin}},\ }\href {\doibase 10.48550/arXiv.2410.10787} {\enquote
  {\bibinfo {title} {Error-{{Detected Quantum Operations}} with {{Neutral Atoms
  Mediated}} by an {{Optical Cavity}}},}\ } (\bibinfo {year} {2024}),\ \Eprint
  {http://arxiv.org/abs/2410.10787} {arXiv:2410.10787 [quant-ph]} \BibitemShut
  {NoStop}%
\bibitem [{\citenamefont {Main}\ \emph {et~al.}(2025)\citenamefont {Main},
  \citenamefont {Drmota}, \citenamefont {Nadlinger}, \citenamefont {Ainley},
  \citenamefont {Agrawal}, \citenamefont {Nichol}, \citenamefont {Srinivas},
  \citenamefont {Araneda},\ and\ \citenamefont
  {Lucas}}]{mainDistributedQuantumComputing2025}%
  \BibitemOpen
  \bibfield  {author} {\bibinfo {author} {\bibfnamefont {D.}~\bibnamefont
  {Main}}, \bibinfo {author} {\bibfnamefont {P.}~\bibnamefont {Drmota}},
  \bibinfo {author} {\bibfnamefont {D.~P.}\ \bibnamefont {Nadlinger}}, \bibinfo
  {author} {\bibfnamefont {E.~M.}\ \bibnamefont {Ainley}}, \bibinfo {author}
  {\bibfnamefont {A.}~\bibnamefont {Agrawal}}, \bibinfo {author} {\bibfnamefont
  {B.~C.}\ \bibnamefont {Nichol}}, \bibinfo {author} {\bibfnamefont
  {R.}~\bibnamefont {Srinivas}}, \bibinfo {author} {\bibfnamefont
  {G.}~\bibnamefont {Araneda}}, \ and\ \bibinfo {author} {\bibfnamefont
  {D.~M.}\ \bibnamefont {Lucas}},\ }\bibfield  {title} {\enquote {\bibinfo
  {title} {Distributed quantum computing across an optical network link},}\
  }\href {\doibase 10.1038/s41586-024-08404-x} {\bibfield  {journal} {\bibinfo
  {journal} {Nature}\ }\textbf {\bibinfo {volume} {638}},\ \bibinfo {pages}
  {383--388} (\bibinfo {year} {2025})}\BibitemShut {NoStop}%
\bibitem [{\citenamefont {Rohde}\ \emph {et~al.}(2025)\citenamefont {Rohde},
  \citenamefont {Huang}, \citenamefont {Ouyang}, \citenamefont {Huang},
  \citenamefont {Su}, \citenamefont {Devitt}, \citenamefont {Ramakrishnan},
  \citenamefont {Mantri}, \citenamefont {Tan}, \citenamefont {Liu},
  \citenamefont {Harrison}, \citenamefont {Radhakrishnan}, \citenamefont
  {Brennen}, \citenamefont {Baragiola}, \citenamefont {Dowling}, \citenamefont
  {Byrnes},\ and\ \citenamefont
  {Munro}}]{rohde2025quantuminternettechnicalversion}%
  \BibitemOpen
  \bibfield  {author} {\bibinfo {author} {\bibfnamefont {Peter~P.}\
  \bibnamefont {Rohde}}, \bibinfo {author} {\bibfnamefont {Zixin}\ \bibnamefont
  {Huang}}, \bibinfo {author} {\bibfnamefont {Yingkai}\ \bibnamefont {Ouyang}},
  \bibinfo {author} {\bibfnamefont {He-Liang}\ \bibnamefont {Huang}}, \bibinfo
  {author} {\bibfnamefont {Zu-En}\ \bibnamefont {Su}}, \bibinfo {author}
  {\bibfnamefont {Simon}\ \bibnamefont {Devitt}}, \bibinfo {author}
  {\bibfnamefont {Rohit}\ \bibnamefont {Ramakrishnan}}, \bibinfo {author}
  {\bibfnamefont {Atul}\ \bibnamefont {Mantri}}, \bibinfo {author}
  {\bibfnamefont {Si-Hui}\ \bibnamefont {Tan}}, \bibinfo {author}
  {\bibfnamefont {Nana}\ \bibnamefont {Liu}}, \bibinfo {author} {\bibfnamefont
  {Scott}\ \bibnamefont {Harrison}}, \bibinfo {author} {\bibfnamefont
  {Chandrashekar}\ \bibnamefont {Radhakrishnan}}, \bibinfo {author}
  {\bibfnamefont {Gavin~K.}\ \bibnamefont {Brennen}}, \bibinfo {author}
  {\bibfnamefont {Ben~Q.}\ \bibnamefont {Baragiola}}, \bibinfo {author}
  {\bibfnamefont {Jonathan~P.}\ \bibnamefont {Dowling}}, \bibinfo {author}
  {\bibfnamefont {Tim}\ \bibnamefont {Byrnes}}, \ and\ \bibinfo {author}
  {\bibfnamefont {William~J.}\ \bibnamefont {Munro}},\ }\href
  {https://arxiv.org/abs/2501.12107} {\enquote {\bibinfo {title} {The quantum
  internet (technical version)},}\ } (\bibinfo {year} {2025}),\ \Eprint
  {http://arxiv.org/abs/2501.12107} {arXiv:2501.12107 [quant-ph]} \BibitemShut
  {NoStop}%
\bibitem [{\citenamefont {Zhang}\ and\ \citenamefont
  {Zhuang}(2021)}]{Zhang_2021}%
  \BibitemOpen
  \bibfield  {author} {\bibinfo {author} {\bibfnamefont {Zheshen}\ \bibnamefont
  {Zhang}}\ and\ \bibinfo {author} {\bibfnamefont {Quntao}\ \bibnamefont
  {Zhuang}},\ }\bibfield  {title} {\enquote {\bibinfo {title} {Distributed
  quantum sensing},}\ }\href {\doibase 10.1088/2058-9565/abd4c3} {\bibfield
  {journal} {\bibinfo  {journal} {Quantum Science and Technology}\ }\textbf
  {\bibinfo {volume} {6}},\ \bibinfo {pages} {043001} (\bibinfo {year}
  {2021})}\BibitemShut {NoStop}%
\bibitem [{\citenamefont {Boulier}\ \emph {et~al.}(2020)\citenamefont
  {Boulier}, \citenamefont {Jacquet}, \citenamefont {Maître}, \citenamefont
  {Lerario}, \citenamefont {Claude}, \citenamefont {Pigeon}, \citenamefont
  {Glorieux}, \citenamefont {Amo}, \citenamefont {Bloch}, \citenamefont
  {Bramati},\ and\ \citenamefont {Giacobino}}]{Boulier2020}%
  \BibitemOpen
  \bibfield  {author} {\bibinfo {author} {\bibfnamefont {Thomas}\ \bibnamefont
  {Boulier}}, \bibinfo {author} {\bibfnamefont {Maxime~J.}\ \bibnamefont
  {Jacquet}}, \bibinfo {author} {\bibfnamefont {Anne}\ \bibnamefont {Maître}},
  \bibinfo {author} {\bibfnamefont {Giovanni}\ \bibnamefont {Lerario}},
  \bibinfo {author} {\bibfnamefont {Ferdinand}\ \bibnamefont {Claude}},
  \bibinfo {author} {\bibfnamefont {Simon}\ \bibnamefont {Pigeon}}, \bibinfo
  {author} {\bibfnamefont {Quentin}\ \bibnamefont {Glorieux}}, \bibinfo
  {author} {\bibfnamefont {Alberto}\ \bibnamefont {Amo}}, \bibinfo {author}
  {\bibfnamefont {Jacqueline}\ \bibnamefont {Bloch}}, \bibinfo {author}
  {\bibfnamefont {Alberto}\ \bibnamefont {Bramati}}, \ and\ \bibinfo {author}
  {\bibfnamefont {Elisabeth}\ \bibnamefont {Giacobino}},\ }\bibfield  {title}
  {\enquote {\bibinfo {title} {Microcavity polaritons for quantum
  simulation},}\ }\href {\doibase https://doi.org/10.1002/qute.202000052}
  {\bibfield  {journal} {\bibinfo  {journal} {Advanced Quantum Technologies}\
  }\textbf {\bibinfo {volume} {3}},\ \bibinfo {pages} {2000052} (\bibinfo
  {year} {2020})},\ \Eprint
  {http://arxiv.org/abs/https://onlinelibrary.wiley.com/doi/pdf/10.1002/qute.202000052}
  {https://onlinelibrary.wiley.com/doi/pdf/10.1002/qute.202000052} \BibitemShut
  {NoStop}%
\bibitem [{\citenamefont {Bloch}\ \emph {et~al.}(2022)\citenamefont {Bloch},
  \citenamefont {Cavalleri}, \citenamefont {Galitski}, \citenamefont {Hafezi},\
  and\ \citenamefont {Rubio}}]{blochStronglyCorrelatedElectron2022}%
  \BibitemOpen
  \bibfield  {author} {\bibinfo {author} {\bibfnamefont {Jacqueline}\
  \bibnamefont {Bloch}}, \bibinfo {author} {\bibfnamefont {Andrea}\
  \bibnamefont {Cavalleri}}, \bibinfo {author} {\bibfnamefont {Victor}\
  \bibnamefont {Galitski}}, \bibinfo {author} {\bibfnamefont {Mohammad}\
  \bibnamefont {Hafezi}}, \ and\ \bibinfo {author} {\bibfnamefont {Angel}\
  \bibnamefont {Rubio}},\ }\bibfield  {title} {\enquote {\bibinfo {title}
  {Strongly correlated electron--photon systems},}\ }\href {\doibase
  10.1038/s41586-022-04726-w} {\bibfield  {journal} {\bibinfo  {journal}
  {Nature}\ }\textbf {\bibinfo {volume} {606}},\ \bibinfo {pages} {41--48}
  (\bibinfo {year} {2022})}\BibitemShut {NoStop}%
\bibitem [{\citenamefont {Bellessa}\ \emph {et~al.}(2024)\citenamefont
  {Bellessa}, \citenamefont {Bloch}, \citenamefont {Deleporte}, \citenamefont
  {Menon}, \citenamefont {Nguyen}, \citenamefont {Ohadi}, \citenamefont
  {Ravets},\ and\ \citenamefont
  {Boulier}}]{bellessaMaterialsExcitonsPolaritons2024}%
  \BibitemOpen
  \bibfield  {author} {\bibinfo {author} {\bibfnamefont {J.}~\bibnamefont
  {Bellessa}}, \bibinfo {author} {\bibfnamefont {J.}~\bibnamefont {Bloch}},
  \bibinfo {author} {\bibfnamefont {E.}~\bibnamefont {Deleporte}}, \bibinfo
  {author} {\bibfnamefont {V.~M.}\ \bibnamefont {Menon}}, \bibinfo {author}
  {\bibfnamefont {H.~S.}\ \bibnamefont {Nguyen}}, \bibinfo {author}
  {\bibfnamefont {H.}~\bibnamefont {Ohadi}}, \bibinfo {author} {\bibfnamefont
  {S.}~\bibnamefont {Ravets}}, \ and\ \bibinfo {author} {\bibfnamefont
  {T.}~\bibnamefont {Boulier}},\ }\bibfield  {title} {\enquote {\bibinfo
  {title} {Materials for excitons--polaritons: {{Exploiting}} the diversity of
  semiconductors},}\ }\href {\doibase 10.1557/s43577-024-00779-6} {\bibfield
  {journal} {\bibinfo  {journal} {MRS Bulletin}\ }\textbf {\bibinfo {volume}
  {49}},\ \bibinfo {pages} {932--947} (\bibinfo {year} {2024})}\BibitemShut
  {NoStop}%
\bibitem [{\citenamefont {Jandura}\ and\ \citenamefont
  {Pupillo}(2022)}]{janduraTimeOptimalTwoThreeQubit2022}%
  \BibitemOpen
  \bibfield  {author} {\bibinfo {author} {\bibfnamefont {Sven}\ \bibnamefont
  {Jandura}}\ and\ \bibinfo {author} {\bibfnamefont {Guido}\ \bibnamefont
  {Pupillo}},\ }\bibfield  {title} {\enquote {\bibinfo {title} {Time-{{Optimal
  Two-}} and {{Three-Qubit Gates}} for {{Rydberg Atoms}}},}\ }\href {\doibase
  10.22331/q-2022-05-13-712} {\bibfield  {journal} {\bibinfo  {journal}
  {Quantum}\ }\textbf {\bibinfo {volume} {6}},\ \bibinfo {pages} {712}
  (\bibinfo {year} {2022})}\BibitemShut {NoStop}%
\bibitem [{\citenamefont {Evered}\ \emph {et~al.}(2023)\citenamefont {Evered},
  \citenamefont {Bluvstein}, \citenamefont {Kalinowski}, \citenamefont {Ebadi},
  \citenamefont {Manovitz}, \citenamefont {Zhou}, \citenamefont {Li},
  \citenamefont {Geim}, \citenamefont {Wang}, \citenamefont {Maskara},
  \citenamefont {Levine}, \citenamefont {Semeghini}, \citenamefont {Greiner},
  \citenamefont {Vuleti{\'c}},\ and\ \citenamefont
  {Lukin}}]{everedHighfidelityParallelEntangling2023}%
  \BibitemOpen
  \bibfield  {author} {\bibinfo {author} {\bibfnamefont {Simon~J.}\
  \bibnamefont {Evered}}, \bibinfo {author} {\bibfnamefont {Dolev}\
  \bibnamefont {Bluvstein}}, \bibinfo {author} {\bibfnamefont {Marcin}\
  \bibnamefont {Kalinowski}}, \bibinfo {author} {\bibfnamefont {Sepehr}\
  \bibnamefont {Ebadi}}, \bibinfo {author} {\bibfnamefont {Tom}\ \bibnamefont
  {Manovitz}}, \bibinfo {author} {\bibfnamefont {Hengyun}\ \bibnamefont
  {Zhou}}, \bibinfo {author} {\bibfnamefont {Sophie~H.}\ \bibnamefont {Li}},
  \bibinfo {author} {\bibfnamefont {Alexandra~A.}\ \bibnamefont {Geim}},
  \bibinfo {author} {\bibfnamefont {Tout~T.}\ \bibnamefont {Wang}}, \bibinfo
  {author} {\bibfnamefont {Nishad}\ \bibnamefont {Maskara}}, \bibinfo {author}
  {\bibfnamefont {Harry}\ \bibnamefont {Levine}}, \bibinfo {author}
  {\bibfnamefont {Giulia}\ \bibnamefont {Semeghini}}, \bibinfo {author}
  {\bibfnamefont {Markus}\ \bibnamefont {Greiner}}, \bibinfo {author}
  {\bibfnamefont {Vladan}\ \bibnamefont {Vuleti{\'c}}}, \ and\ \bibinfo
  {author} {\bibfnamefont {Mikhail~D.}\ \bibnamefont {Lukin}},\ }\bibfield
  {title} {\enquote {\bibinfo {title} {High-fidelity parallel entangling gates
  on a neutral-atom quantum computer},}\ }\href {\doibase
  10.1038/s41586-023-06481-y} {\bibfield  {journal} {\bibinfo  {journal}
  {Nature}\ }\textbf {\bibinfo {volume} {622}},\ \bibinfo {pages} {268--272}
  (\bibinfo {year} {2023})}\BibitemShut {NoStop}%
\bibitem [{\citenamefont {Tavis}\ and\ \citenamefont
  {Cummings}(1968)}]{tavisExactSolutionNMoleculeRadiationField1968}%
  \BibitemOpen
  \bibfield  {author} {\bibinfo {author} {\bibfnamefont {Michael}\ \bibnamefont
  {Tavis}}\ and\ \bibinfo {author} {\bibfnamefont {Frederick~W.}\ \bibnamefont
  {Cummings}},\ }\bibfield  {title} {\enquote {\bibinfo {title} {Exact
  {{Solution}} for an \${{N}}\$-{{Molecule---Radiation-Field Hamiltonian}}},}\
  }\href {\doibase 10.1103/PhysRev.170.379} {\bibfield  {journal} {\bibinfo
  {journal} {Physical Review}\ }\textbf {\bibinfo {volume} {170}},\ \bibinfo
  {pages} {379--384} (\bibinfo {year} {1968})}\BibitemShut {NoStop}%
\bibitem [{\citenamefont {Hunger}\ \emph {et~al.}(2010)\citenamefont {Hunger},
  \citenamefont {Steinmetz}, \citenamefont {Colombe}, \citenamefont {Deutsch},
  \citenamefont {H{\"a}nsch},\ and\ \citenamefont
  {Reichel}}]{hungerFiberFabryPerot2010}%
  \BibitemOpen
  \bibfield  {author} {\bibinfo {author} {\bibfnamefont {D}~\bibnamefont
  {Hunger}}, \bibinfo {author} {\bibfnamefont {T}~\bibnamefont {Steinmetz}},
  \bibinfo {author} {\bibfnamefont {Y}~\bibnamefont {Colombe}}, \bibinfo
  {author} {\bibfnamefont {C}~\bibnamefont {Deutsch}}, \bibinfo {author}
  {\bibfnamefont {T~W}\ \bibnamefont {H{\"a}nsch}}, \ and\ \bibinfo {author}
  {\bibfnamefont {J}~\bibnamefont {Reichel}},\ }\bibfield  {title} {\enquote
  {\bibinfo {title} {A fiber {{Fabry}}--{{Perot}} cavity with high finesse},}\
  }\href {\doibase 10.1088/1367-2630/12/6/065038} {\bibfield  {journal}
  {\bibinfo  {journal} {New Journal of Physics}\ }\textbf {\bibinfo {volume}
  {12}},\ \bibinfo {pages} {065038} (\bibinfo {year} {2010})}\BibitemShut
  {NoStop}%
\bibitem [{\citenamefont {Uphoff}\ \emph {et~al.}(2015)\citenamefont {Uphoff},
  \citenamefont {Brekenfeld}, \citenamefont {Rempe},\ and\ \citenamefont
  {Ritter}}]{uphoffFrequencySplittingPolarization2015}%
  \BibitemOpen
  \bibfield  {author} {\bibinfo {author} {\bibfnamefont {Manuel}\ \bibnamefont
  {Uphoff}}, \bibinfo {author} {\bibfnamefont {Manuel}\ \bibnamefont
  {Brekenfeld}}, \bibinfo {author} {\bibfnamefont {Gerhard}\ \bibnamefont
  {Rempe}}, \ and\ \bibinfo {author} {\bibfnamefont {Stephan}\ \bibnamefont
  {Ritter}},\ }\bibfield  {title} {\enquote {\bibinfo {title} {Frequency
  splitting of polarization eigenmodes in microscopic {{Fabry}}--{{Perot}}
  cavities},}\ }\href {\doibase 10.1088/1367-2630/17/1/013053} {\bibfield
  {journal} {\bibinfo  {journal} {New Journal of Physics}\ }\textbf {\bibinfo
  {volume} {17}},\ \bibinfo {pages} {013053} (\bibinfo {year}
  {2015})}\BibitemShut {NoStop}%
\bibitem [{\citenamefont {Kaiser}\ \emph {et~al.}(2022)\citenamefont {Kaiser},
  \citenamefont {Glaser}, \citenamefont {Ley}, \citenamefont {Grimmel},
  \citenamefont {Hattermann}, \citenamefont {Bothner}, \citenamefont {Koelle},
  \citenamefont {Kleiner}, \citenamefont {Petrosyan}, \citenamefont
  {G{\"u}nther},\ and\ \citenamefont
  {Fort{\'a}gh}}]{kaiserCavitydrivenRabiOscillations2022}%
  \BibitemOpen
  \bibfield  {author} {\bibinfo {author} {\bibfnamefont {Manuel}\ \bibnamefont
  {Kaiser}}, \bibinfo {author} {\bibfnamefont {Conny}\ \bibnamefont {Glaser}},
  \bibinfo {author} {\bibfnamefont {Li~Yuan}\ \bibnamefont {Ley}}, \bibinfo
  {author} {\bibfnamefont {Jens}\ \bibnamefont {Grimmel}}, \bibinfo {author}
  {\bibfnamefont {Helge}\ \bibnamefont {Hattermann}}, \bibinfo {author}
  {\bibfnamefont {Daniel}\ \bibnamefont {Bothner}}, \bibinfo {author}
  {\bibfnamefont {Dieter}\ \bibnamefont {Koelle}}, \bibinfo {author}
  {\bibfnamefont {Reinhold}\ \bibnamefont {Kleiner}}, \bibinfo {author}
  {\bibfnamefont {David}\ \bibnamefont {Petrosyan}}, \bibinfo {author}
  {\bibfnamefont {Andreas}\ \bibnamefont {G{\"u}nther}}, \ and\ \bibinfo
  {author} {\bibfnamefont {J{\'o}zsef}\ \bibnamefont {Fort{\'a}gh}},\
  }\bibfield  {title} {\enquote {\bibinfo {title} {Cavity-driven {{Rabi}}
  oscillations between {{Rydberg}} states of atoms trapped on a superconducting
  atom chip},}\ }\href {\doibase 10.1103/PhysRevResearch.4.013207} {\bibfield
  {journal} {\bibinfo  {journal} {Physical Review Research}\ }\textbf {\bibinfo
  {volume} {4}},\ \bibinfo {pages} {013207} (\bibinfo {year}
  {2022})}\BibitemShut {NoStop}%
\bibitem [{\citenamefont {Pritchard}\ \emph {et~al.}(2014)\citenamefont
  {Pritchard}, \citenamefont {Isaacs}, \citenamefont {Beck}, \citenamefont
  {McDermott},\ and\ \citenamefont
  {Saffman}}]{pritchardHybridAtomphotonQuantum2014}%
  \BibitemOpen
  \bibfield  {author} {\bibinfo {author} {\bibfnamefont {J.~D.}\ \bibnamefont
  {Pritchard}}, \bibinfo {author} {\bibfnamefont {J.~A.}\ \bibnamefont
  {Isaacs}}, \bibinfo {author} {\bibfnamefont {M.~A.}\ \bibnamefont {Beck}},
  \bibinfo {author} {\bibfnamefont {R.}~\bibnamefont {McDermott}}, \ and\
  \bibinfo {author} {\bibfnamefont {M.}~\bibnamefont {Saffman}},\ }\bibfield
  {title} {\enquote {\bibinfo {title} {Hybrid atom-photon quantum gate in a
  superconducting microwave resonator},}\ }\href {\doibase
  10.1103/PhysRevA.89.010301} {\bibfield  {journal} {\bibinfo  {journal}
  {Physical Review A}\ }\textbf {\bibinfo {volume} {89}},\ \bibinfo {pages}
  {010301} (\bibinfo {year} {2014})}\BibitemShut {NoStop}%
\bibitem [{\citenamefont {Lei}\ \emph {et~al.}(2020)\citenamefont {Lei},
  \citenamefont {Krayzman}, \citenamefont {Ganjam}, \citenamefont {Frunzio},\
  and\ \citenamefont {Schoelkopf}}]{leiHighCoherenceSuperconducting2020}%
  \BibitemOpen
  \bibfield  {author} {\bibinfo {author} {\bibfnamefont {Chan~U}\ \bibnamefont
  {Lei}}, \bibinfo {author} {\bibfnamefont {Lev}\ \bibnamefont {Krayzman}},
  \bibinfo {author} {\bibfnamefont {Suhas}\ \bibnamefont {Ganjam}}, \bibinfo
  {author} {\bibfnamefont {Luigi}\ \bibnamefont {Frunzio}}, \ and\ \bibinfo
  {author} {\bibfnamefont {Robert~J.}\ \bibnamefont {Schoelkopf}},\ }\bibfield
  {title} {\enquote {\bibinfo {title} {High coherence superconducting microwave
  cavities with indium bump bonding},}\ }\href {\doibase 10.1063/5.0003907}
  {\bibfield  {journal} {\bibinfo  {journal} {Applied Physics Letters}\
  }\textbf {\bibinfo {volume} {116}},\ \bibinfo {pages} {154002} (\bibinfo
  {year} {2020})}\BibitemShut {NoStop}%
\bibitem [{\citenamefont {Brune}\ \emph {et~al.}(1996)\citenamefont {Brune},
  \citenamefont {Hagley}, \citenamefont {Dreyer}, \citenamefont {Ma{\^\i}tre},
  \citenamefont {Maali}, \citenamefont {Wunderlich}, \citenamefont {Raimond},\
  and\ \citenamefont {Haroche}}]{bruneObservingProgressiveDecoherence1996}%
  \BibitemOpen
  \bibfield  {author} {\bibinfo {author} {\bibfnamefont {M.}~\bibnamefont
  {Brune}}, \bibinfo {author} {\bibfnamefont {E.}~\bibnamefont {Hagley}},
  \bibinfo {author} {\bibfnamefont {J.}~\bibnamefont {Dreyer}}, \bibinfo
  {author} {\bibfnamefont {X.}~\bibnamefont {Ma{\^\i}tre}}, \bibinfo {author}
  {\bibfnamefont {A.}~\bibnamefont {Maali}}, \bibinfo {author} {\bibfnamefont
  {C.}~\bibnamefont {Wunderlich}}, \bibinfo {author} {\bibfnamefont {J.~M.}\
  \bibnamefont {Raimond}}, \ and\ \bibinfo {author} {\bibfnamefont
  {S.}~\bibnamefont {Haroche}},\ }\bibfield  {title} {\enquote {\bibinfo
  {title} {Observing the {{Progressive Decoherence}} of the ``{{Meter}}'' in a
  {{Quantum Measurement}}},}\ }\href {\doibase 10.1103/PhysRevLett.77.4887}
  {\bibfield  {journal} {\bibinfo  {journal} {Physical Review Letters}\
  }\textbf {\bibinfo {volume} {77}},\ \bibinfo {pages} {4887--4890} (\bibinfo
  {year} {1996})}\BibitemShut {NoStop}%
\bibitem [{\citenamefont {Signoles}\ \emph {et~al.}(2017)\citenamefont
  {Signoles}, \citenamefont {Dietsche}, \citenamefont {Facon}, \citenamefont
  {Grosso}, \citenamefont {Haroche}, \citenamefont {Raimond}, \citenamefont
  {Brune},\ and\ \citenamefont
  {Gleyzes}}]{signolesCoherentTransferLowAngularMomentum2017}%
  \BibitemOpen
  \bibfield  {author} {\bibinfo {author} {\bibfnamefont {A.}~\bibnamefont
  {Signoles}}, \bibinfo {author} {\bibfnamefont {E.~K.}\ \bibnamefont
  {Dietsche}}, \bibinfo {author} {\bibfnamefont {A.}~\bibnamefont {Facon}},
  \bibinfo {author} {\bibfnamefont {D.}~\bibnamefont {Grosso}}, \bibinfo
  {author} {\bibfnamefont {S.}~\bibnamefont {Haroche}}, \bibinfo {author}
  {\bibfnamefont {J.~M.}\ \bibnamefont {Raimond}}, \bibinfo {author}
  {\bibfnamefont {M.}~\bibnamefont {Brune}}, \ and\ \bibinfo {author}
  {\bibfnamefont {S.}~\bibnamefont {Gleyzes}},\ }\bibfield  {title} {\enquote
  {\bibinfo {title} {Coherent {{Transfer}} between {{Low-Angular-Momentum}} and
  {{Circular Rydberg States}}},}\ }\href {\doibase
  10.1103/PhysRevLett.118.253603} {\bibfield  {journal} {\bibinfo  {journal}
  {Physical Review Letters}\ }\textbf {\bibinfo {volume} {118}},\ \bibinfo
  {pages} {253603} (\bibinfo {year} {2017})}\BibitemShut {NoStop}%
\bibitem [{\citenamefont {Mozley}\ \emph {et~al.}(2005)\citenamefont {Mozley},
  \citenamefont {Hyafil}, \citenamefont {Nogues}, \citenamefont {Brune},
  \citenamefont {Raimond},\ and\ \citenamefont
  {Haroche}}]{mozleyTrappingCoherentManipulation2005}%
  \BibitemOpen
  \bibfield  {author} {\bibinfo {author} {\bibfnamefont {J.}~\bibnamefont
  {Mozley}}, \bibinfo {author} {\bibfnamefont {P.}~\bibnamefont {Hyafil}},
  \bibinfo {author} {\bibfnamefont {G.}~\bibnamefont {Nogues}}, \bibinfo
  {author} {\bibfnamefont {M.}~\bibnamefont {Brune}}, \bibinfo {author}
  {\bibfnamefont {J.-M.}\ \bibnamefont {Raimond}}, \ and\ \bibinfo {author}
  {\bibfnamefont {S.}~\bibnamefont {Haroche}},\ }\bibfield  {title} {\enquote
  {\bibinfo {title} {Trapping and coherent manipulation of a {{Rydberg}} atom
  on a microfabricated device: A proposal},}\ }\href {\doibase
  10.1140/epjd/e2005-00184-7} {\bibfield  {journal} {\bibinfo  {journal} {The
  European Physical Journal D}\ }\textbf {\bibinfo {volume} {35}},\ \bibinfo
  {pages} {43--57} (\bibinfo {year} {2005})}\BibitemShut {NoStop}%
\bibitem [{\citenamefont {Cohen}\ and\ \citenamefont
  {Thompson}(2021)}]{cohenQuantumComputingCircular2021}%
  \BibitemOpen
  \bibfield  {author} {\bibinfo {author} {\bibfnamefont {Sam~R.}\ \bibnamefont
  {Cohen}}\ and\ \bibinfo {author} {\bibfnamefont {Jeff~D.}\ \bibnamefont
  {Thompson}},\ }\bibfield  {title} {\enquote {\bibinfo {title} {Quantum
  {{Computing}} with {{Circular Rydberg Atoms}}},}\ }\href {\doibase
  10.1103/PRXQuantum.2.030322} {\bibfield  {journal} {\bibinfo  {journal} {PRX
  Quantum}\ }\textbf {\bibinfo {volume} {2}},\ \bibinfo {pages} {030322}
  (\bibinfo {year} {2021})}\BibitemShut {NoStop}%
\bibitem [{\citenamefont {Buhmann}\ \emph {et~al.}(2008)\citenamefont
  {Buhmann}, \citenamefont {Tarbutt}, \citenamefont {Scheel},\ and\
  \citenamefont {Hinds}}]{buhmannSurfaceinducedHeatingCold2008}%
  \BibitemOpen
  \bibfield  {author} {\bibinfo {author} {\bibfnamefont {Stefan~Yoshi}\
  \bibnamefont {Buhmann}}, \bibinfo {author} {\bibfnamefont {M.~R.}\
  \bibnamefont {Tarbutt}}, \bibinfo {author} {\bibfnamefont {Stefan}\
  \bibnamefont {Scheel}}, \ and\ \bibinfo {author} {\bibfnamefont {E.~A.}\
  \bibnamefont {Hinds}},\ }\bibfield  {title} {\enquote {\bibinfo {title}
  {Surface-induced heating of cold polar molecules},}\ }\href {\doibase
  10.1103/PhysRevA.78.052901} {\bibfield  {journal} {\bibinfo  {journal}
  {Physical Review A}\ }\textbf {\bibinfo {volume} {78}},\ \bibinfo {pages}
  {052901} (\bibinfo {year} {2008})}\BibitemShut {NoStop}%
\bibitem [{\citenamefont {Rabl}\ \emph {et~al.}(2006)\citenamefont {Rabl},
  \citenamefont {DeMille}, \citenamefont {Doyle}, \citenamefont {Lukin},
  \citenamefont {Schoelkopf},\ and\ \citenamefont
  {Zoller}}]{rablHybridQuantumProcessors2006}%
  \BibitemOpen
  \bibfield  {author} {\bibinfo {author} {\bibfnamefont {P.}~\bibnamefont
  {Rabl}}, \bibinfo {author} {\bibfnamefont {D.}~\bibnamefont {DeMille}},
  \bibinfo {author} {\bibfnamefont {J.~M.}\ \bibnamefont {Doyle}}, \bibinfo
  {author} {\bibfnamefont {M.~D.}\ \bibnamefont {Lukin}}, \bibinfo {author}
  {\bibfnamefont {R.~J.}\ \bibnamefont {Schoelkopf}}, \ and\ \bibinfo {author}
  {\bibfnamefont {P.}~\bibnamefont {Zoller}},\ }\bibfield  {title} {\enquote
  {\bibinfo {title} {Hybrid {{Quantum Processors}}: {{Molecular Ensembles}} as
  {{Quantum Memory}} for {{Solid State Circuits}}},}\ }\href {\doibase
  10.1103/PhysRevLett.97.033003} {\bibfield  {journal} {\bibinfo  {journal}
  {Physical Review Letters}\ }\textbf {\bibinfo {volume} {97}},\ \bibinfo
  {pages} {033003} (\bibinfo {year} {2006})}\BibitemShut {NoStop}%
\end{thebibliography}%
\end{document}